\documentclass[preprint2]{aastex}

\shorttitle{Microlensing in Spectra of Multiply Lensed QSO}
\shortauthors{Motta et al.}

\usepackage{epsfig}

\begin{document}

\title{Measuring Microlensing using Spectra of Multiply Lensed Quasars}

\author{V. Motta}
\affil{Departamento de F\'{\i}sica y Astronom\'{\i}a, 
Universidad de Valpara\'{\i}so, 
Avda. Gran Breta\~na 1111, Playa Ancha, Valpara\'{\i}so 2360102, Chile} 
\email{vmotta@dfa.uv.cl}

\author{E. Mediavilla}
\affil{Instituto de Astrof\'{\i}sica de Canarias, 
Avda. V\'{\i}a L\'actea s/n, 
La Laguna, 
Tenerife 38200, Spain}
\affil{Departamento de Astrof\'{\i}sica, 
Universidad de La Laguna, 
La Laguna, 
Tenerife 38205, Spain}
\email{emg@iac.es}

\author{E. Falco}
\affil{Whipple Observatory, 
Smithsonian Institution, 
670 Mt. Hopkins Road, P.O. Box 97,
Amado, Arizona 85645, USA}
\email{falco@cfa.harvard.edu}

\and

\author{J.A. Mu\~noz}
\affil{Departamento de Astronom\'{\i}a y Astrof\'{\i}sica, 
Universidad de Valencia, 
46100-Burjassot,
Valencia, Spain}
\email{jmunoz@uv.es}

\begin{abstract}

We report on a program of spectroscopic observations of
gravitationally-lensed QSOs with multiple images. We seek to establish
whether microlensing is occurring in each QSO image using only  
single-epoch observations.  We calculate flux ratios for the cores of
emission lines in image pairs to set a baseline for no
microlensing.  The offset of the continuum flux ratios relative to
this baseline yields the microlensing magnification free from
extinction, as extinction affects the continuum and the lines equally.
When we find chromatic microlensing,
we attempt to constrain the size of the QSO accretion disk. 
SDSSJ1004+4112 and HE1104-1805 show chromatic microlensing with
amplitudes $0.2< |\Delta m| < 0.6$ and $0.2< |\Delta m| < 0.4$ mag,
respectively. Modeling the accretion disk with a Gaussian source
($I\propto \exp(-R^2/2r_s^2)$) of size $r_s\propto \lambda^p$ and
using magnification maps to simulate microlensing we find 
$r_s(\lambda 3363)=7\pm3 \, \rm light-days 
(18.1\pm7.8 \times 10^{15} \, \rm cm$) 
and $p=1.1\pm 0.4$ for SDSS1004+4112, and
$r_s(\lambda 3363)=6\pm2 \, \rm light-days 
(15.5\pm5.2 \times 10^{15} \, \rm cm$) 
and $p=0.7\pm0.1$ for HE1104-1805. 
For SDSSJ1029+2623 we find strong chromaticity of $\sim 0.4$ mag in the
continuum flux ratio, which probably arises from
microlensing although not all the available data fit within this
explanation. 
For Q0957+561 we measure B-A magnitude differences of 0.4 mag, 
much greater than the $\sim$0.05 mag
amplitude usually inferred from lightcurve variability. It may 
substantially modify the current interpretations of microlensing in
this system, likely favoring the hypothesis of smaller sources and/or
larger microdeflectors.
For HS0818+1227, our data yield posible evidence of microlensing.

 \end{abstract}



\section{Introduction}

Gravitational lenses are a powerful tool to study not only the
structure of the lensed quasar but also the composition of the lens
galaxy \citep{schneider92,kochanek04a,wambsganss06}.  Simple lens
models are usually sufficient to reproduce the positions of lensed QSO
images, but they can fail to reproduce the optical fluxes of these
images. The so-called flux ratio anomalies are thought to be produced
by small-scale structures in the gravitational potential of lens
galaxies \citep{witt95,mao98,chiba02,metcalf01,dalal02,
 schechter02,keeton02,bradac02,metcalf02,moustakas03}. These
structures are either dark matter subhalos or stars and the effects
they produce are referred to as millilensing and microlensing
respectively.

A substructure is able to produce a flux anomaly if the radius of its
Einstein ring is large compared to the emitting region. Since the
sizes of the quasar continuum emitting regions depend on the
wavelength, microlensing by stars in a lens galaxy will yield a
wavelength-dependent magnification of the continuum
\citep{wambsganss91,wisotzki95,mosquera09,mediavilla11} that can be
strong for the UV and the optical but is negligible for the
IR. Microlensing could also affect the high ionization broad emission
lines (BEL) that are expected to arise from the inner part of the
broad-line region (BLR).  Specifically, microlensing would affect the
broad wings of the profiles of the high ionization lines that
correspond to high velocity emitters, leaving unchanged the core
\citep{popovic01,abajas02,richards04,lewis04,gomez06}. Low ionization
BEL and narrow emission lines (NEL) arise from considerably larger
regions and are supposed to be insensitive to microlensing
\citep{abajas02} although the low ionization BEL of some 
low-luminosity lensed AGNs may be slightly affected by microlensing.

Observational studies aimed at measuring the microlensing effect in
lensed QSOs usually consist of broadband observations 
repeated over extended
periods, longer than the time delays for image pairs
\citep{wozniak00,oscoz01,colley02,schechter03,fohlmeister08}, leading 
to several years of monitoring.

On the other hand, most of the microlensing searches using optical
imaging have been concentrated on quadruple lenses because the effect
of substructure is more important at high magnification
\citep{witt95,schechter02,pooley07}. They provide enough
constraints to fit the simplest singular isothermal (SIS) model
\citep{schechter02,kochanek04} and find the
flux anomalies. Double lenses, however, do not provide enough constraints to
produce such a model unless the fluxes are used as additional
constraints. In these cases, the flux ratio of emission lines
\citep{wisotzki93,mediavilla09,mediavilla11} or in the infrared
\citep{agol00} has been used assuming that the emission regions are
larger than the microlensing source and dust extinction is negligible.

In the present paper, we use spectra of lensed quasars as an alternative
approach to photometric monitoring, to study microlensing in quadruple
or double lenses.  For each pair of images of a lensed quasar ($A$ and
$B$), we base our analysis on the measurement of the offsets of the
flux ratio of the core of the emission lines $(F_B/F_A)_{core}$
compared to the flux ratio of the continuum $(F_B/F_A)_{cont}$. This
analysis allows us to distinguish between microlensing and dust
extinction without assuming a model for the lensed system
\cite[see][and references therein]{mediavilla11}. In this way, a
single-epoch spectroscopic observation can suffice --through the
measurement of microlensing-- to estimate physical parameters of
interest of the lens galaxy (like the fraction of mass in compact
  objects, Mediavilla et al. 2009) or of the unresolved quasar source
(like the size or the radial temperature profile).

In section 2 we present the data for 5 gravitationally-lensed quasars
with multiple images for which we have obtained low-resolution spectra
with signal-to-noise greater than 40. The systems, HS0818+1227,
Q0957+561, SDSS1004+4112, SDSS1029+2623, and HE1104-1805, were
selected because the separations between the images were larger
than 3$\arcsec$. Section 3 is devoted to present the data analysis methodology. 
We discuss our results in \S4 and give some concluding remarks in section \S5.

\section{Observations and data reduction}

Microlensing detection using spectra of lensed QSOs has stringent
requirements.  First of all, we need high signal-to-noise ratio in the
spectra ($SNR\simeq40$) and sufficient spectral resolution
($v_{res}<200$ km s$^{-}$) to resolve the shape of the line
profiles. Second, we need to obtain simultaneous, spatially-separated spectra
of pairs of lensed QSO images to compare their continuum and emission
lines at different wavelengths, which requires good seeing conditions
($\le 0.8 \arcsec$). These requirements are achieved using 6-8 m class
telescopes under good seeing conditions.

We observed the sample on 11 and 12 January 2008 with the Blue 
Channel spectrograph on the MMT. Table \ref{obs} shows the 
log of observations. 
We also observed HE1104-1805 on 07 April 
2008 with the FORS2 spectrograph at the Very Large Telescope (VLT).
Our ground-based observations were acquired under excellent atmospheric
conditions (Table \ref{obs}).  
For Q0957+561, we used archival data\footnote{archive data 
were obtained at the Space Telescope Science Institute, operated by the 
Association of Universities for Research in Astronomy, Inc., under NASA 
contract NAS 5-26555} obtained with the STIS spectrograph on the Hubble 
Space Telescope (HST). Components $A$ and $B$ were observed with HST at 
different 
epochs to account for time delay variations in the continuum spectra (see 
Table \ref{obs}). A detailed description of these observations and the 
spectrum analysis can be found in \cite{hutchings03}.

We performed the data reduction with IRAF\footnote{IRAF is 
distributed by the National Optical Astronomy Observatory, which is operated 
by the Association of Universities for Research in Astronomy, Inc., under 
cooperative agreement with the National Science Foundation} tasks. These
included bias subtraction, flat fielding, extraction of 1-D spectra and 
wavelength calibration. As we are currently interested only in flux 
ratios (i.e. magnitude difference $m_B-m_A=-2.5 \log (F_B/F_A)$), 
we did not flux-calibrate our data. 
Cosmic-ray rejection was 
carried out in those cases where we had at least three exposures. 
The data obtained from the HST archive are already fully reduced.

In spite of the careful data reduction process, some systematic errors 
can affect our measurements. We discuss these in the following paragraphs.

\subsection{Spectrum Cross-Contamination}

To avoid cross-contamination between the spectra of lensed QSO image pairs we 
selected pairs with separation much wider than the average seeing 
($\le 0\farcs7$), which was larger in turn than our typical seeing.
Our pairs have separations ranging from $2\farcs6$ to 
$22\farcs6$ as shown in Table \ref{obs}; we estimate that 
cross-contamination is negligible in our observations. 

\subsection{Long-Slit Losses}

To obtain simultaneous pairs of spectra with a single slit, 
we did not observe our targets  
at the parallactic angle; in each case we used the position 
angle defined by the two components. In these ground-based observations,
we lost a small amount of the blue part 
of the spectra because part of the blue quasar light may fall outside the slit.
Our airmass range was $1.02 - 1.77$; we used a 
$1\farcs0$ slit width. 
We used a program developed by E. Marchetti at 
ESO\footnote{http://www.eso.org/gen-fac/pubs/astclim/lasilla/} 
to calculate the differential 
atmospheric refraction (DAR) for each given wavelength and airmass. 
The atmospheric parameters (temperature, humidity, and pressure) we used are 
$T=11.5$ $^o$C, $H=14.5$ \%, $P=743$ mbar for VLT and 
$T=12.0$ $^o$C, $H=13.4$ \%, $P=741.6$ mbar for MMT. 
The focus for FORS2 is set at 5000 \AA \, which is our 
reference wavelength to calculate the displacement due to DAR. 
We estimated  
the relative loss in the flux of each pair of 
images at 3500 \AA \, and 8000 \AA ,  
and at 4500 \AA \, and 9500 \AA \, for the MMT and VLT data respectively. We 
found relative losses of $<1$\% in all MMT spectra; for the 
VLT spectra, we found losses of $<9$\% 
due to an error in the position angle used in the observations.
Considering 
the separation between pairs of images ($2\farcs6$ to $22\farcs6$), 
the losses are nearly identical for both spectra in each case, 
and as we are concerned with flux ratio changes with wavelength, those 
losses do not affect our results.

\section{Data Analysis Methods and Uncertainties}

\subsection{Continuum microlensing measurement} \label{mulens}

The method we use to untangle microlensing and extinction is based on
the measurement of the offsets between the continuum and the emission
line flux ratios \citep[see,
  e.g.,][]{vanderriest90,motta02,wucknitz03,wisotzki03,mediavilla09,mediavilla11,sluse11}.
The multicomponent nature of quasar emission-lines imply that the
emission-line spectra are produced over a wide range of distances from
the central continuum \cite[see e.g.][]{sulentic00}.  According to
\cite{marziani10} the low ionization lines (LIL) and the core of the
high ionization lines (HIL) will be dominated by a component (FWHM
$\sim 600-5000 \, \rm km \, \rm s^{-1}$) that corresponds to the
region of reverberation mapping typically large enough as to be
insensitive to microlensing by solar mass objects. The broad wings of
the emission lines (FWHM $\sim 10000 \, \rm km \, \rm s^{-1}$),
however, could arise from the inner parts of the BLR and may be
microlensed. For this reason, we prefer to use exclusively the line
cores (dominated by the NLR and the outer regions of the BLR) as
reference to set the baseline for no microlensing. To compute the core
flux without attempting an analytical decomposition into several
components \citep{marziani10}, we have used a narrow band
decomposition similar to that used by \cite{sluse11}.  Specifically,
we define as core flux the continuum subtracted flux integrated in a
relatively narrow velocity interval (from 25 to 90~\AA \ depending on
the line profile shape for the different sources) centered on the peak
of the line.  
To accommodate the varying widths of the lines, 
the continuum estimate for each line requires
windows with varying width as indicated below for each lens system.

For each component and each emission line we used DIPSO
\citep{howarth04} in STARLINK\footnote{Support provided by the
  Starlink Project which is run by CCLRC on behalf of PPARC.} to fit a
function $y_c = a \lambda + b$ to the continuum on either side of the
emission line, given a total wavelength range ($\lambda _A$, $\lambda
_B$). The task also gives the error coefficients ($\Delta a$, $\Delta
b$) in the continuum fitting.  This error is largest in the bluest and
reddest ends of the continuum, because it is affected by the response
of the CCD.  The flux under the continuum is then obtained as the
integral below the fitted function $y_c$, i.e. $F_c=(a/2) (\lambda_B -
\lambda _A)^2 + b (\lambda_B - \lambda_A)$. The error in the flux is
estimated as $\Delta F_c = (\Delta a /2)(\lambda_B - \lambda_A)^2 +
\Delta b (\lambda_B - \lambda_A)$.

The emission line flux is obtained by integrating the emission line
profiles in each continuum-subtracted emission line using DIPSO. As
commented above, we have separated the line core from the wings which
could be affected by microlensing. The error in the narrow emission
line is estimated as the error in the continuum fitting. In those
cases in which the emission line is affected by absorption lines, a
narrower integration window was chosen (10 to 15~\AA \ in the case of
SDSS1029+2623).  In most of the cases in which these absorptions are
mild, they can be successfully avoided.  However, when the absorptions
are broad and affect the central part of the emission lines
(e.g. SDSS1029+2623) the measurements have correspondingly larger
uncertainties.

\subsection{Impact of microlensing in the BEL}

\cite{nemiroff88} and \cite{schneider90} suggested that, depending on the
structure of the BLR, microlensing could modify the broad line profiles. 
\cite{abajas02} have estimated which gravitational lens systems are more likely
to show BLR changes due to microlensing. Some examples of these variations in
the BLR have been presented by 
\cite{filippenko89,chartas02,chartas04,richards04,gomez06,sluse11} 
in the cases of MG0414+0534, H1413+117, SDSS1004+4112,
and Q2237+0305.

Superposition of the spectra for each image pair (see Figures 1,
  6, 7, 11, 13, and 16) shows excellent matches between the emission
  lines profiles of HE1104-1805, SDSS1029+2623, Q0957+561 and
  HS0818+1227. Significant microlensing of BELs is detected only in
the wings of one of the systems, SDSS1004+4112 (see \S
\ref{sdss1004}). A slight enhancement of the red wings that may be
tentatively related to microlensing has been also detected in
HE1104-1805 (see \S \ref{he1104}).  Thus, the impact of microlensing
on the BEL looks negligible except in SDSS1004+4112.  The
  excellent matches between the emission line profiles for each image
  pair imply that, except for fluctuations due to absorptions or
  noise, the choice of the size of the line core has no impact on the
  results. To show this explicitly, we have compared the A-B magnitudes
  (averaged on all the lines for each system) computed using only the
  line core or the whole line (as estimated from the factor used to
  match the line profiles). We find differences of
  $|(A-B)_{cores}-(A-B)_{whole-lines}|$=0.02, 0.03, 0.01, and 0.06 for
  HE1104, SDSS1029, QSO0957, and HE0818, respectively. Thus, we find
  that for all the systems except SDSS1004+4112 the A-B emission line
  ratios do not significantly depend on the choice of the core width.
In any case, we have used the core of the emission lines exclusively 
to estimate the emission line flux ratios (see section \ref{mulens}).

\subsection{Estimate of accretion disk parameters } \label{disk-size}

For those cases in which chromatic microlensing is detected, we can study the
structure of the accretion disk in the lensed quasar by estimating its size and
temperature profile. We model the accretion disk as a Gaussian, $I\propto
\exp(-R^2/2r_s^2)$, with radius variable with wavelength, $r_s\propto
\lambda^p$. To estimate the probability of reproducing the measured microlensing
magnifications we have randomly placed a Gaussian source on microlensing
magnification maps 
of $30 \times 30$ Einstein Radii squared ($1000
\times 1000$ pixels) for SDSS1004+4112 and $58.8 \times 58.8$ Einstein Radii squared 
($2000 \times 2000$ pixels) for HE1104-1805 
computed for each image using the Inverse Polygon Mapping
method \citep{mediavilla06}. The convergence ($\kappa$) and shear ($\gamma$) for
each image are selected from available models in the literature \cite[see
e.g.][]{kochanek06,mediavilla09}.  We take $\alpha=0.1$ for the fraction of mass
in compact objects, a reasonable value according to current estimates \cite[see
e.g.][]{schechter02,mediavilla09,pooley09}.  We consider $1 \ M_{\sun}$ 
microlenses. Following a Bayesian approach as in \cite{mediavilla11}, we
estimate the probability of $r_s$ and $p$ conditioned on the measured
microlensing magnifications for both uniform and logarithmic priors on $r_s$. We
have considered these two priors to analyze the sensitivity of our study to the
treatment of the size prior \cite[see][]{morgan10,mediavilla11}. 
We consider a range of 1 to 15 light-days ($2.6-38.9\times10^{15} \, \rm cm$) for $r_s$ and a range of
0 to 3 for $p$. In \S \ref{results} we will apply this method to SDSS1004+4112 and
HE1104-1805.
The results for $r_s$ and $p$ are given with $1 \sigma$ errors.

\subsection{Dust extinction fitting}

Each lensed QSO image follows a different path through the lens
galaxy, encountering different amounts of dust and gas that produce
differential extinction. \cite{falco99} measured the mean differential
extinction, $\Delta E (B-V)$, in 23 lens galaxies using HST broad-band
filters. This extinction can affect not only the continuum flux ratio
but also the emission-line fluxes
\citep{motta02,mediavilla05,mediavilla09,mediavilla11}.  Thus,
considering that the cores of the emission lines are affected neither
by microlensing nor by intrinsic variability, measuring the emission
lines flux ratio in several wavelengths provides us with a method to
determine the existence of dust extinction in the system. We fitted
the extinction curve to the magnitude difference in emission lines for
images 1 and 2 using the equation
\citep{falco99,munoz04} $$m_1(\lambda) - m_2(\lambda)=-2.5 \log \left(
\frac{M_1}{M_2} \right) + (E_1-E_2) R_V\left( \frac{\lambda}{1+z_L}
\right),$$ where $M_1/M_2$ is the constant magnification ratio,
$E_1-E_2=\Delta E$ is the extinction difference, and $R_V[
  \lambda/(1+z_L) ]$ is the extinction curve in the lens rest frame.
We minimized $\chi^2$ per number of degree of freedom
($\chi^2_{DOF}$).  In the majority of the systems (except Q0957+561)
we have only a few narrow emission lines in the optical part of the
spectra. We make our estimates with the \cite{cardelli89} extinction
curve of the Milky Way (i.e. we fixed the parameter $R_V=3.1$) at the
redshift of the lens galaxy. As is standard and to facilitate
comparison of our results with those of other authors, magnitude
differences are shown as a function of inverse wavelength in microns
in the lens galaxy rest frame.

\subsection{Contamination from other sources of chromaticity}

There are two effects, intrinsic variability and contamination by the lens
galaxy, that can
produce chromatic variations in the flux of lensed QSOs and, hence,  
mimic microlensing. 

The continuum flux variation in QSOs is a well-know effect that does
not significantly affect the NEL fluxes \citep{peterson93}. Intrinsic
continuum variability combined with the time delay between images can
produce a change in the flux ratios between images that can be
wavelength dependent, thus inducing changes in the chromaticity. These
changes should be avoided if possible or at least estimated. In two of
the objects, Q0957+561 and HE1104-1805 we can use data taken at two
different epochs separated by the time delay to avoid the problem of
intrinsic variability.

In all the objects, we can estimate the effects of intrinsic
variability \cite[following][]{yonehara08} using the structure
function inferred from the SDSS imaging data of quasars
\citep{vandenberk04,ivezic04}. We will consider the less favorable
case; an intrinsic magnitude of $\rm M_I=-21$ for the quasar
\citep{yonehara08}, the bluest photometric band to measure variability
and the two bands with the largest separation in wavelength to
estimate the chromaticity variation.  In the case of SDSS1004+4112A,B
with a measured time-delay of about 40 days, the expected intrinsic
variability is $\lesssim 0.1$ mag and the chromaticity change is
$\lesssim 0.03$ mag. For HE1104-1805 with a measured time delay of
about 150 days and for HS0818+1227 with a comparable theoretical
delay, variability of $\sim 0.1$ mag and chromaticity change $\lesssim
0.05$ mag are predicted. Finally, for the largest separation systems,
Q0957+561 and SDSS1029+2623, variability of $\lesssim 0.2$ mag and
chromaticity change of $\lesssim 0.08$ mag are expected. Thus, changes
in chromaticity, that are most significant to study the quasar
structure, are rather small.

The expected values of the intrinsic variability are in reasonable
agreement with the analysis of lightcurves for Q0957+561
\citep{goicoechea08,goicoechea02,ovaldsen03a,ovaldsen03b},
SDSS1004+4112 \citep{fohlmeister08} and HE1104-1805
\citep{poindexter07}. In the case of HE1104-1805, \cite{poindexter07}
specifically studied the effect of intrinsic variability on flux
ratios finding a global displacement of 0.1 mag in magnitude
differences due to the time delay without apparent changes in
chromaticity in the optical (from the J to B photometric bands).

In summary, we used photometry corrected for known time delays 
to avoid the
effects induced by intrinsic variability in Q0957+561 and
HE1104-1805. These effects are within the uncertainties for
SDSS1004+4112 and, likely, for HS0818+1227 (although we lack on a
measured time delay for this object). Finally, SDSS1029+2623 can
potentially have relatively strong effects ($\sim 0.1$ mag change in
chromaticity) induced by intrinsic variability.

On the other hand, in the cases of HS0818+1227 and Q0957+561 the lens galaxy is
bright and very close to one of the components on the sky ($0\farcs6$ and
$1\farcs0$ respectively) and some of the spectra may  suffer
contamination from the continuum of the lens galaxy.  This continuum
contamination is stronger at longer wavelengths, but it does not affect the
emission line fluxes. In these two cases, to avoid the continuum flux
contamination we have considered the broad-band flux ratio obtained by 
CASTLES\footnote{CfA-Arizona Space Telescope LEns Survey, 
Kochanek, C.S., Falco, E.E., Impey, C., Lehar, J., McLeod, B., 
Rix H.-W., http://www.cfa.harvard.edu/glensdata/}
using HST imaging, in which the lens galaxy was modeled and subtracted.

\section{Results} \label{results}

\subsection{SDSS1004+4112} \label{sdss1004}

SDSS1004+4112 is a five-image lens system at $z_s=1.734$ discovered by
\cite{inada03} with distances between components ranging from
$3\farcs7$ to $14\farcs6$. The lens is a cluster at $z_l=0.68$
\citep{oguri04,inada08}, which has also been studied in X-rays
\citep{ota06}. This system has known CIV broad-line profile variations
\citep{richards04} that are argued to arise either from microlensing
\citep{richards04, gomez06,abajas07} or due to small line-of-sight
differences through the quasar absorbing outflows \citep{green06}.
Recently, \cite{fohlmeister08} have measured a time delay of
$40.6\pm1.8$ days for images $A$ and $B$, and $822\pm2$ days for $C$
and $D$, detecting microlensing variability with an amplitude of 
the order of $0.15$~mag between $A$ and $B$ \citep{fohlmeister08}.

\begin{figure*}
\begin{center}
\epsfig{file=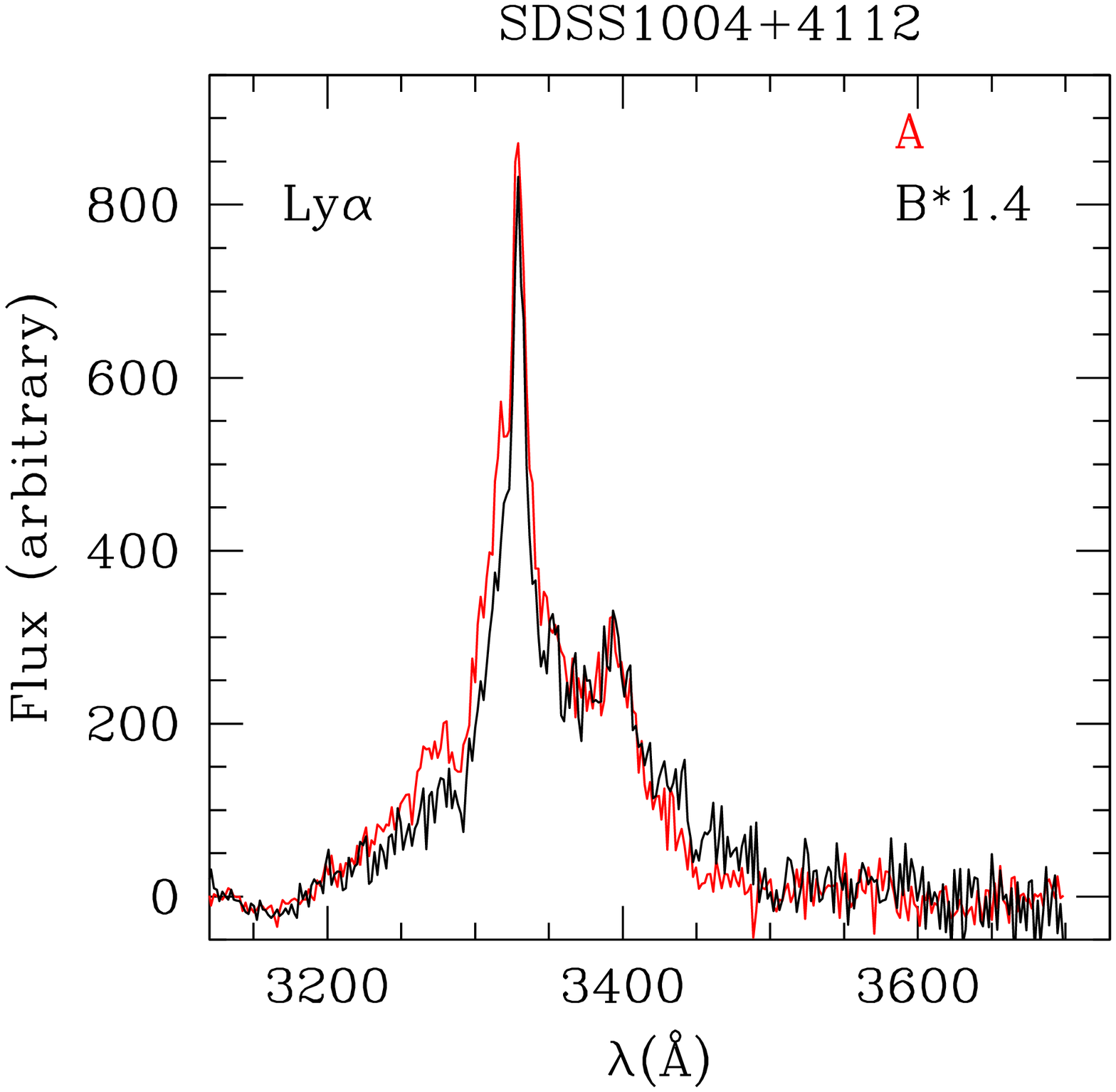,width=5cm,angle=0}
\epsfig{file=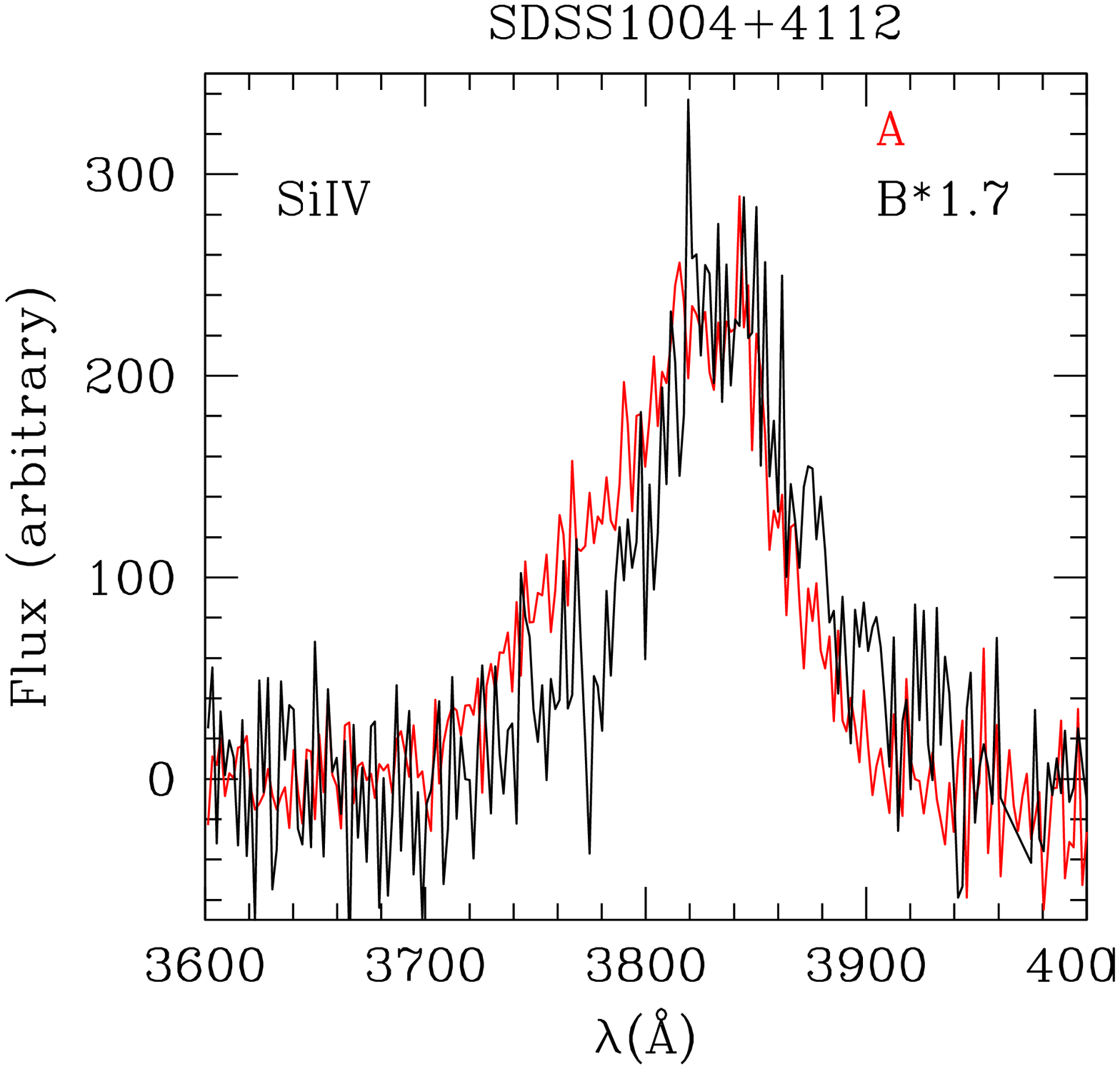,width=5cm,angle=0}
\epsfig{file=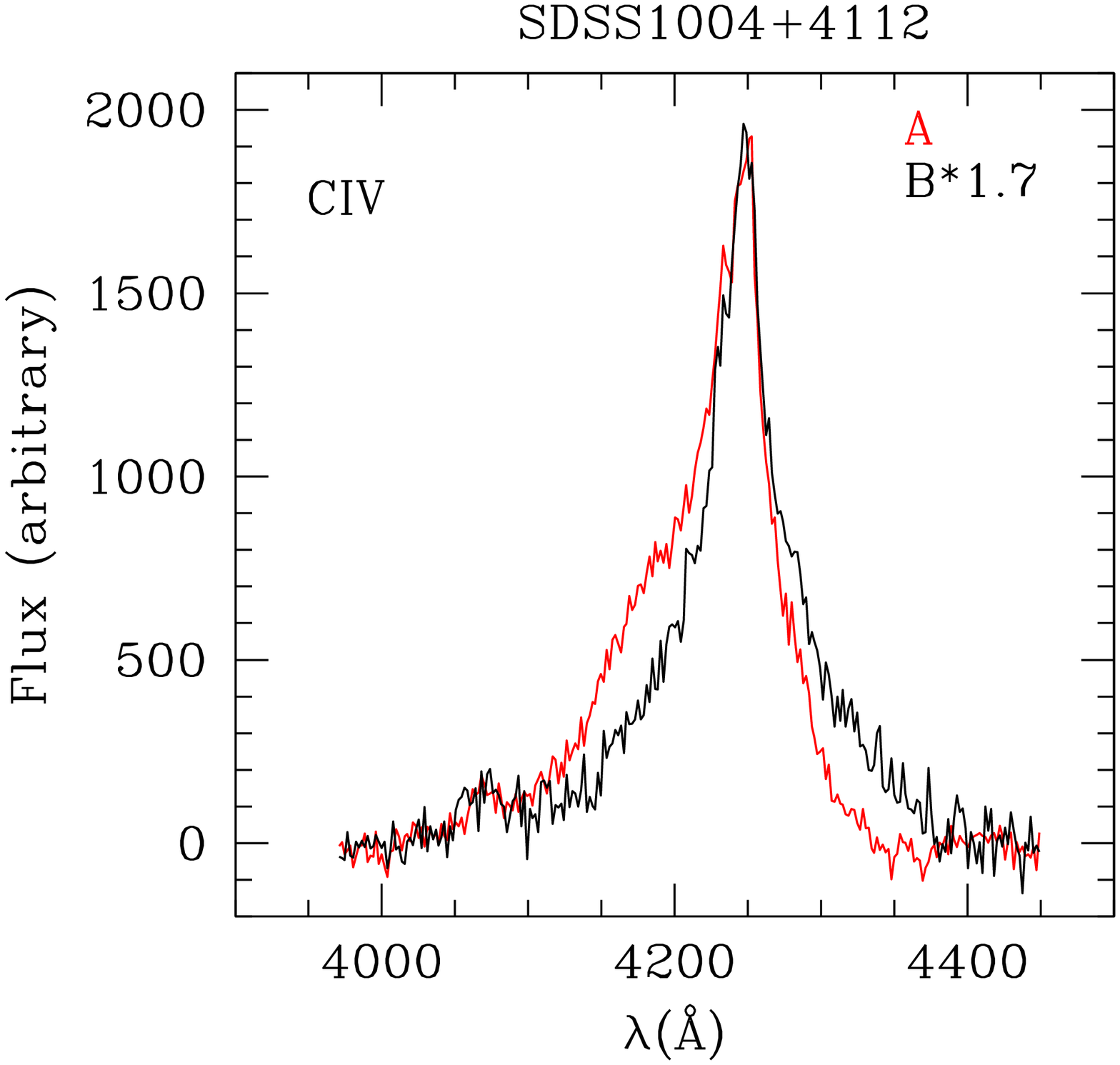,width=5cm,angle=0}
\epsfig{file=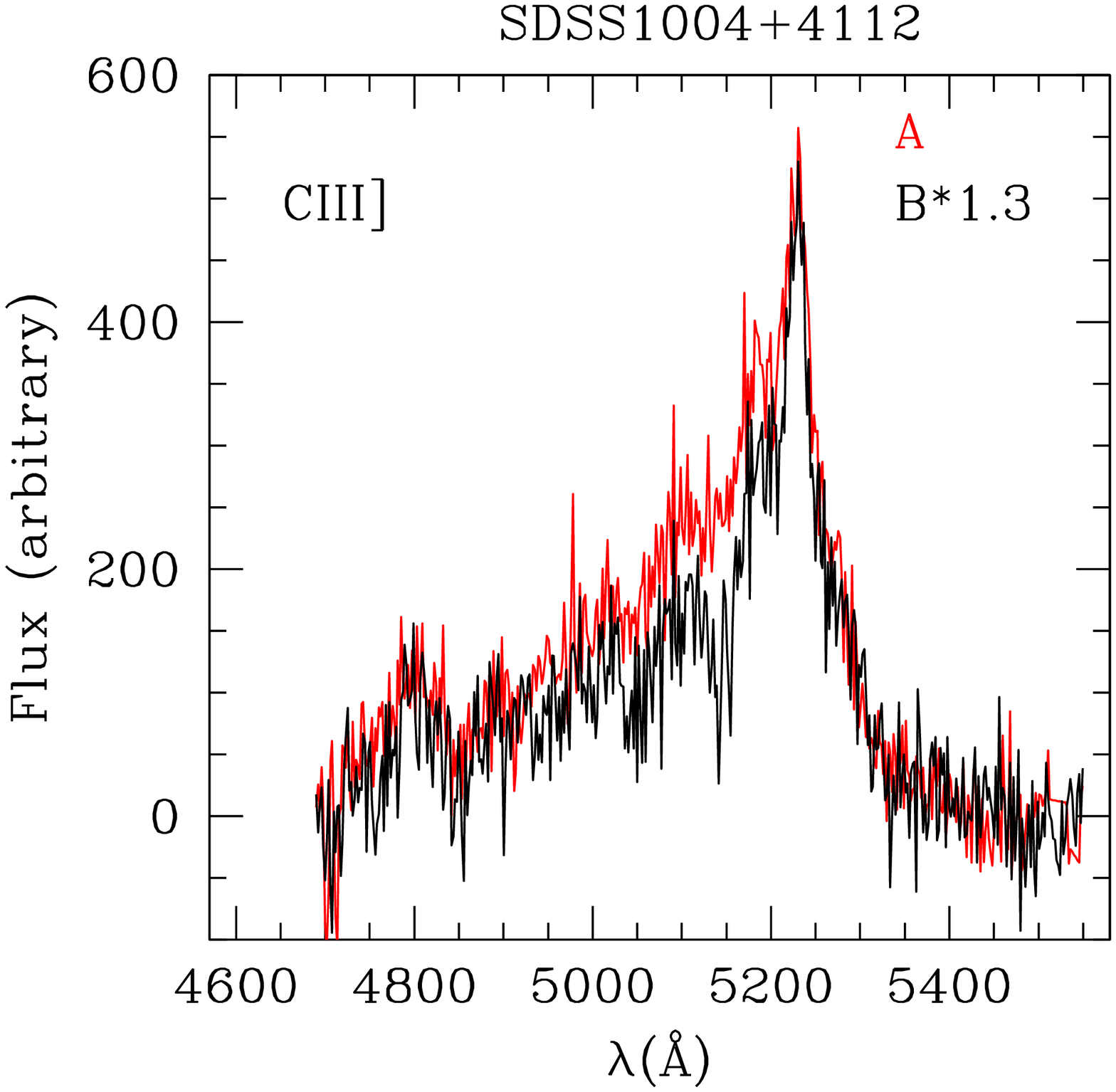,width=5cm,angle=0}\\
\caption{Ly$\alpha$, SiIV, CIV, CIII] emission line profiles for 
SDSS1004+4112 vs. observed 
$\lambda$. The {\em red line} represents the continuum subtracted emission 
lines for $A$. The {\em black line} represents the continuum 
subtracted emission line for $B$ multiplied by a factor to match 
the peak of $A$. The factors are shown in each panel. 
\label{prof_sdss1004}}
\end{center}
\end{figure*}

\begin{figure*}
\begin{center}
\plotone{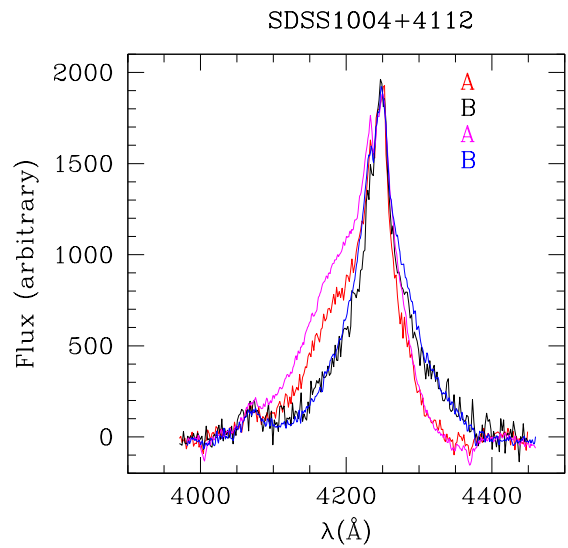}\\
\caption{CIV emission line profile comparison for SDSS1004+4112. {\em
    Red} and {\em black} lines represent A and B MMT spectra
  respectively, {\em magenta} and {\em blue} represent A and B Keck
  spectra obtained by \cite{richards04} respectively.
\label{civprofile}}
\end{center}
\end{figure*}

Comparing the $A$ and $B$ spectra taken with the MMT we notice an
enhancement in the blue wing and a decrement in the red wing of the
CIV and SIV emission lines (Figure \ref{prof_sdss1004}).  Ly$\alpha$
and CIII] emission lines show smaller differences.  Hence, our results
  are consistent with \cite{richards04,gomez06} and \cite{lamer06} although
  the amplitude of the enhancement of the blue wing is smaller than
  that observed previously. This can be appreciated in
  Figure~\ref{civprofile} where the $A$ and $B$ CIV emission line
  profiles taken in 2004 with the Keck telescope are presented (data kindly provided
  by G.T. Richards).  While the $B$ component and the red part of the
  A component are basically the same in both epochs, the blue wing
  enhancement of component $A$ is significantly smaller in 2008. This
  variability is the kind of gradual change in the line profile
  expected from microlensing. According to \S\ref{mulens}, in what follows we will
  use the cores of the emission lines to compute flux ratios avoiding
  the effects of microlensing in the blue wings.

\begin{figure*}
\begin{center}
\plotone{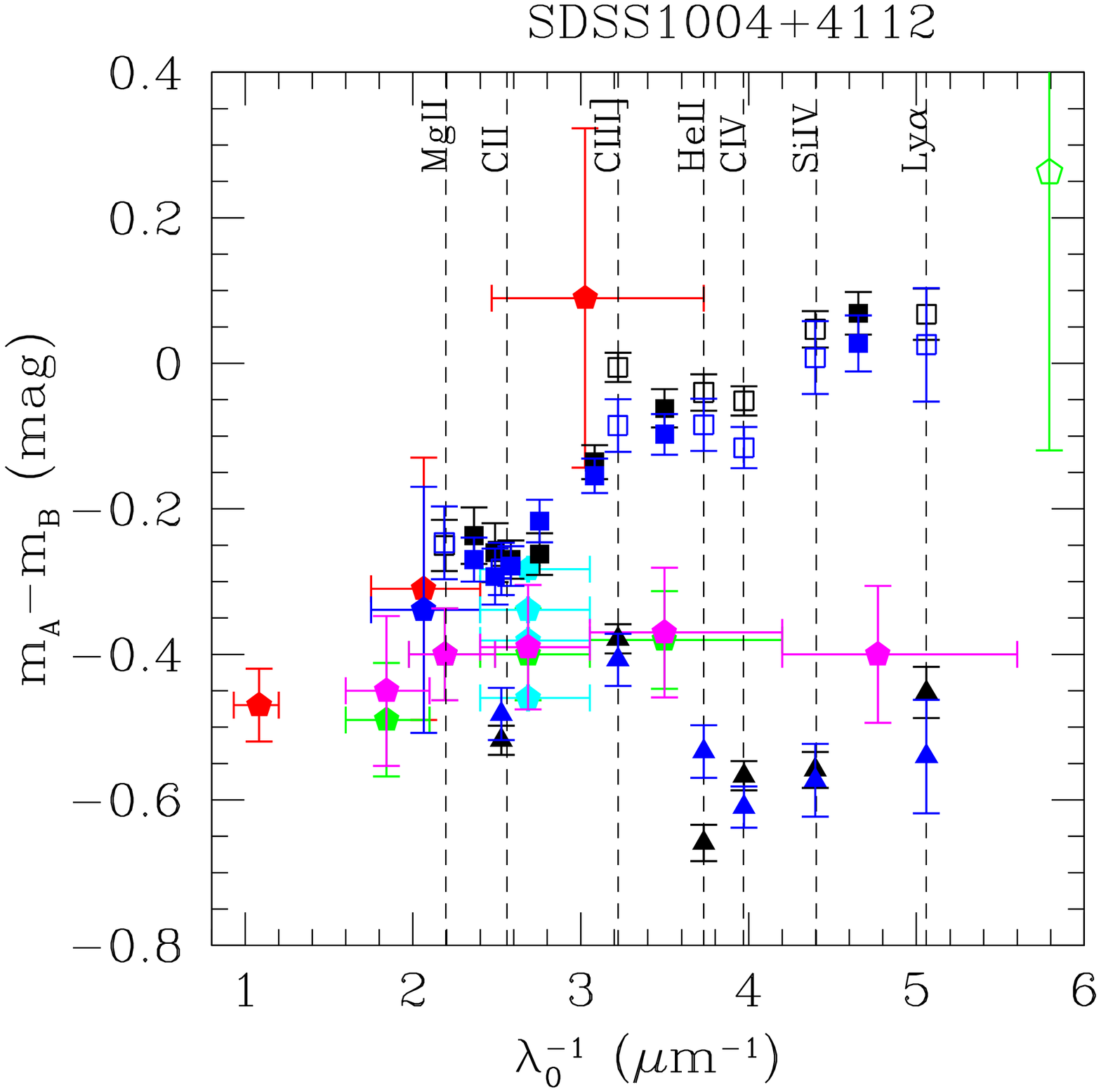}
\caption{\footnotesize{Magnitude differences $m_A-m_B$ vs
    $\lambda_0^{-1}$ ($\lambda$ in the lens galaxy restframe) for
    SDSS1004+4112.  We use the standard units of $\mu$m$^{-1}$ for
extinction studies, which are convenient to cover the range of 
observed $\lambda$. 
{\em Solid pentagons} represent the integrated
    continuum obtained from (broad-band) CASTLES ({\em red}),
    \cite{inada03} ({\em green}), \cite{inada05} ({\em blue}),
    \cite{oguri04} ({\em magenta}), and \cite{fohlmeister08} ({\em
      cyan}). The {\em green open pentagon} represents the X-ray data
    obtained by \cite{ota06} (for display convenience, we shifted it
    in wavelength from 60 to 290 $\mu$m; i.e from 28 to
    5.8~$\mu$m$^{-1}$ in the rest frame).  The {\em black} and {\em
      blue squares} represent the magnitude differences from the
    integrated continuum in our spectra ({\em solid}) and from the
    integrated fitted continuum under the emission lines ({\em open})
    for two different exposures. {\em Black} and {\em blue triangles} are
    the magnitude difference in emission line core. }
\label{diff_sdss1004a}}
\end{center}
\end{figure*}

$A-B$ magnitude differences in the continuum and in the emission lines
  estimated from our spectra or obtained from the literature are shown
  in Figure \ref{diff_sdss1004a} (see also Tables \ref{lit} and
  \ref{mag_sdss1004}).  The $A-B$ magnitude differences corresponding
  to the emission lines show no trend with wavelength (within
  uncertainties) and are distributed around $<A-B>=-0.52\pm0.07$~mag 
  supporting the absence of dust extinction and defining the baseline
  for no microlensing magnification.
In 2004 the continuum difference curve obtained from the spectra
matched within errors the zero microlensing baseline defined from the
low ionization emission lines. With small offsets the broad-band based
continuum data from \cite{oguri04} and \cite{inada03,inada05} also
match the baseline for no microlensing. This lack of microlensing
evidence in the continuum in 2004 (as the counterpart of the blue
wing enhancements) was considered a serious drawback to interpret the
enhancements in terms of microlensing \citep{gomez06}.

On the contrary, our $A-B$ continuum difference measurements (see
Figure \ref{diff_sdss1004a}) based on spectra taken in 2008, strongly
depart from the zero microlensing baseline with an increasing trend
towards the blue that would include the X-ray measurements obtained by
\cite{ota06}. The magnitude difference in the continuum is consistent
with CASTLES broad-band data.

The A-B continuum differences corrected for 
the time delay measured by \cite{fohlmeister08}
change in the sequence: $-0.460\pm 0.005$~mag (2003-04), 
$-0.283\pm 0.007$~mag (2004-05), 
$-0.339\pm0.005$~mag (2005-06), and $-0.381\pm 0.007$~mag (2006-07).  
The lowest value, $-0.46\pm0.005$~mag, is close
to the mean magnitude difference in the emission lines, 
$-0.52\pm0.07$~mag, likely indicating that
at this epoch (2003-04) the system showed little microlensing.

\begin{figure*}
\begin{center}
\plotone{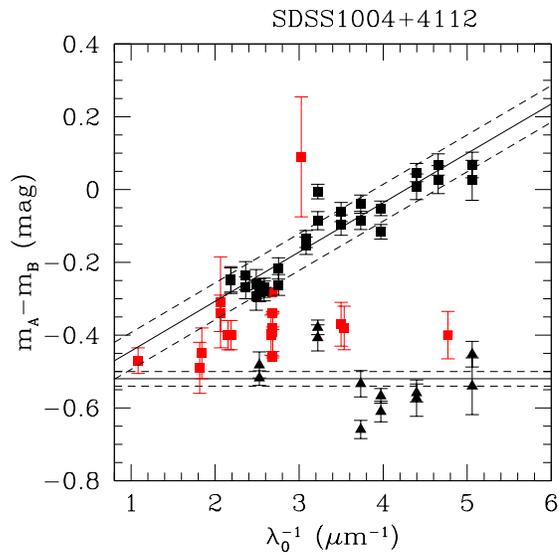}
\caption{Model fitted to the data shown in Figure
  \ref{diff_sdss1004a}.  {\em Squares} and {\em triangles} represent
  continuum and NEL data respectively.  {\em Black lines} represent
  the function fitted to the continua and the average of the emission
  line cores. {\em Dashed lines} are the standard deviations
  for the continuum fits and the standard error of the mean for the
  emission line cores.
\label{diff_sdss1004b}}
\end{center}
\end{figure*}

Figure \ref{diff_sdss1004b} shows a linear fit to the continuum data
and the average of the emission line data.  The magnitude difference
variation in the continuum data (with a slope of $0.13\pm0.04\,\rm
mag\, \mu m^{-1}$) implies differences with respect to the emission
lines of $\sim 0.2$ and $\sim 0.5$~mag at 7680 and 3320 \AA \,
respectively.  Our results are consistent with the trend indicated by
the X-ray continuum data \citep{ota06}. In summary, our data indicate
negligible dust extinction and evidence of chromatic microlensing
affecting the continuum.  These results and the variability detected
in the emission line profile give strong support to the hypothesis of
microlensing to explain the enhancement in the blue wings.

\begin{figure*}
\begin{center}
\epsfig{file=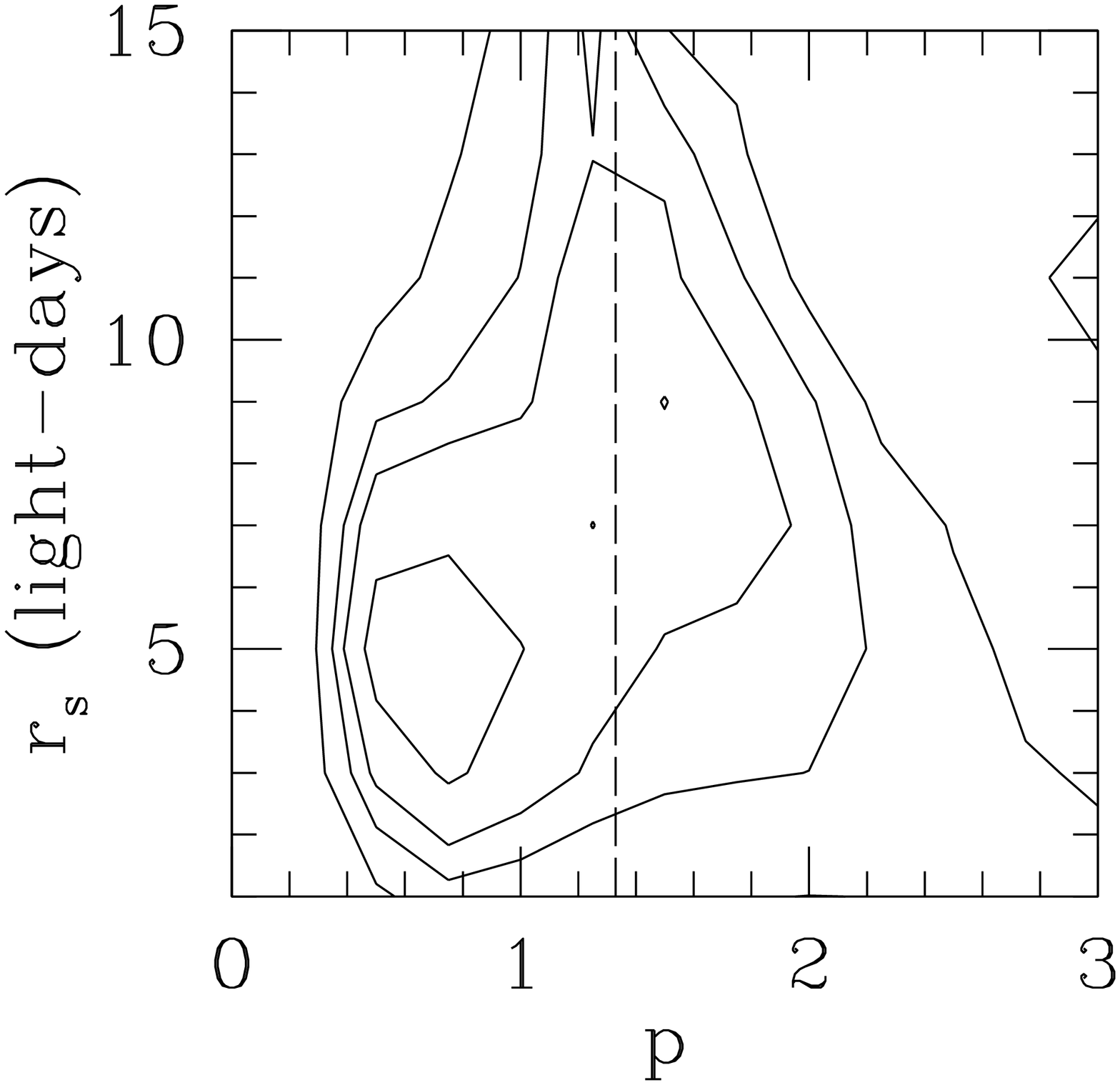,width=8cm,angle=0}
\epsfig{file=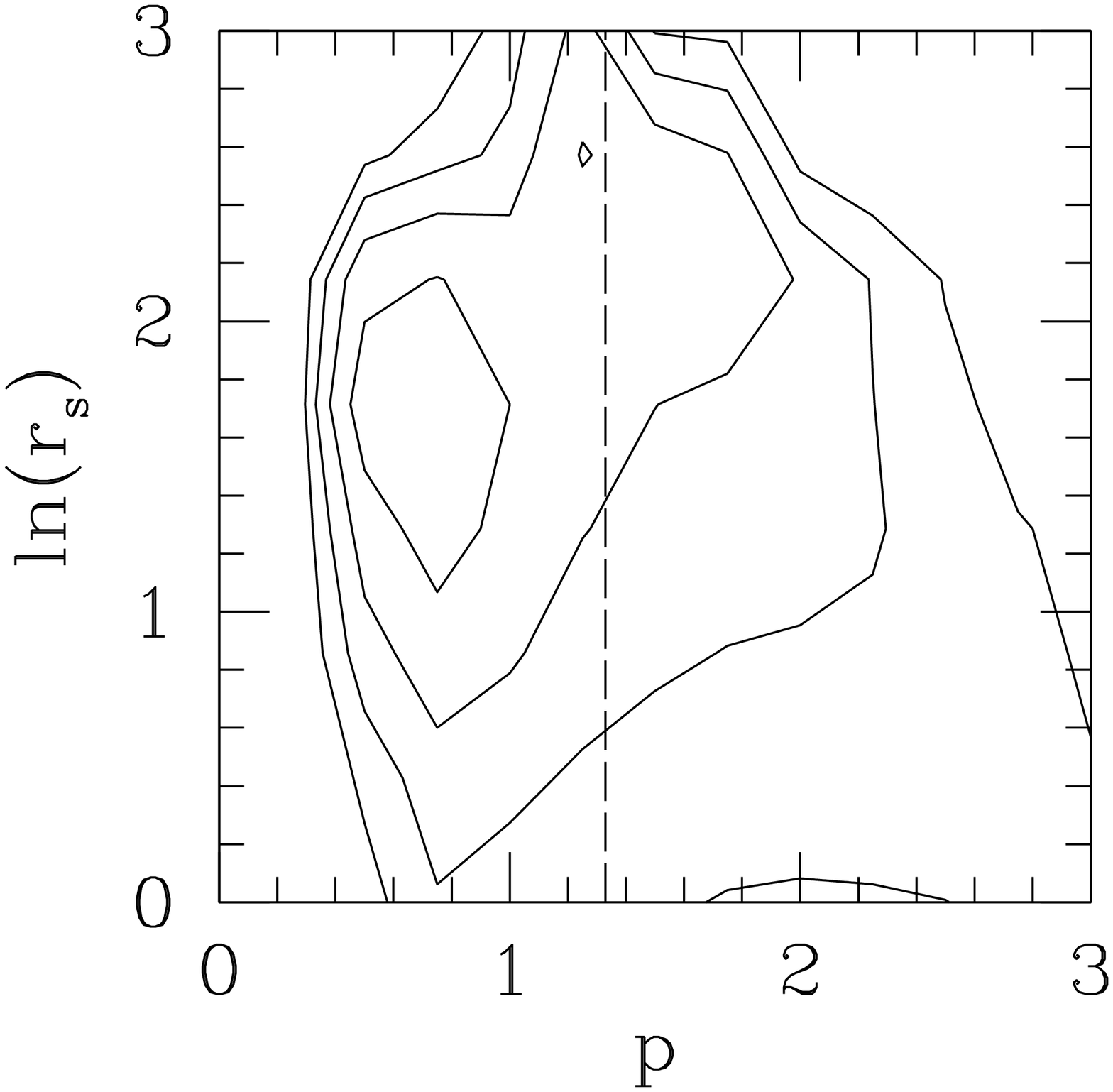,width=8cm,angle=0}\\
\caption{Two-dimensional pdfs obtained using the measured chromatic
  microlensing for SDSS1004+4112 (Table \ref{map_sdss1004}) for both
  linear ({\em left}) and logarithmic ({\em right}) grids in
  $r_s$. Contours correspond to $0.5 \sigma$, $1 \sigma$, $1.5
  \sigma$, and $2 \sigma$ respectively.  We estimate $r_s=7\pm3 \, \rm
  light-days (18.1\pm7.8 \times 10^{15} \rm cm$) and $p=1.1\pm0.4$ for
  the linear prior and $r_s=6^{+4}_{-3} \, \rm light-days
  (15.5^{+10.4}_{-7.8} \times 10^{15} \rm cm$) and $p=1.0\pm0.4$ for the
  logarithmic prior.  The {\em dashed line} corresponds to the value
  predicted by the thin disk model ($p=4/3$)
\label{sdss1004size}}
\end{center}
\end{figure*}

The structure of the accretion disk was studied using the procedure
explained in section \ref{disk-size}. We used the values
($\kappa_A=0.48$, $\gamma_A=0.59$) and ($\kappa_B=0.48$,
$\gamma_B=0.48$) taken from \cite{mediavilla09} to obtain the
magnification maps for the $A$ and $B$ images respectively.  Applying
this procedure to microlensing measurements at three different
wavelengths corresponding to our MMT data (see Table
\ref{map_sdss1004}), we obtained the 2D probability density functions
(pdfs) shown in Figure \ref{sdss1004size} for both linear and
logarithmic grids in $r_s$. From these distributions we obtain
estimates $r_s=7\pm3 \, \rm light-days (18.1\pm7.8 \times 10^{15} \,
\rm cm$) and $p=1.1\pm0.4$ for the linear prior and $r_s=6^{+4}_{-3} \,
\rm light-days (15.5^{+10.4}_{-7.8} \times 10^{15} \, \rm cm$) and
$p=1.0\pm0.4$ for the logarithmic prior.  Although the value of $p$ is
consistent within uncertainties with the thin disk theory it is
interesting to mention the trend in this and in other objects to have
$p<4/3$ \cite[see][]{mediavilla11,blackburne11}.  The microlensing
estimate for the size also exceeds substantially the estimate obtained
from thin-disk theory \cite[$r_s \sim 0.3 \rm light days = 0.78 \times
  10^{15} \, \rm cm$,][]{mosquera11}.

To study the impact of intrinsic variability in these results, we
  can compare the A-B difference we measured using the emission lines or the
  continuum at 12500\AA\ (see Table 4), where microlensing and dust
  extinction should be less significant. We find a difference between
  both measurements (which is a conservative upper bound to
  continuum variability) of 0.08 mag, for an insignificant impact
  on the estimate of $r_s$ and $p$.

\subsection{HE1104-1805} \label{he1104}

HE1104-1805 was discovered by \cite{wisotzki93}; it consists of two
lensed images $A$ and $B$ separated by $3\farcs15$ at $z_s=2.319$. The
lens galaxy was detected by \cite{courbin98} at $z_l=0.729$. Image $A$
is $1\farcs1$ from the main lens galaxy.  Variability in the continuum
was detected in spectra taken by \cite{wisotzki95} (optical),
\cite{courbin98} (near infrared), and \cite{chartas09} (X-ray).
\cite{poindexter07} monitored the system between 2003 and 2006,
concluding that the magnitude difference in the optical bands has
changed from $-1.7$, when the lens was discovered, to $-1.2$ in their
optical data (2006).  These authors also provide a time delay
estimation of $152.2^{+2.8}_{-3.0}$ (1 $\sigma$) days.

\begin{figure*}
\begin{center}
\epsfig{file=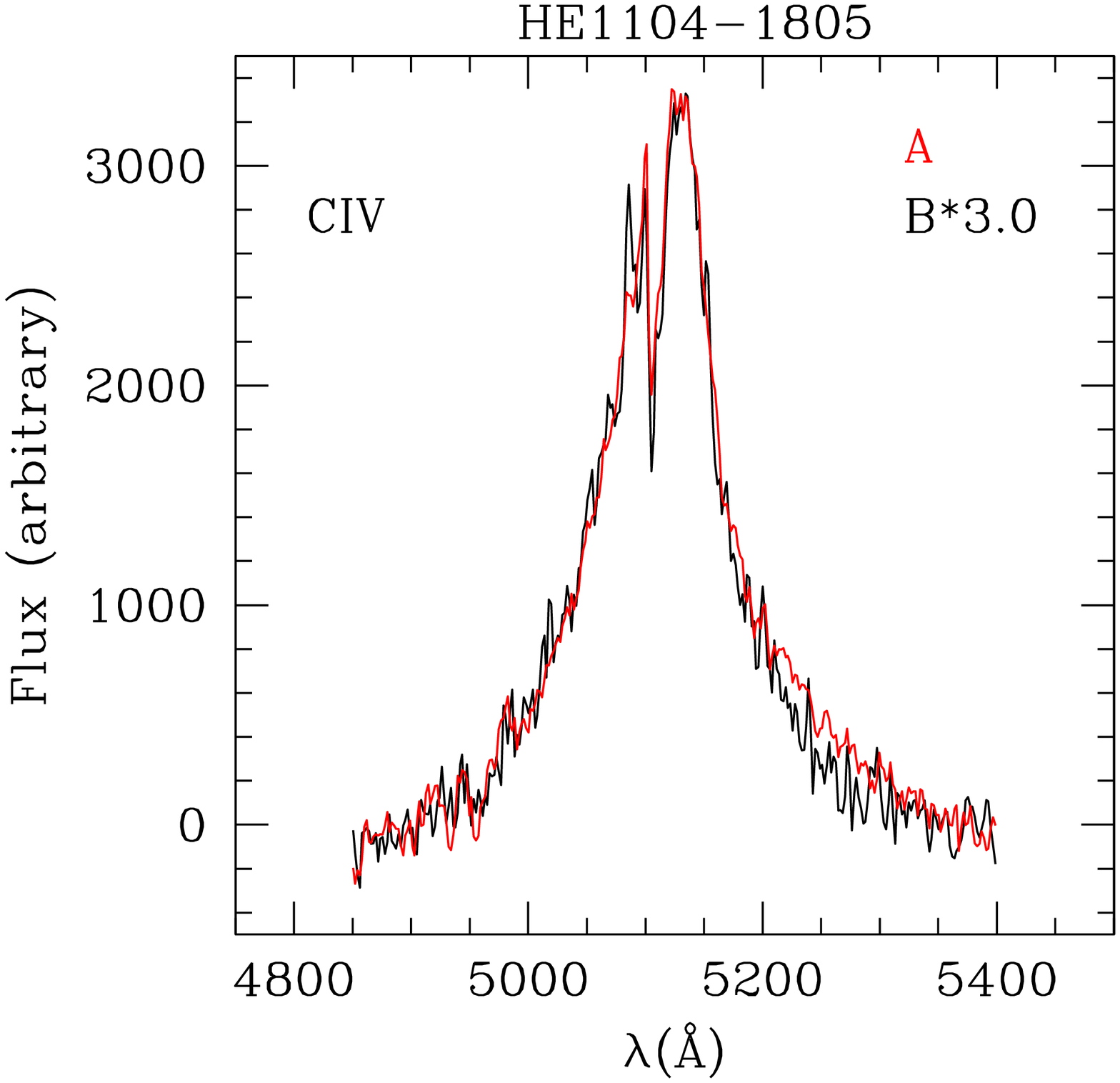,width=5cm,angle=0}
\epsfig{file=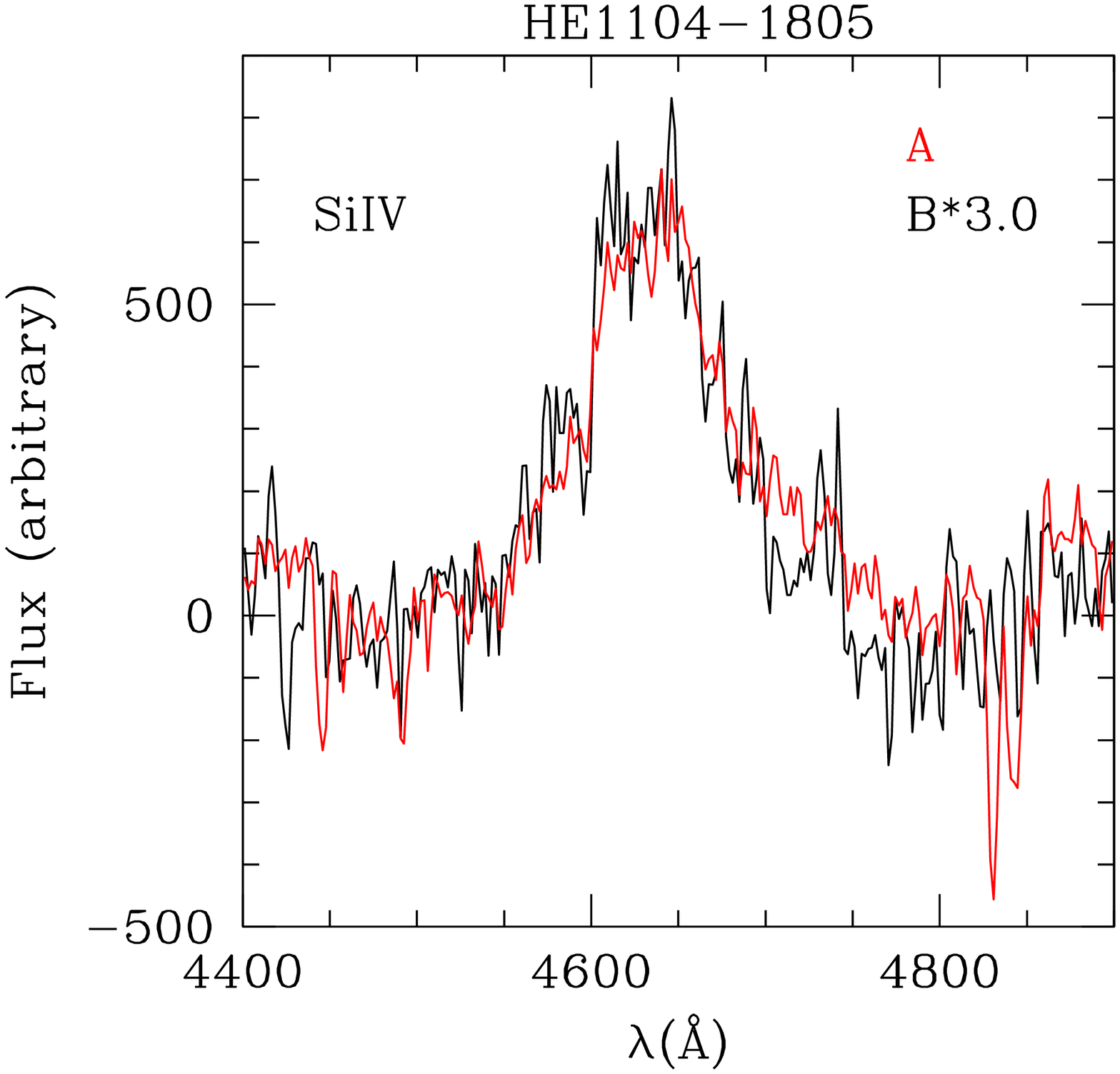,width=5cm,angle=0}
\epsfig{file=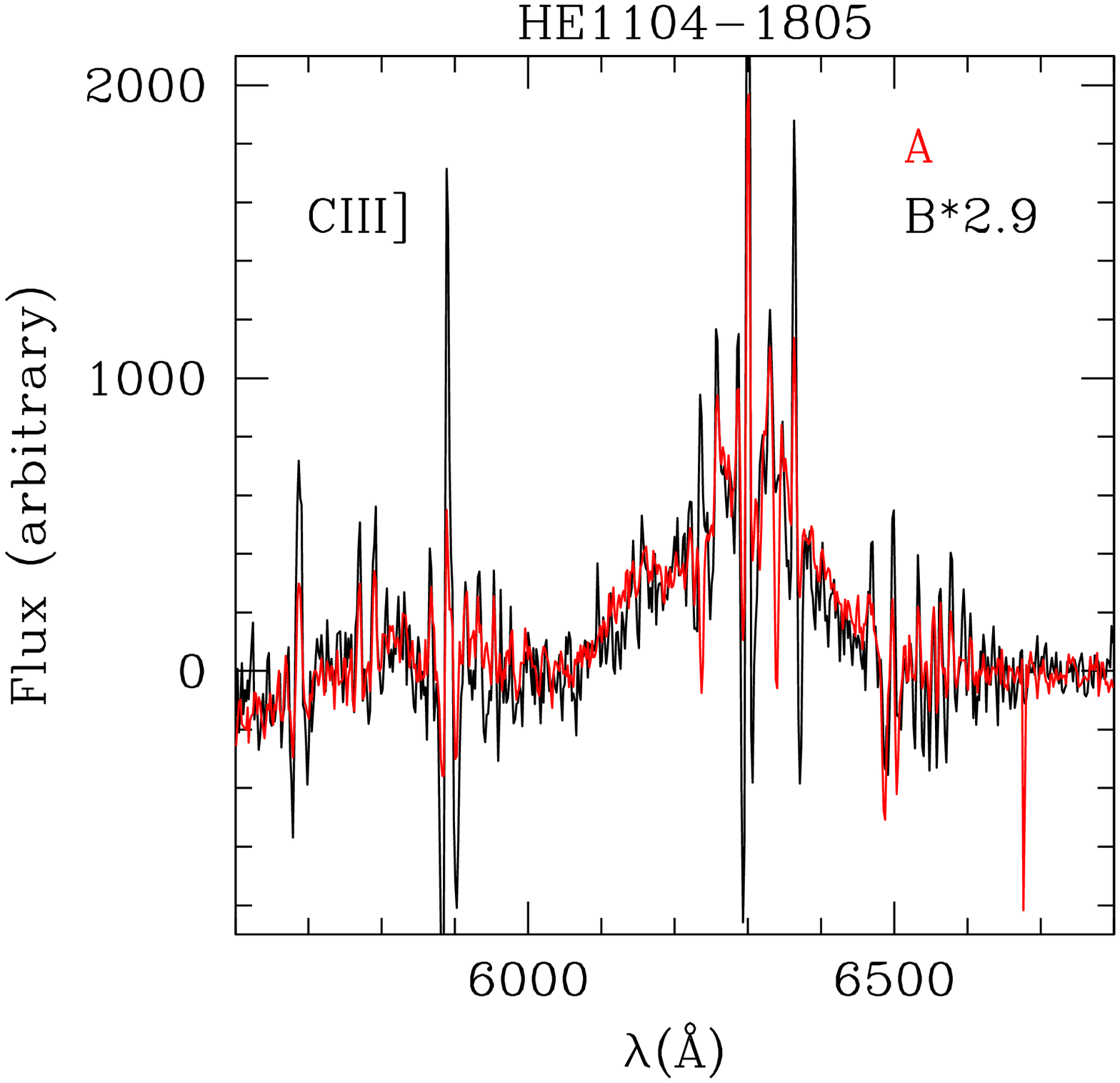,width=5cm,angle=0}
\epsfig{file=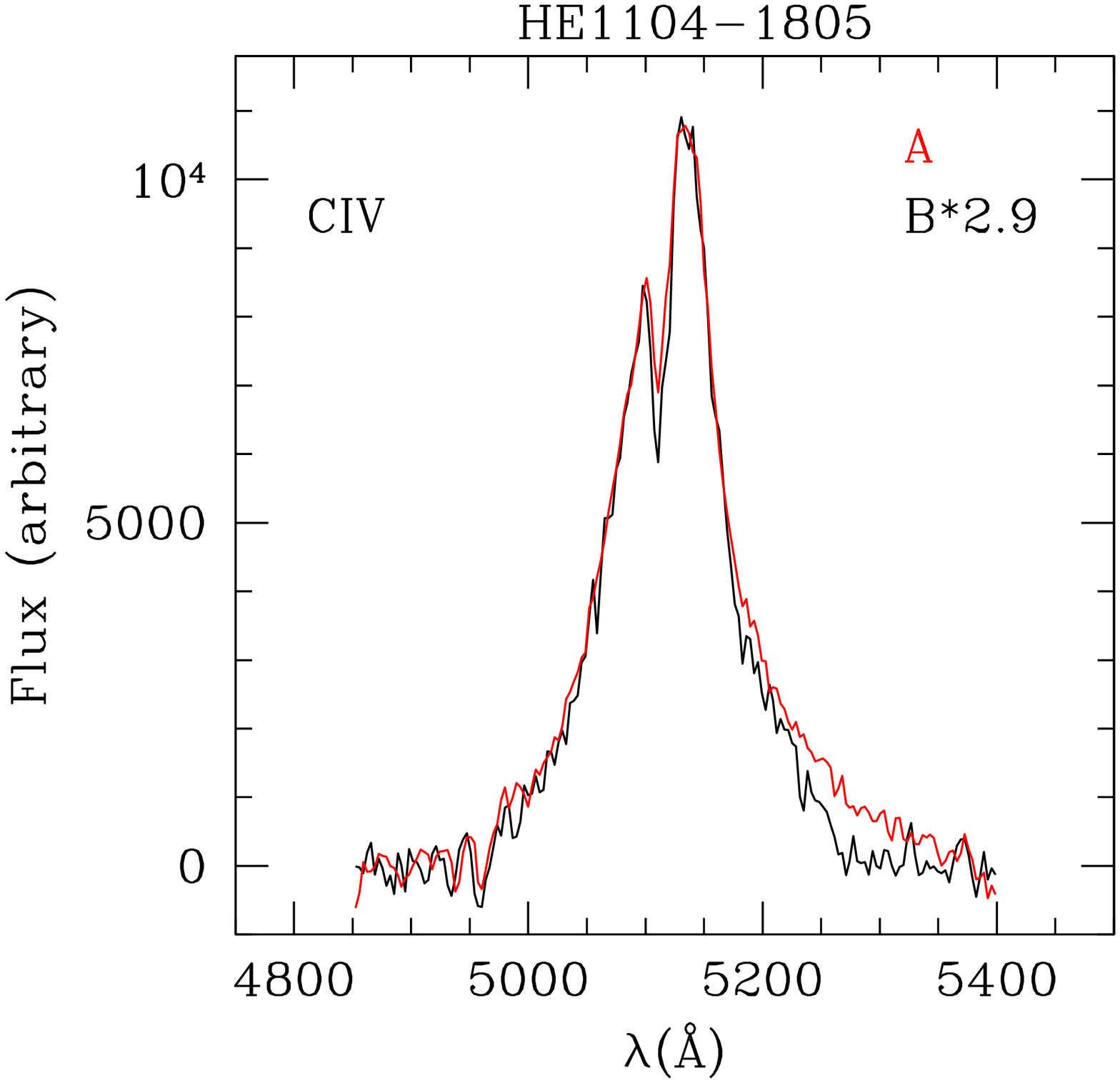,width=5cm,angle=0}
\epsfig{file=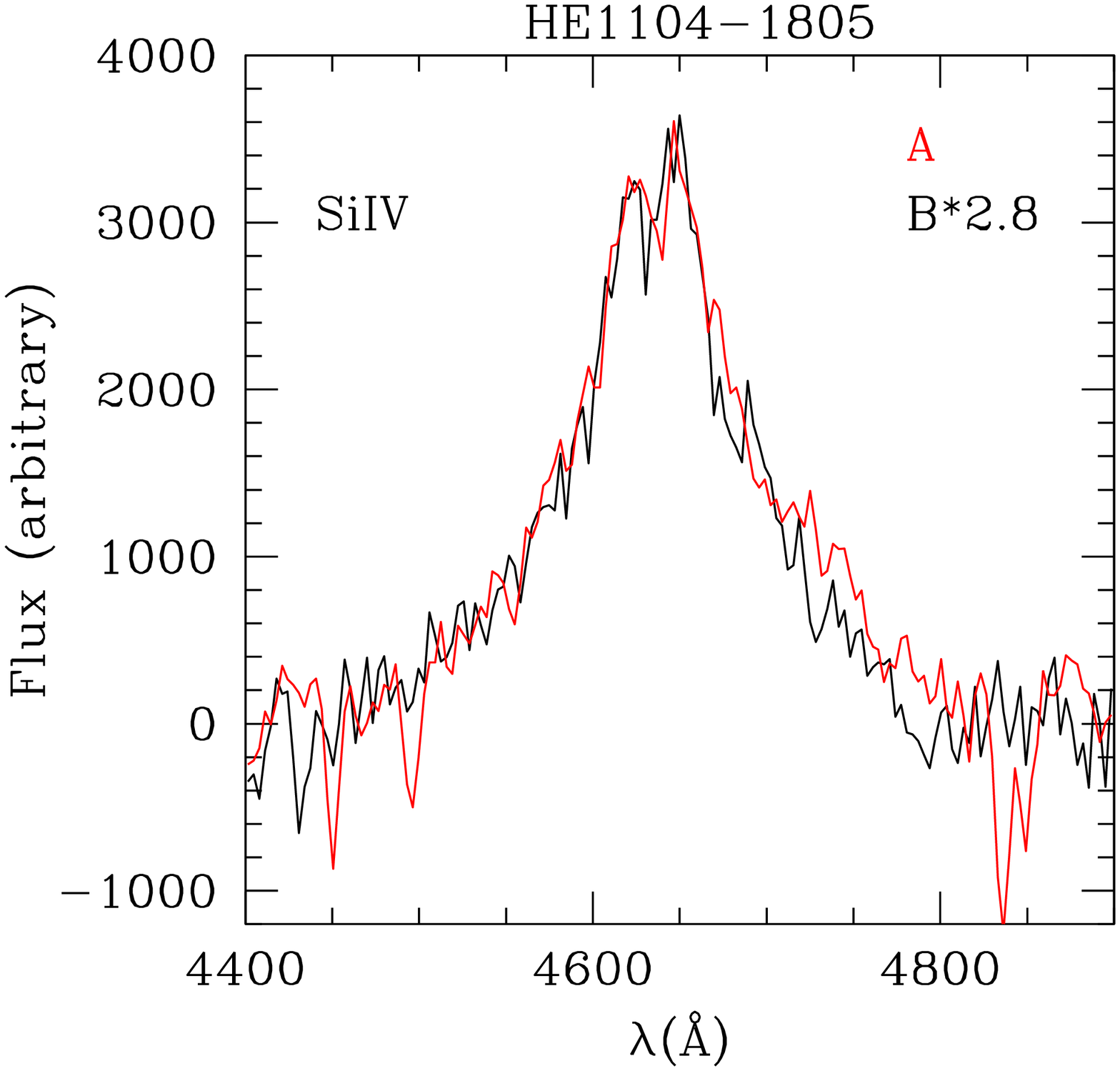,width=5cm,angle=0}
\epsfig{file=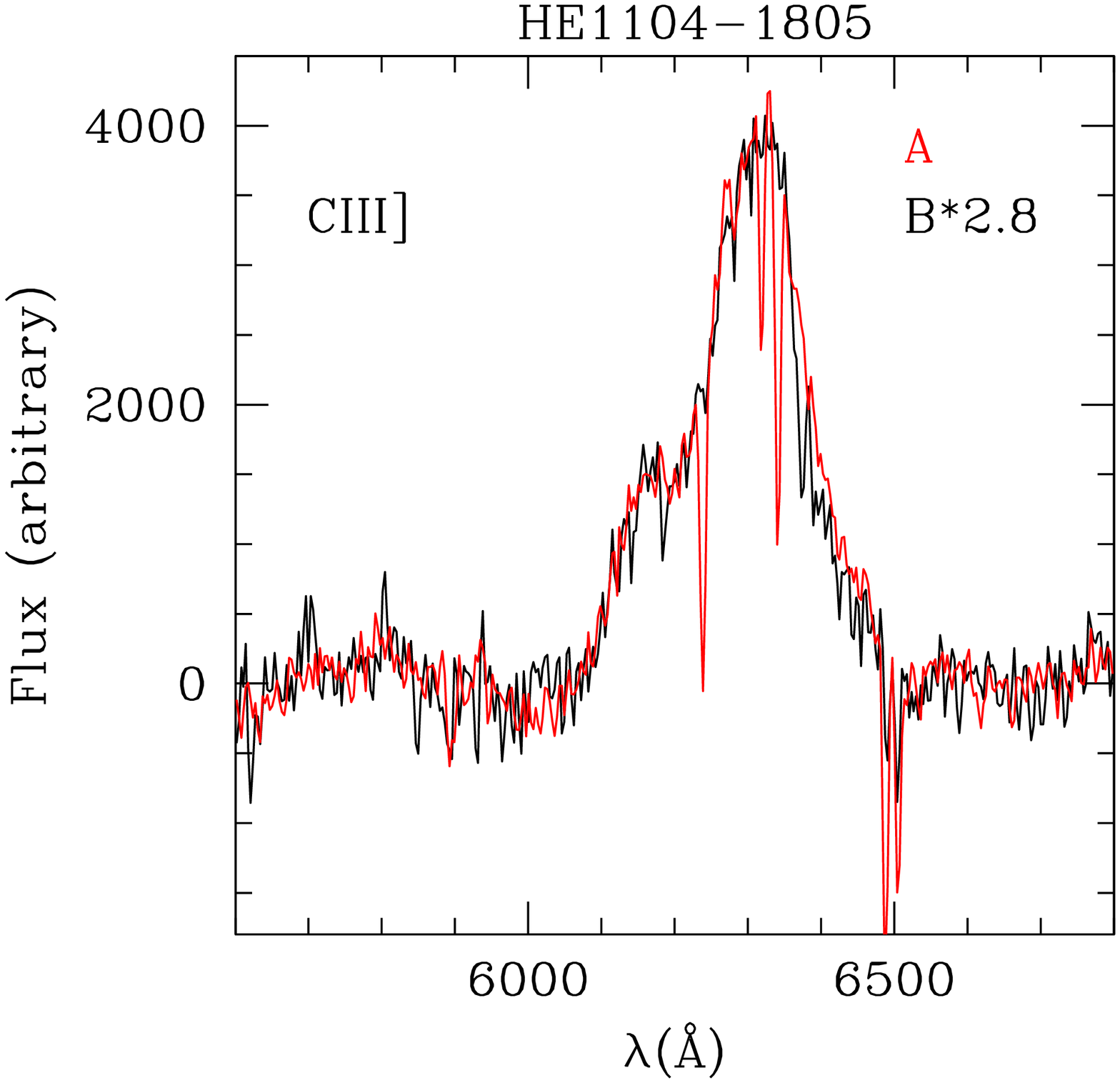,width=5cm,angle=0}
\caption{CIV, SiIV, CIII] emission line profiles for HE1104-1805 vs.
  observed $\lambda$.  {\em Upper panel} MMT spectra. The {\em red
    line} represents the continuum-subtracted emission lines for
  $A$. The {\em black line} represents the continuum subtracted
  emission line for $B$ multiplied by a factor to match the peak of
  $A$.  The factors are shown in each panel.
  {\em Bottom panel} same as upper panel but for VLT spectra. 
\label{prof3_he1104}}
\end{center}
\end{figure*}

\begin{figure*}
\begin{center}
\epsfig{file=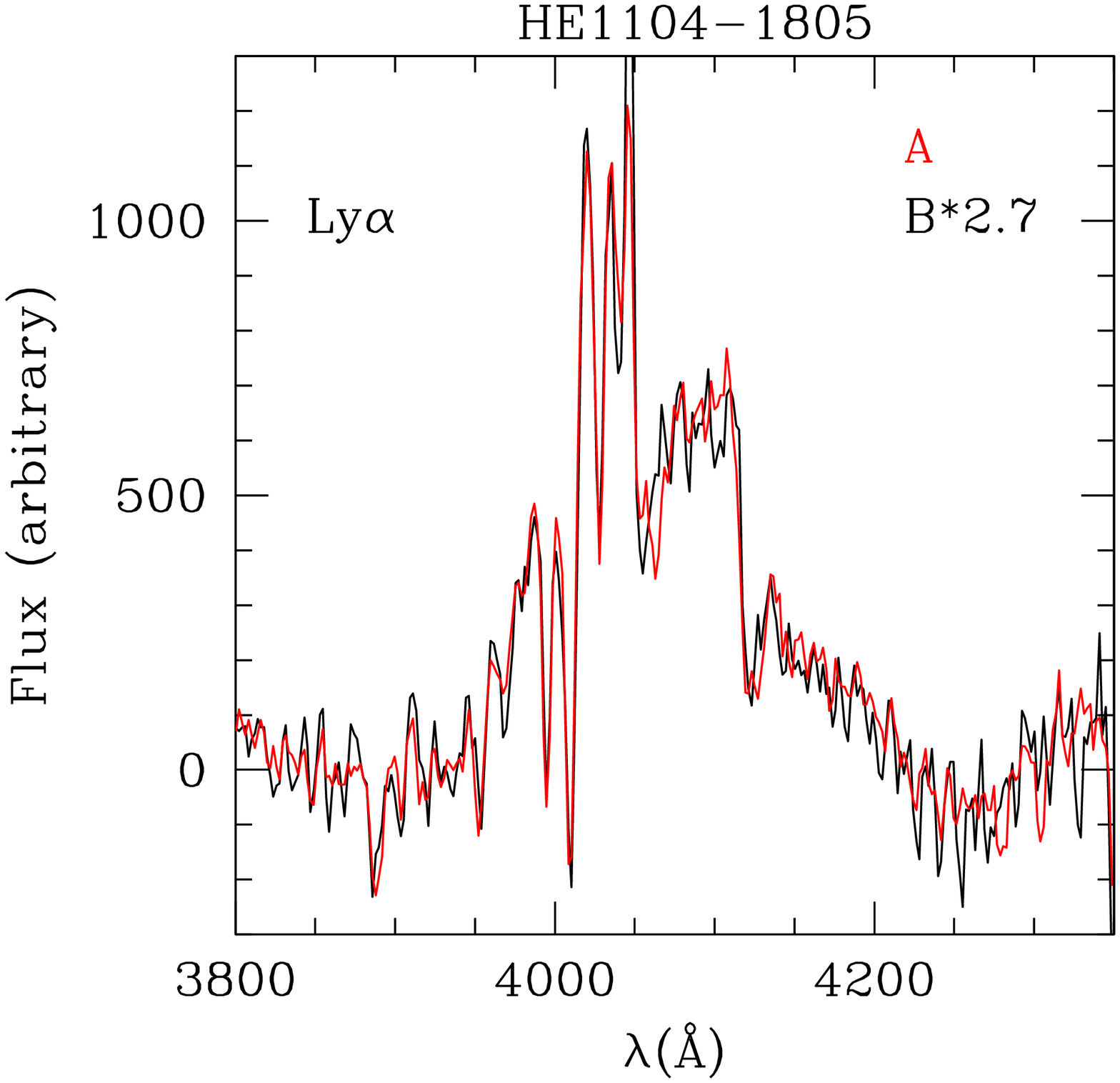,width=5cm,angle=0}
\epsfig{file=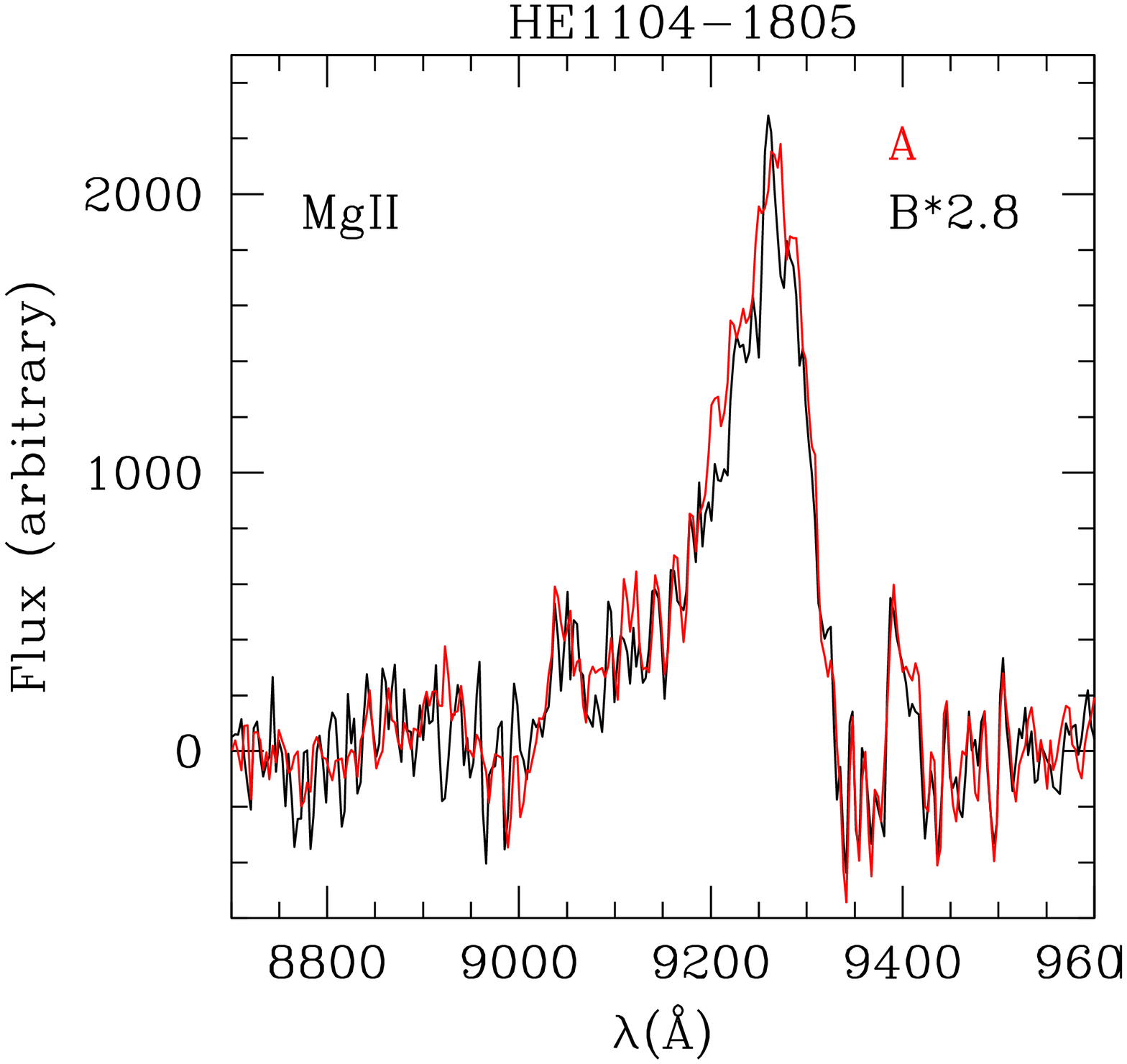,width=5cm,angle=0}
\caption{Ly$\alpha$, MgII emission line profiles for HE1104-1805 vs.
  observed $\lambda$.  The {\em red line} represents the continuum
  subtracted emission lines for $A$. {\em Black line} represents the
  continuum subtracted Ly$\alpha$ emission line for $B$ multiplied by
  2.7 to match the peak of $A$ ({\em red line}).
\label{prof4_he1104}}
\end{center}
\end{figure*}

The data obtained with the MMT and VLT show that the emission line
profiles of both images, A and B, are very similar, although some
slight but interesting differences can be found in the broad
components of CIV and SiIV (Figures \ref{prof3_he1104} and
\ref{prof4_he1104}).  Ly$\alpha$ is only seen in our MMT spectra. The
profile of the MgII emission line is asymmetric both in $A$ and
$B$. CIII] presents heavy absorption lines both in the BEL and NEL.
  In the higher SNR data obtained from VLT it is clearly seen that the
  $A$ spectrum shows several absorption lines, none of them present in
  the $B$ spectrum.  The profiles of CIV and SiIV emission lines show
  a slight enhancement in the red wing of $A$ compared to those of $B$
  both in MMT and VLT data.  These wing enhancements present only in
  high ionization lines might be evidence of microlensing.

\begin{figure*}
\begin{center}
\plotone{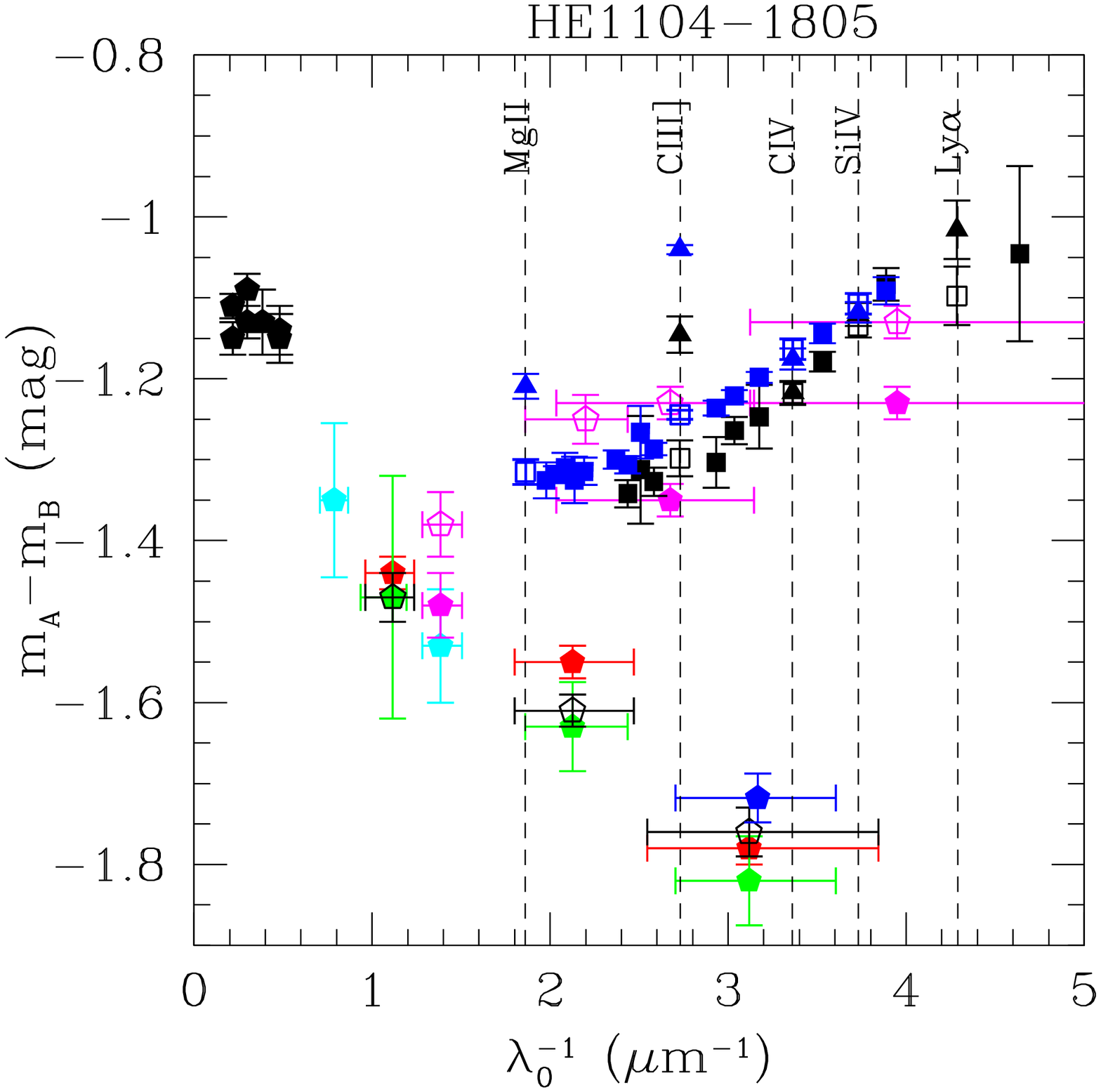}
\caption{\footnotesize{Magnitude differences $m_A-m_B$ vs
    $\lambda_0^{-1}$ ($\lambda$ in the lens galaxy restframe) for
    HE1104-1805. {\em Black and blue} represent the magnitude
    differences obtained from MMT and VLT spectra respectively.  {\em
      Solid squares} are the magnitude differences in the continuum,
    {\em open squares} in the integrated continuum under the emission
    line, and {\em solid triangles} in the emission line core.  The
    broadband data obtained from other authors are plotted as {\em
      pentagons} in different colors representing: CASTLES ({\em
      red}), \cite{lehar00} ({\em green}), \cite{courbin98} ({\em
      cyan}), \cite{schechter03} ({\em blue}), \cite{falco99} ({\em
      open black}), \cite{poindexter07} Spitzer IRAC ({\em solid
      black}). {\em Magenta pentagons} represent the optical broadband
    data obtained by \cite{poindexter07} with ({\em solid}) and
    without ({\em open}) time-delay correction. }
\label{diff_he1104a}}
\end{center}
\end{figure*}

Figure \ref{diff_he1104a} (see also Table \ref{mag_he1104}) presents
the magnitude differences in the continuum and in the emission lines.
We have also included data from the literature (Table \ref{lit}). The
mean B-A magnitude difference ($<B-A>=-1.13\pm0.02$~mag) 
corresponding to the
emission lines obtained from MMT and VLT spectra is consistent with
the values derived by \cite{wisotzki95} ($\sim -1.14$~mag) and
\cite{courbin00} ($-1.16\pm0.04$~mag). These values are also in agreement
with the value estimated from infrared data $-1.13\pm0.03$~mag 
\citep{poindexter07}. These results confirm that the cores of the
emission lines are not affected by microlensing and that little
extinction is present.

The $B-A$ magnitude differences in the continuum obtained from the MMT
and VLT spectra show a slope that is in agreement with optical
broad-band data obtained in 2006 \citep{poindexter07}\footnote{As we
  cannot correct our data for time delay, we have considered both the
  time-delay corrected and uncorrected optical data obtained by
  \cite{poindexter07} (magenta pentagons).  Notice also that the lens
  galaxy continuum is very faint, so it cannot contaminate our
  spectra.}.  Broad-band data obtained several years before
\citep{falco99,lehar00,courbin00,schechter03} are all consistent
(slope $-0.16\pm0.03\,\rm mag\, \mu m^{-1}$) 
but are very different from our own recent data and that
obtained by \cite{poindexter07}.

\begin{figure*}
\begin{center}
\plotone{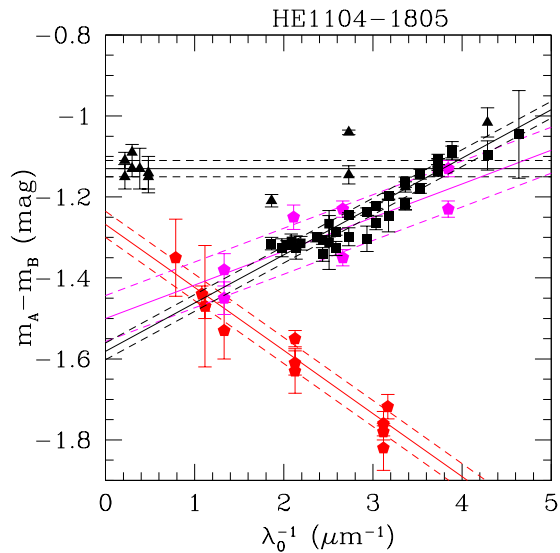}
\caption{Model fitted to the data shown in Figure \ref{diff_he1104a}.
  {\em Black lines} represent the fitted function to the continua
  ({\em squares}) for MMT and VLT data and the average of the emission
  line cores ({\em triangles}) respectively. The {\em red line} represents
  the fitted function to the broadband data in the literature
  \citep[CASTLES;][]{lehar00,courbin98,falco99,schechter03} at the
  same epoch.  The {\em magenta line} represents the fitted function
  to the broadband data obtained by \cite{poindexter07} with and
  without time-delay correction.  The data obtained in the infrared by
  \cite{poindexter07} are plotted as {\em black triangles}.  {\em
    Dashed lines} are the standard deviation for each fit (continua
  and broadband data) and the error of the mean for the emission
  line cores.
\label{diff_he1104b}}
\end{center}
\end{figure*}

Linear fits to the magnitude differences of continua are shown in
Figure \ref{diff_he1104b}.  The slope for the magnitude differences in
the emission lines and the IR data is $0.00\pm0.06\,\rm mag\, \mu
m^{-1}$ which is consistent with no extinction and it is in good
agreement with results obtained from near-infrared spectra by
\cite{courbin00} ($\Delta E < 0.01$) and those found by \cite{falco99}
($\Delta E=0.07\pm0.1$) using broadband data.  The continuum data from
the literature are fitted in two separate sets: 1992-1994 data with a
slope of $-0.16\pm0.03\,\rm mag\, \mu m^{-1}$ and the more recent data
from \cite{poindexter07} with a slope of $0.08\pm0.06\,\rm mag\, \mu
m^{-1}$.  The slope of the linear fit to our 2008 continuum data
(MMT+VLT), $0.12\pm0.02\,\rm mag\, \mu m^{-1}$, is thus in good
agreement with the slope of $0.08\pm0.06$ corresponding to the 2006
data of \cite{poindexter07}, but remarkably different from the value
corresponding to 1992-1994 broadband data $-0.16\pm0.03\,\rm mag\, \mu
m^{-1}$.  Thus, microlensing in HE1104-1805 has induced an extreme
change in continuum slope that needs explanation. The chromaticity
during the 1992-1994 epoch with an increasing amplitude towards the
blue leads to an straightforward interpretation in terms of the
magnification of the dominant component $A$. To explain the slope of
the continuum corresponding to 2006-2008 epoch (under common
assumptions about the unresolved source structure) we need to combine
chromatic microlensing in both $A$ and $B$. For instance, we can
consider the combination of two events of magnification in both $A$
and $B$ with a progressive increase in the strength of the $B$ event
from 1994 to 2008. This is only a qualitative example and simulations
are needed to consistently reproduce microlensing chromaticity in each
of the two epochs.

Following the procedure described in section \ref{disk-size} we have
used the detected microlensing chromaticity to study the structure of
the accretion disk in HE1104-1805. We have done this for three sets of
data: our VLT continuum data from 2008, the \cite{poindexter07} data
corrected for time delay and the average of the data from
\cite{courbin98}, \cite{falco99}, \cite{lehar00}, and
\cite{schechter03} that consistently follow a common trend with
wavelength. In Table \ref{map_he1104} we present the microlensing
measurements for each dataset. To compute the microlensing maps we
have used the following projected densities and shears for each lens
image ($\kappa_A=0.64$, $\gamma_A=0.52$) and ($\kappa_B=0.33$,
$\gamma_B=0.21$) according to \cite{mediavilla09}

\begin{figure*}
\begin{center}
\epsfig{file=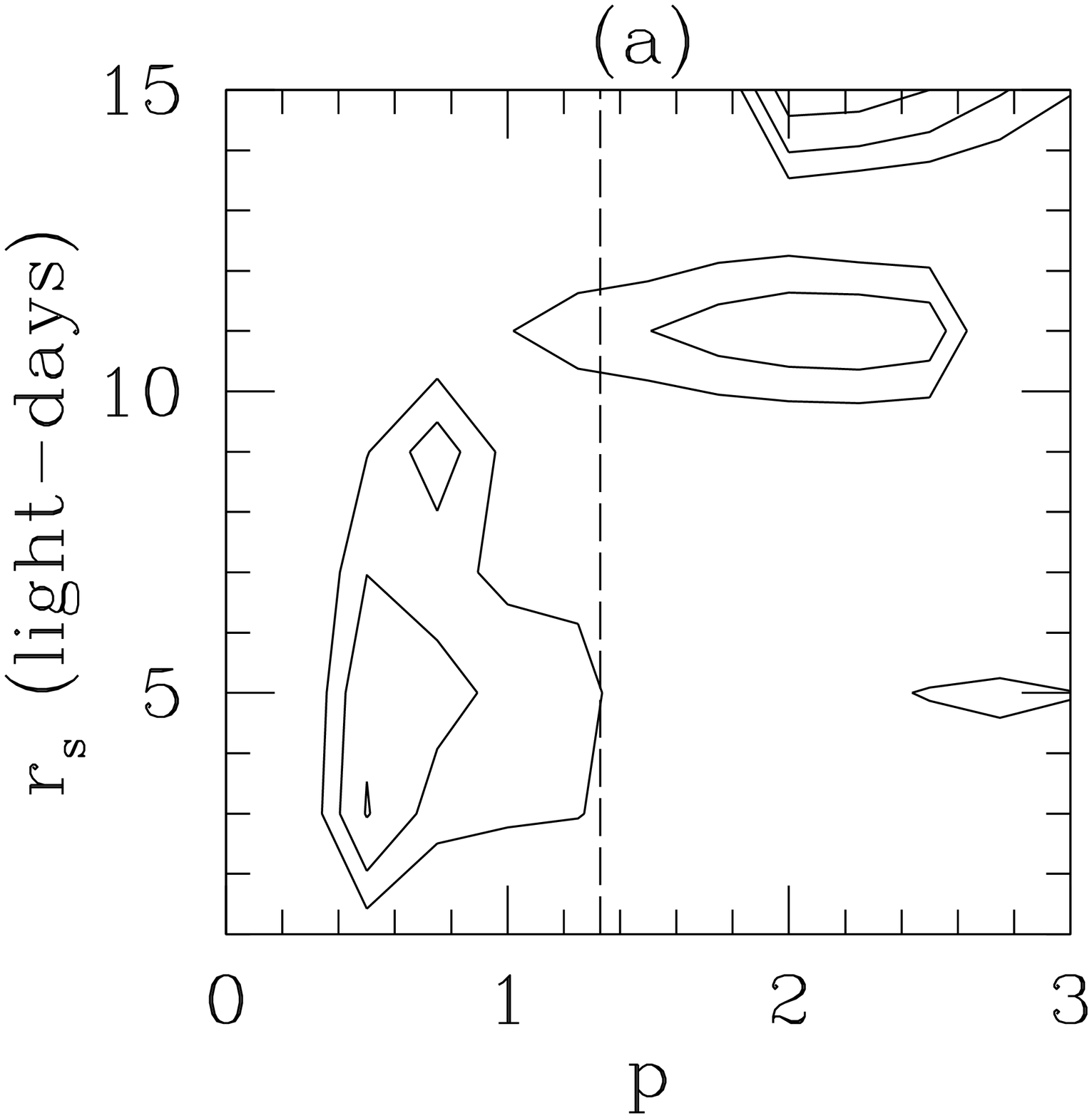,width=4.5cm,angle=0}
\epsfig{file=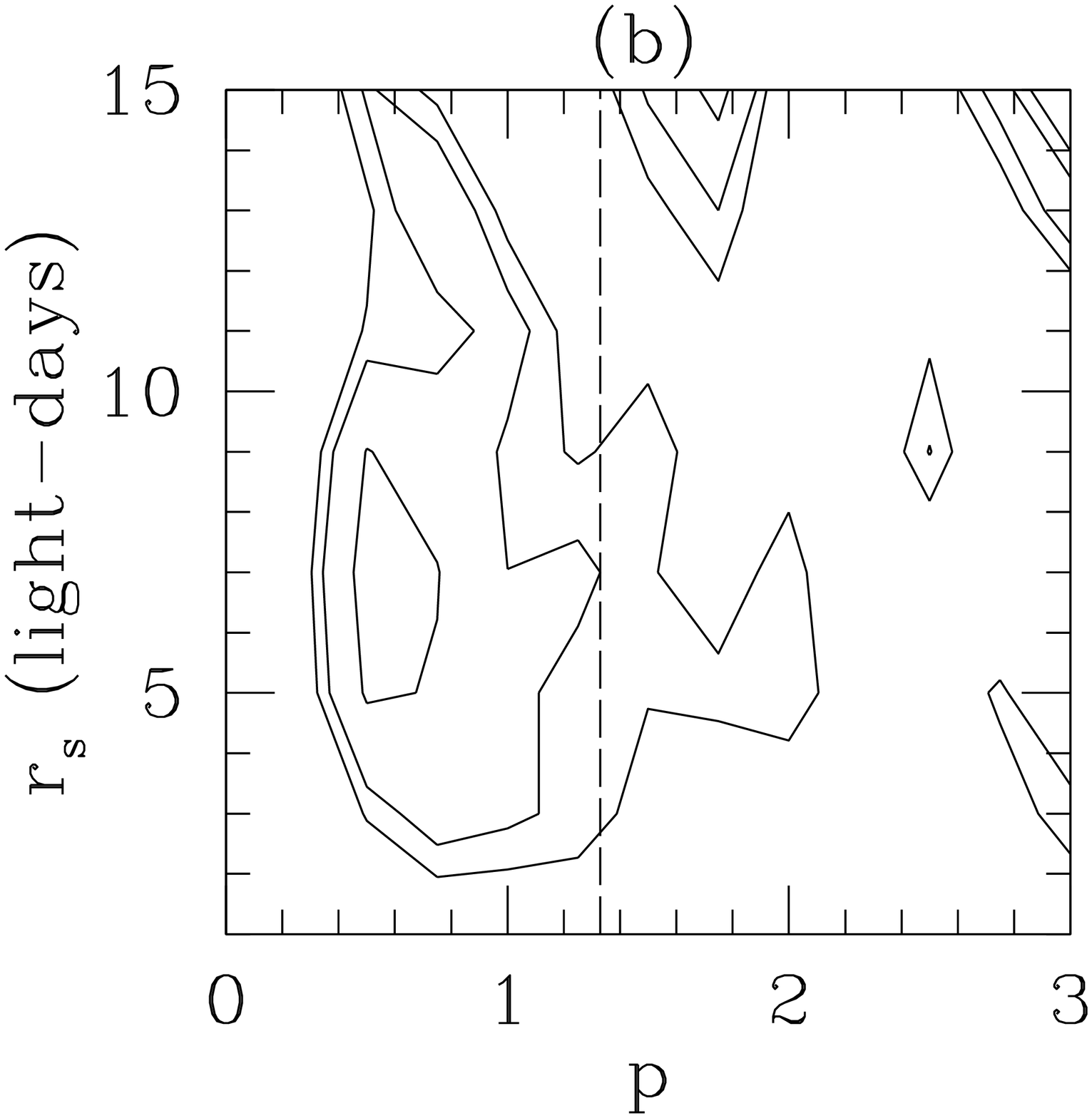,width=4.5cm,angle=0}
\epsfig{file=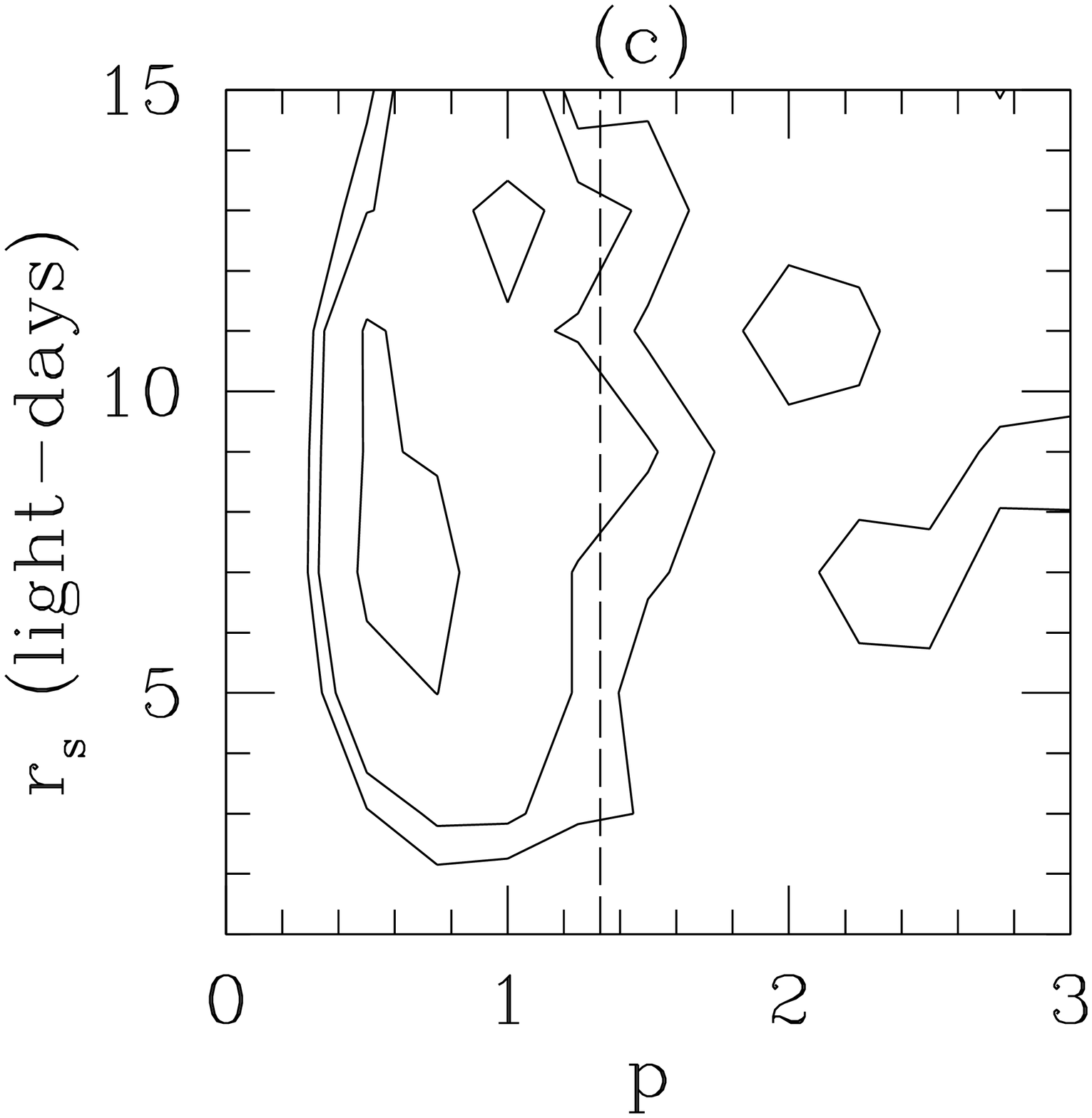,width=4.5cm,angle=0}
\epsfig{file=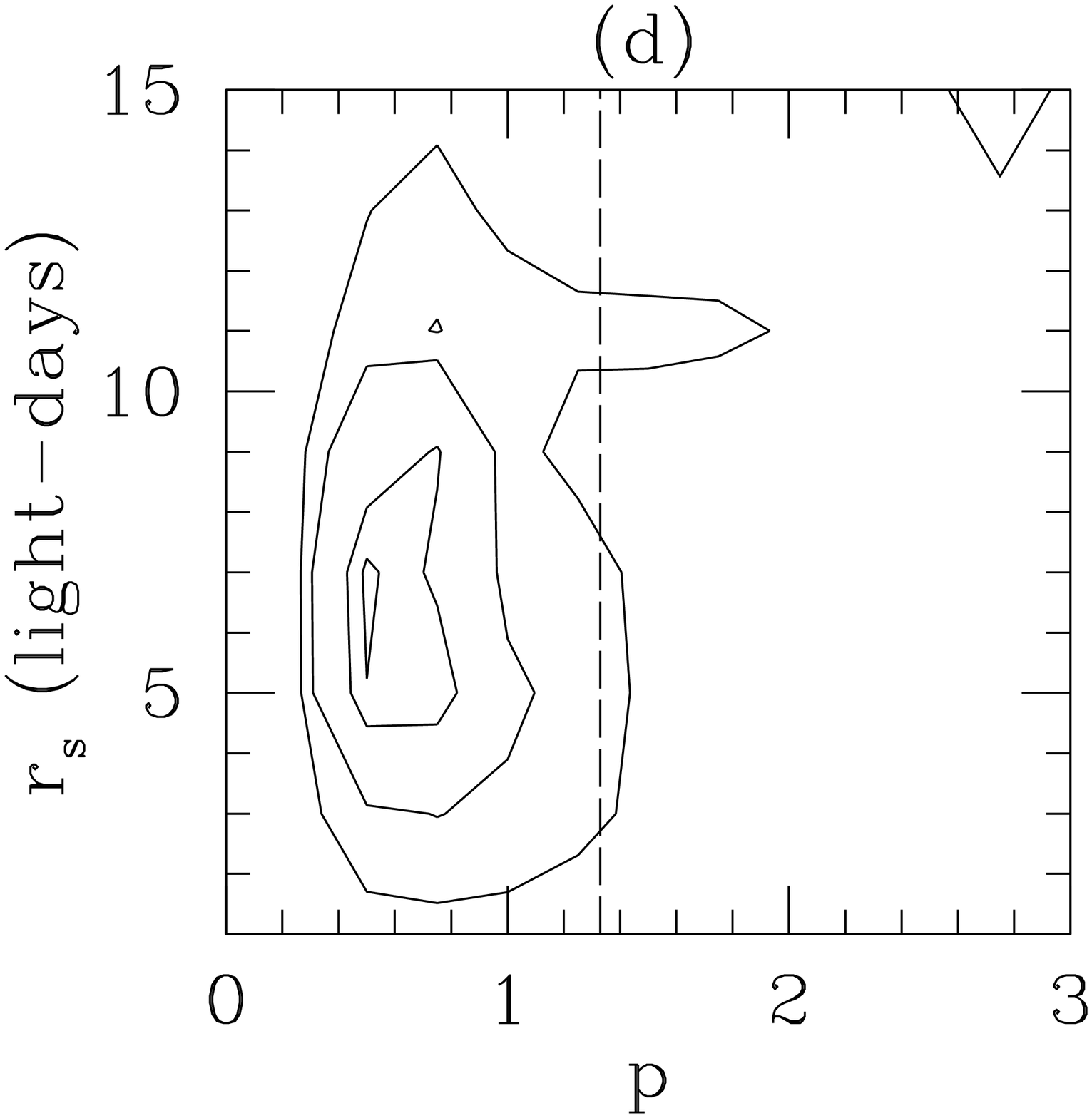,width=4.5cm,angle=0}\\
\epsfig{file=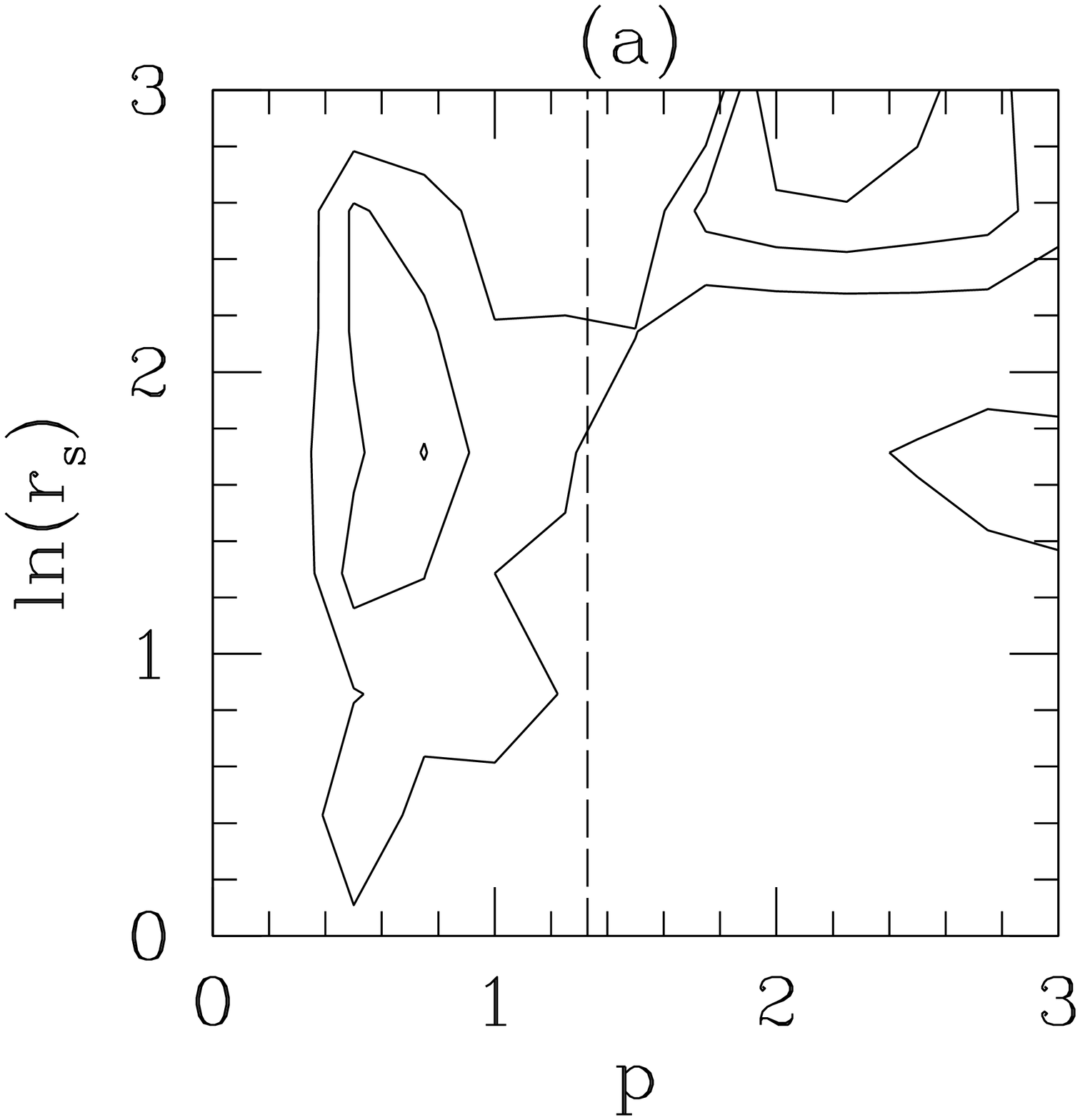,width=4.5cm,angle=0}
\epsfig{file=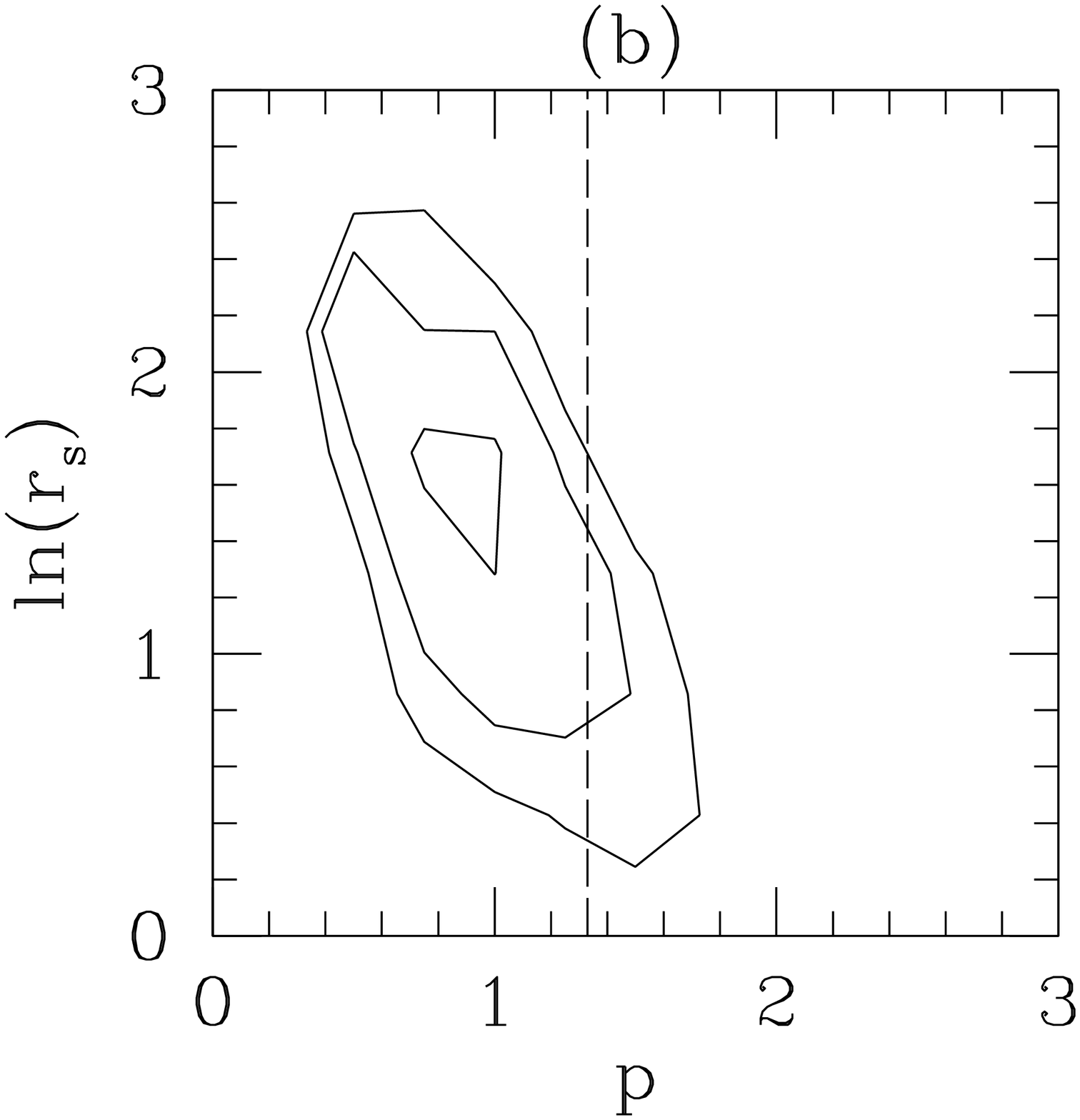,width=4.5cm,angle=0}
\epsfig{file=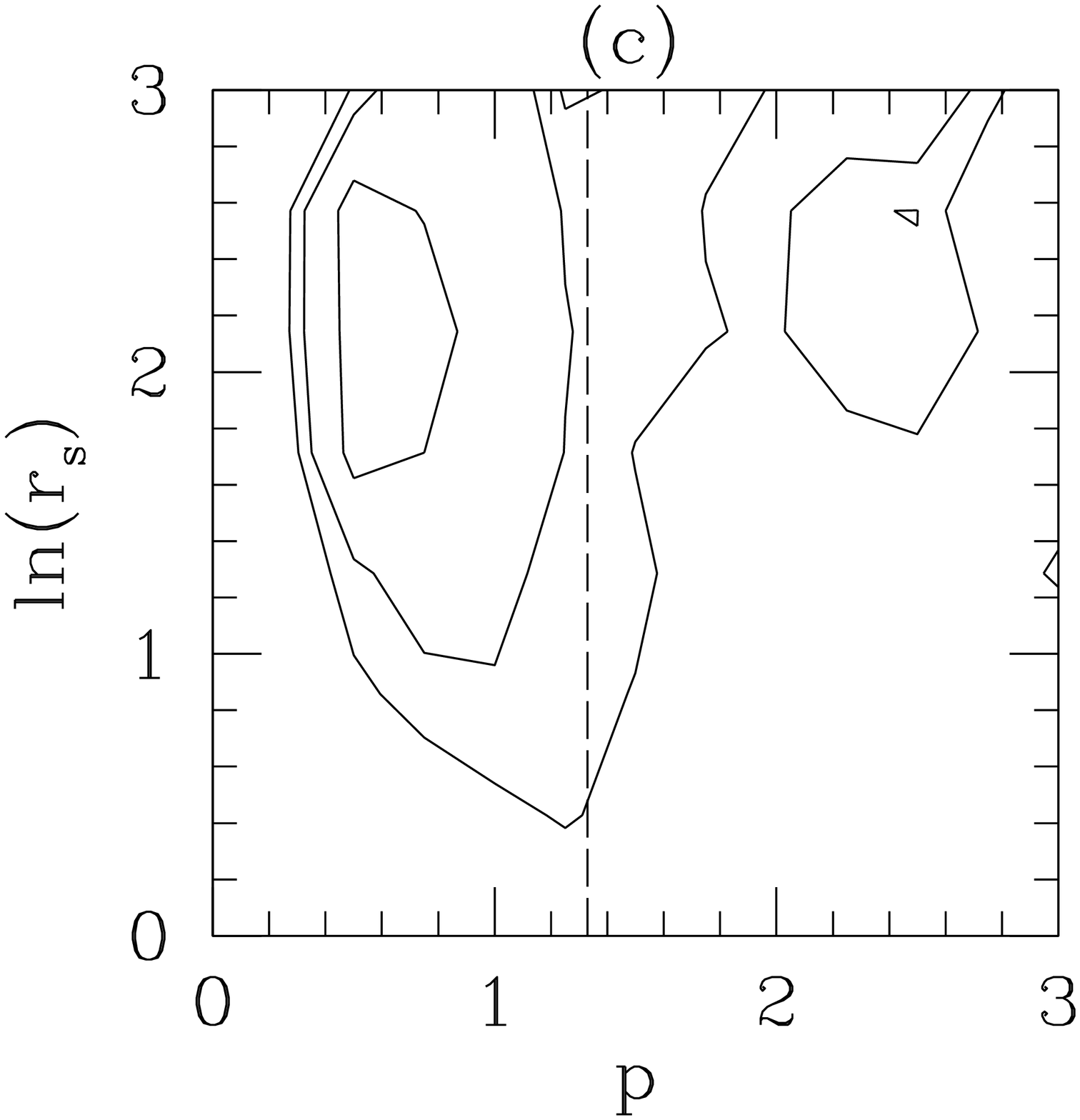,width=4.5cm,angle=0}\\
\epsfig{file=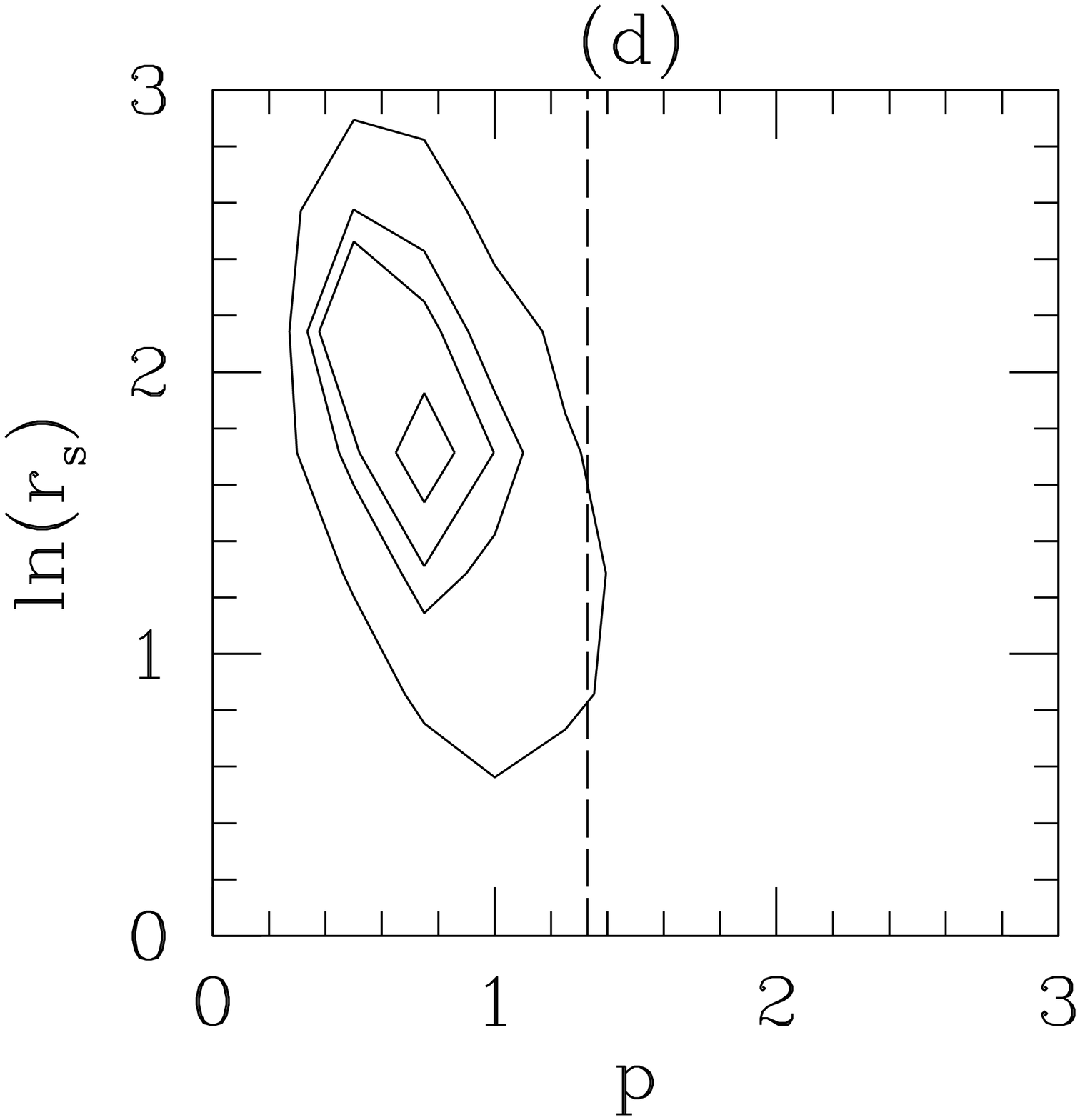,width=4.5cm,angle=0}
\caption{\footnotesize{Two-dimensional pdfs obtained using the
    measured chromatic microlensing for HE1104-1805 (Table
    \ref{map_he1104}) for both linear ({\em top}) and logarithmic
    ({\em bottom}) grids in $r_s$. Contours are $0.5 \sigma$, $1
    \sigma$, and $1.5 \sigma$ confidence levels respectively.  From
          {\em left} to {\em right} pdfs for: our MMT/VLT data (a),
          average of broadband data previous to 2003 (b),
          \cite{poindexter07} data corrected by time delay (c), and
          the intersection among the three previous maps (d).  In the
          intersection maps we also show the contour corresponding to
          $2 \sigma$ confidence level.}
\label{he1104size}}
\end{center}
\end{figure*}

The resulting pdfs are plotted in Figure \ref{he1104size}, and the
expected values and uncertainties in Table \ref{pdfs_he1104}. The
  pdfs corresponding to the MMT/VLT (Fig.\ref{he1104size}a) and to the
  \cite{poindexter08} data (Fig.\ref{he1104size}c) are not as 
concentrated near the maximum of the pdf as in the case of the
  broad-band data (Fig \ref{he1104size}b). These pdfs may present a
secondary maximum (perhaps due to the complexity of the microlensing
phenomenon corresponding to this epoch) and, individually
  considered, are not very conclusive. However, the product pdf
  strongly increases the concentration of the probability near the
  maximum and the significance of the estimates: $r_s=6\pm 2 \, \rm
light-days (15.5\pm5.2 \times 10^{15} \, \rm cm$), $p=0.7\pm0.1$ for
the linear prior and $r_s=6^{+2}_{-1} \, \rm light-days 
(15.5^{+5.2}_{-2.6} \times 10^{15} \, \rm cm$), $p=0.7\pm0.1$ for the
logarithmic prior.  Our $r_s$ estimates correspond to one half light
radius at the central wavelength of B filter, $R_{1/2}(\lambda4311)=
8\pm 2 \, \rm light days (76\pm 19 \times 10^{17} \, \rm cm$).  This
value is in good agreement with the results obtained by \cite{munoz11}
with HST data and by \cite{poindexter08} from photometric monitoring.
The values of $p$ are, however, considerably smaller.

The microlensing-based size estimates are significantly larger than those
inferred from the black-hole mass or from the observed I-band flux
\cite[see][]{poindexter08}.

\subsection{SDSS1029+2623}

SDSS1029+2623 was discovered by \cite{inada06}; it consists of two
images $A$ and $B$ separated by $22.5 \arcsec$ at $z_s=2.197$ and a
cluster lens galaxies at $z_L \sim 0.55$. Recently \cite{oguri08}
found a third image $C$ $1.8 \arcsec$ from $B$ and several complex
absorption systems in the emission lines.

\begin{figure*}
\begin{center}
\epsfig{file=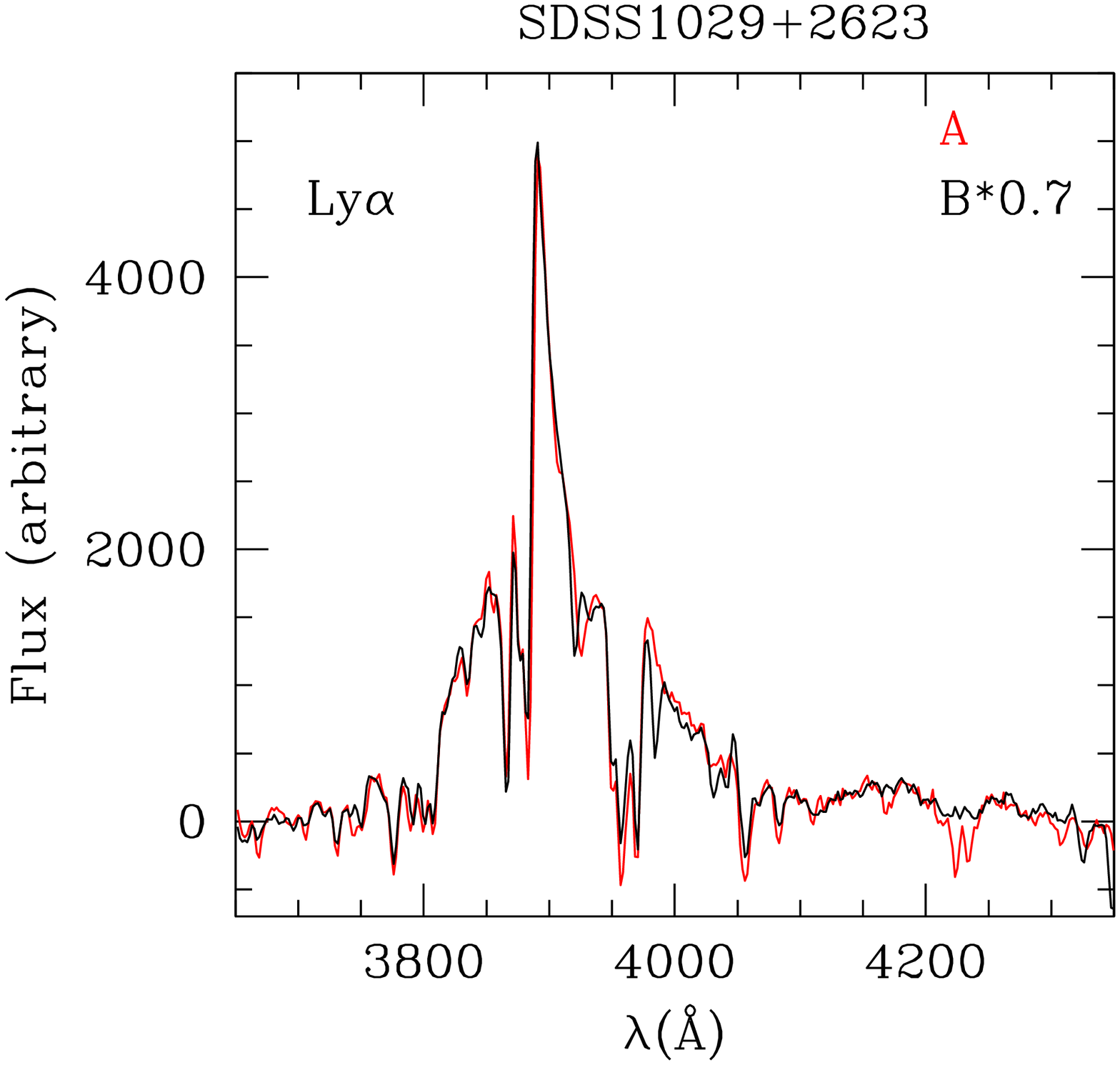,width=5cm,angle=0}
\epsfig{file=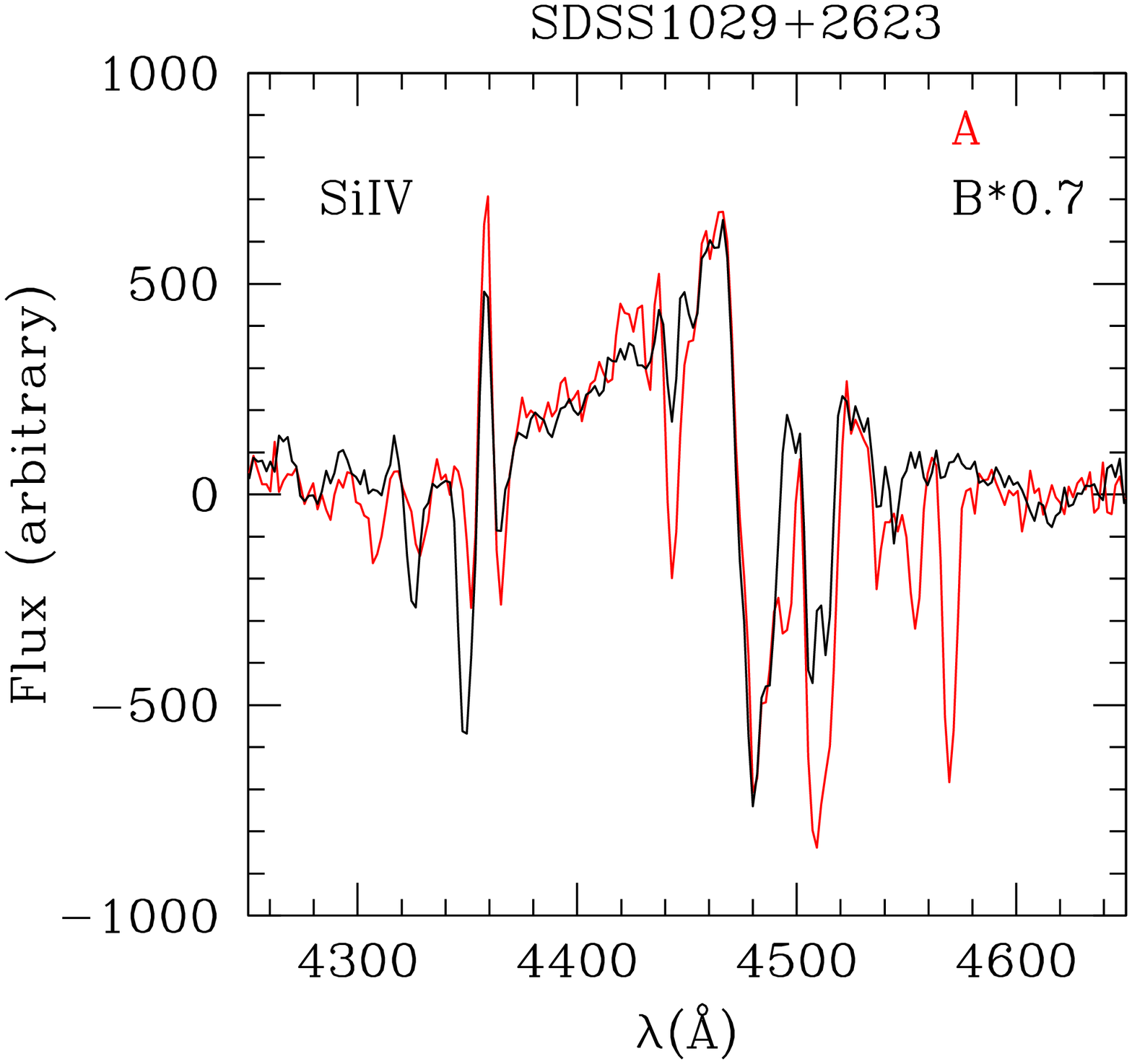,width=5cm,angle=0}
\epsfig{file=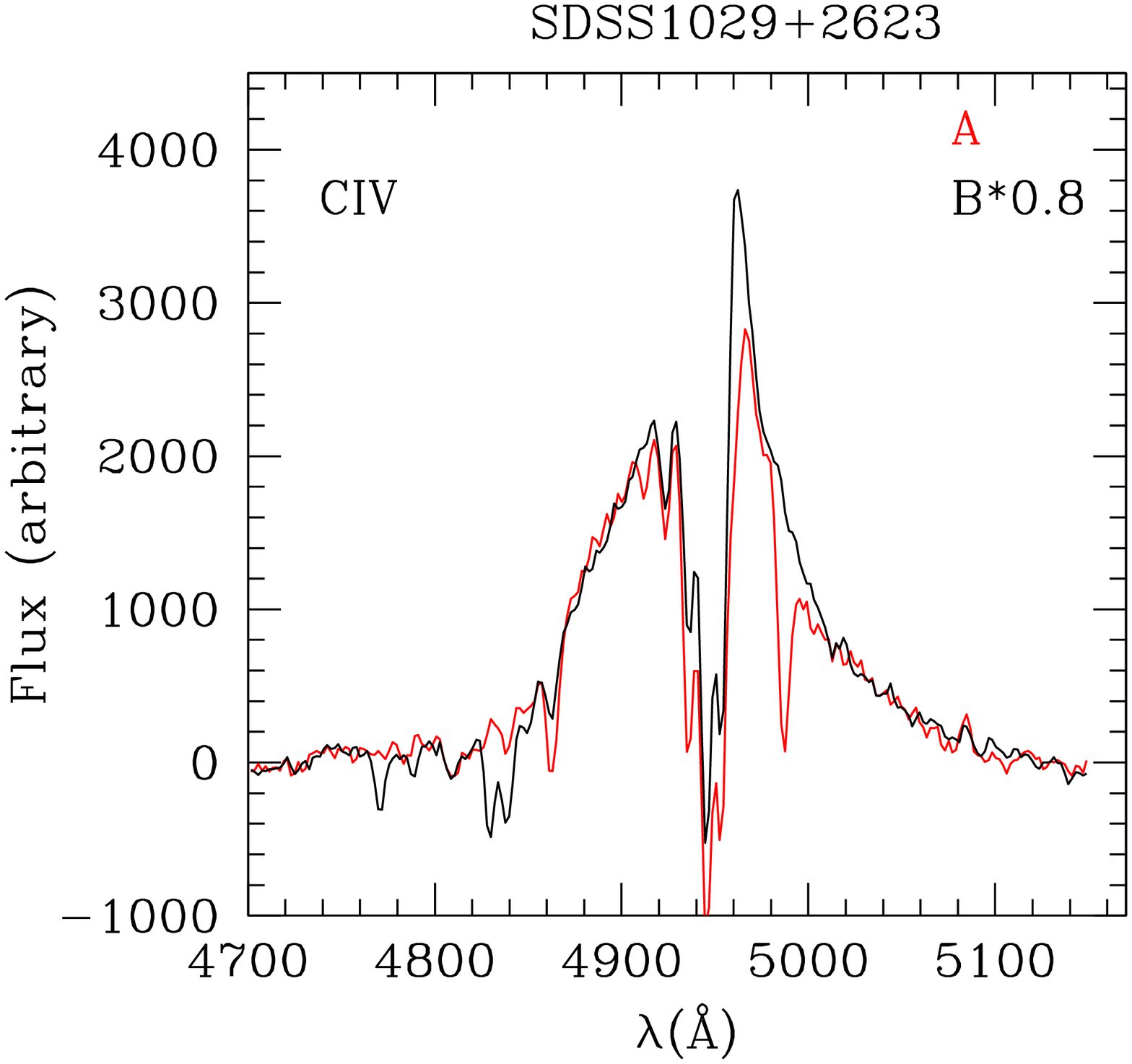,width=5cm,angle=0}
\epsfig{file=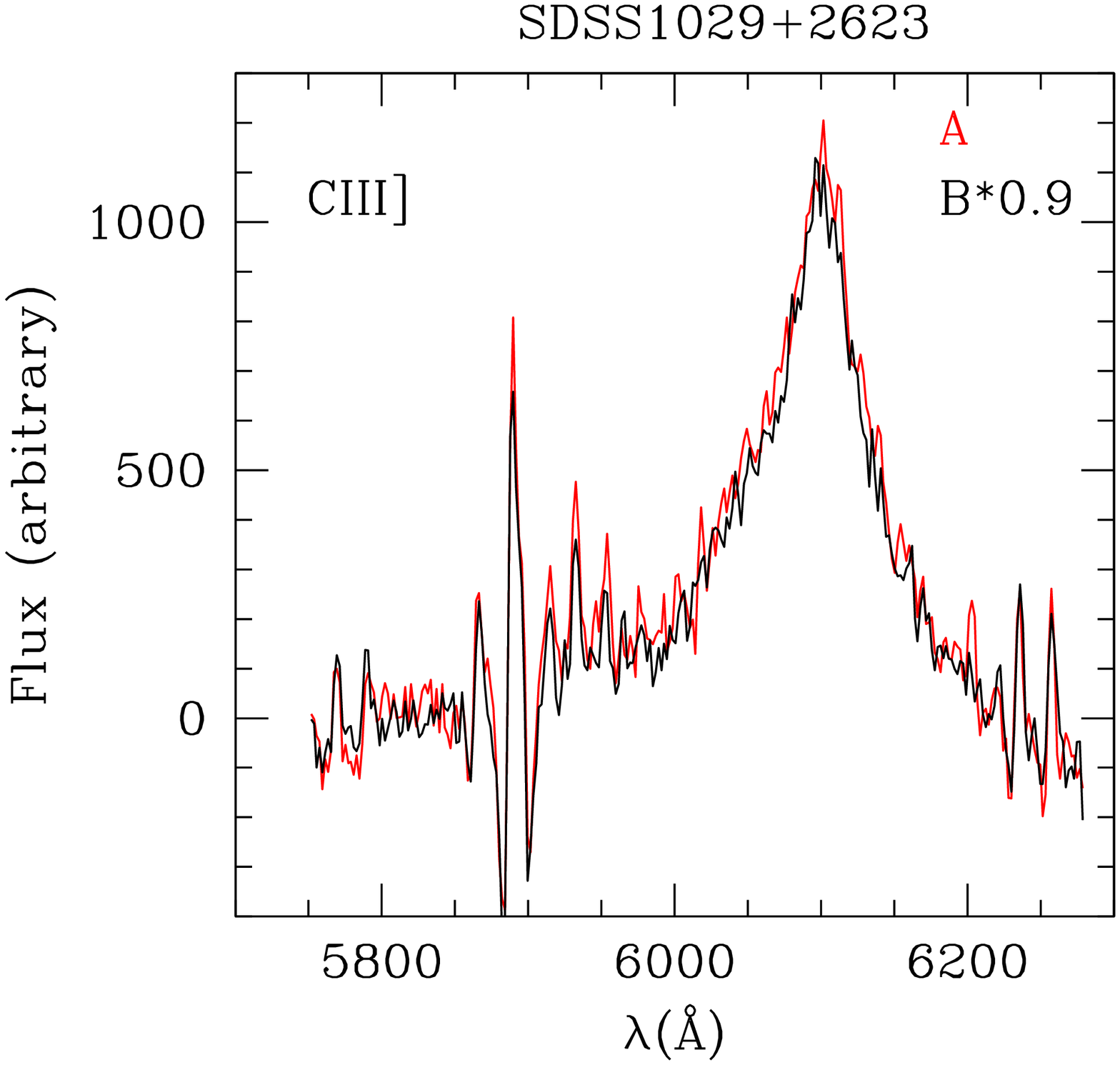,width=5cm,angle=0}
\caption{Ly$\alpha$, SiIV, CIV, CIII emission line profiles for
  SDSS1029+2623 vs. observer $\lambda$.  The {\em red line} represents
  the continuum subtracted emission lines for $A$.  The {\em black
    line} represents the continuum subtracted emission line for $B$
  multiplied by a factor to match the peak of $A$.  The factors are
  shown in each panel.  Sky lines are seen on both sides of CIII].
\label{prof2_sdss1029}}
\end{center}
\end{figure*}

Although the emission line profiles are similar for $A$ and $B$ (Figure
\ref{prof2_sdss1029}), there are several groups of absorption line
systems affecting Ly$\alpha$ and CIV that are associated with
MgI/MgII/FeII absorption systems, as found by \cite{oguri08}.  There
are also self-absorption systems associated with Ly$\alpha$, SiIV, and
CIV lines that are present in both components but with significant
differences.  In spite of this we have attempted to determine flux
ratios by defining suitable integration windows to avoid the
absorptions. In the case of CIII], the emission line profiles are
  almost identical in both components and show no absorption
  lines. Thus, the results derived from CIII] should be more reliable
    than the results inferred from the other lines.

\begin{figure*}
\begin{center}
\plotone{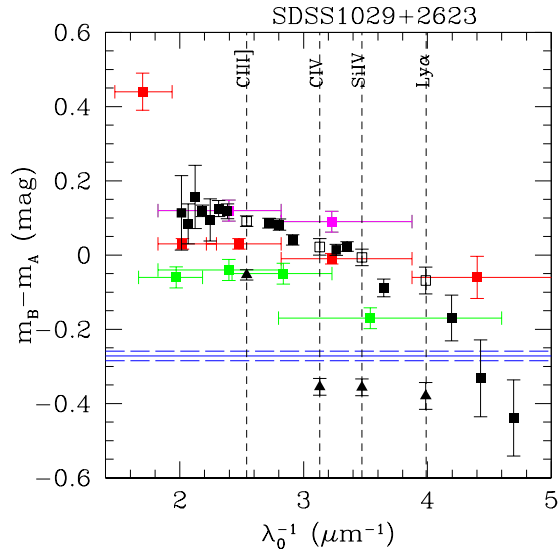}
\caption{Magnitude differences $m_B-m_A$ vs $\lambda_0^{-1}$
  ($\lambda$ in the lens galaxy restframe) for SDSS1029+2623.  {\em
    Solid squares} represent the integrated continua, in color those
  obtained by \cite{inada06} ({\em red}) and \cite{oguri08} ({\em
    green} and {\em magenta} represent data obtained in 2007 and in
  2008 respectively), and in {\em black} those obtained from our
  spectra. {\em Open black squares} represent the difference in the
  integrated fitted continua under the emission lines. {\em Black
    triangles} are the magnitude differences in the emission
  line cores. The {\em blue line} represents the magnitude difference and
  its error ({\em blue dashed lines}) at radio wavelengths
  \citep{kratzer11}.
\label{diff_sdss1029a}}
\end{center}
\end{figure*}

The $B-A$ magnitude differences obtained from our data (continuum and
emission lines) compared to those obtained by \cite{inada06} and
\cite{oguri08} are shown in Figure \ref{diff_sdss1029a}
(Table~\ref{mag_sdss1029}). Our continuum flux ratio agrees well with
the data corresponding to the $g$ and $K$ broadband filters from
\cite{oguri08} that were taken with the Keck at the same
epoch. However, there is a difference of $\sim 0.1$ mag with the data
taken at other epochs. This is explained by variability in the
continuum between 2007 and 2008 (another peculiar feature is that the
measurement in the $z$ band \citep{inada06} is $\sim 0.3$ mag above
all the other broad-band measurements). In principle the variability
could be attributed to microlensing or intrinsic variability of the
quasar continuum combined with a time lag between both
components. However, the strong chromaticity of the continuum flux
ratio (of about 0.4 mag) that exceeds the $\sim 0.1$ mag global offset
between continuum flux ratios at different epochs excludes the
explanation based on intrinsic continuum variability. Dust extinction,
on the other hand, cannot explain the chromaticity for the flux ratio
inferred from radio observations agrees with the flux ratio of the
bluest continuum contrary to the expectations under this hypothesis.
Thus, microlensing is the more likely
explanation and is supported by the agreement of the flux ratios
inferred from three of the lines, CIV, SiIV, and Ly$\alpha$ with the
radio flux ratio. 
However, the flux ratio inferred from the other emission line, CIII] 
which presents the smoothest line profile, shows a
large offset with respect to the baseline defined by the radio data
that disagrees with the microlensing hypothesis.  Under the
hypothesis of chromatic microlensing we could follow the same steps
as in the case of SDSS1004+4112 to estimate the size and temperature
profile of the quasar source in SDSS1029+2623. However, lens
modeling in this system is complex \cite[see e.g.][]{kratzer11} and
we defer this study to future work.

\subsection{Q0957+561}

\begin{figure*}
\begin{center}
\epsfig{file=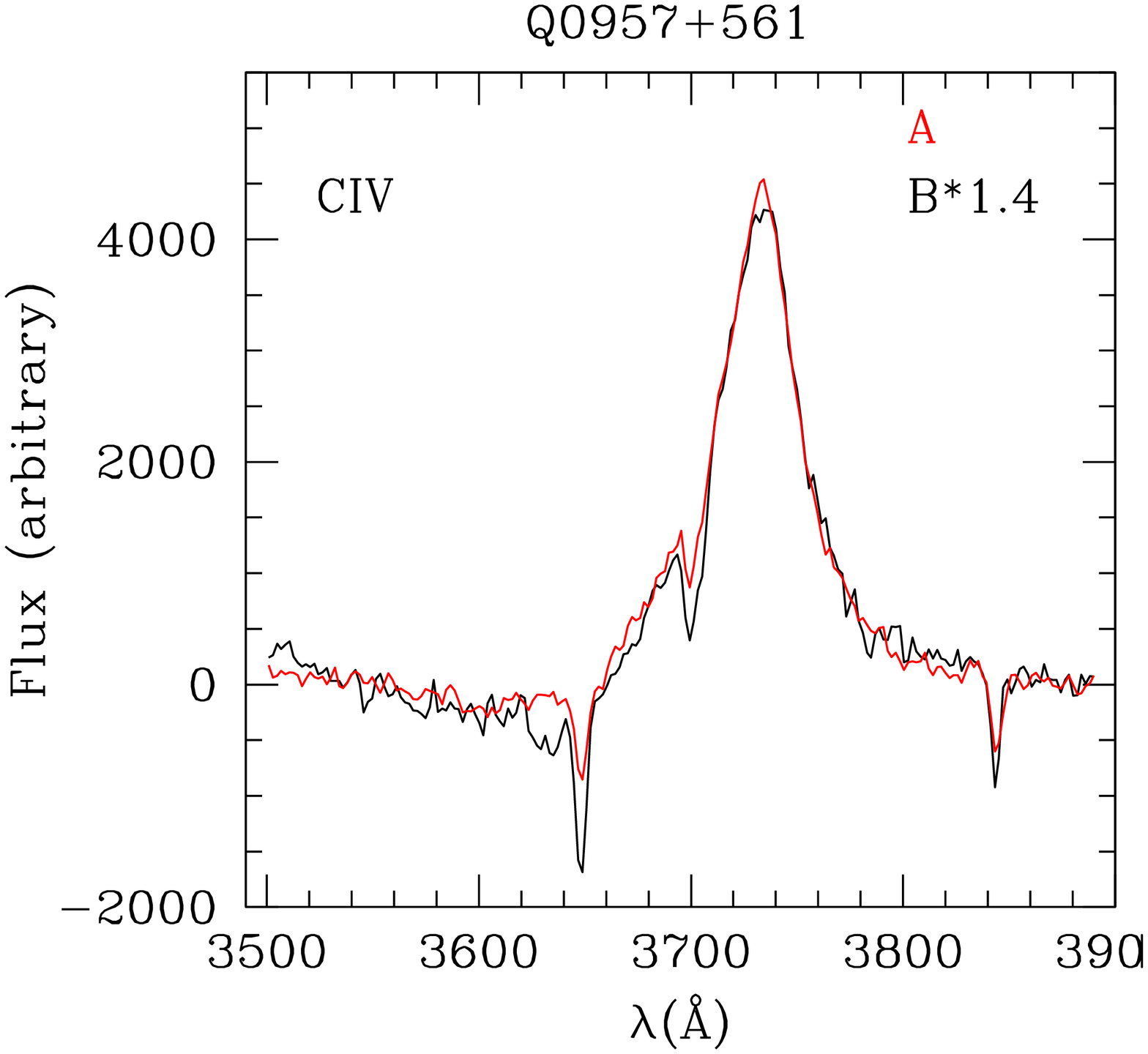,width=5cm,angle=0}
\epsfig{file=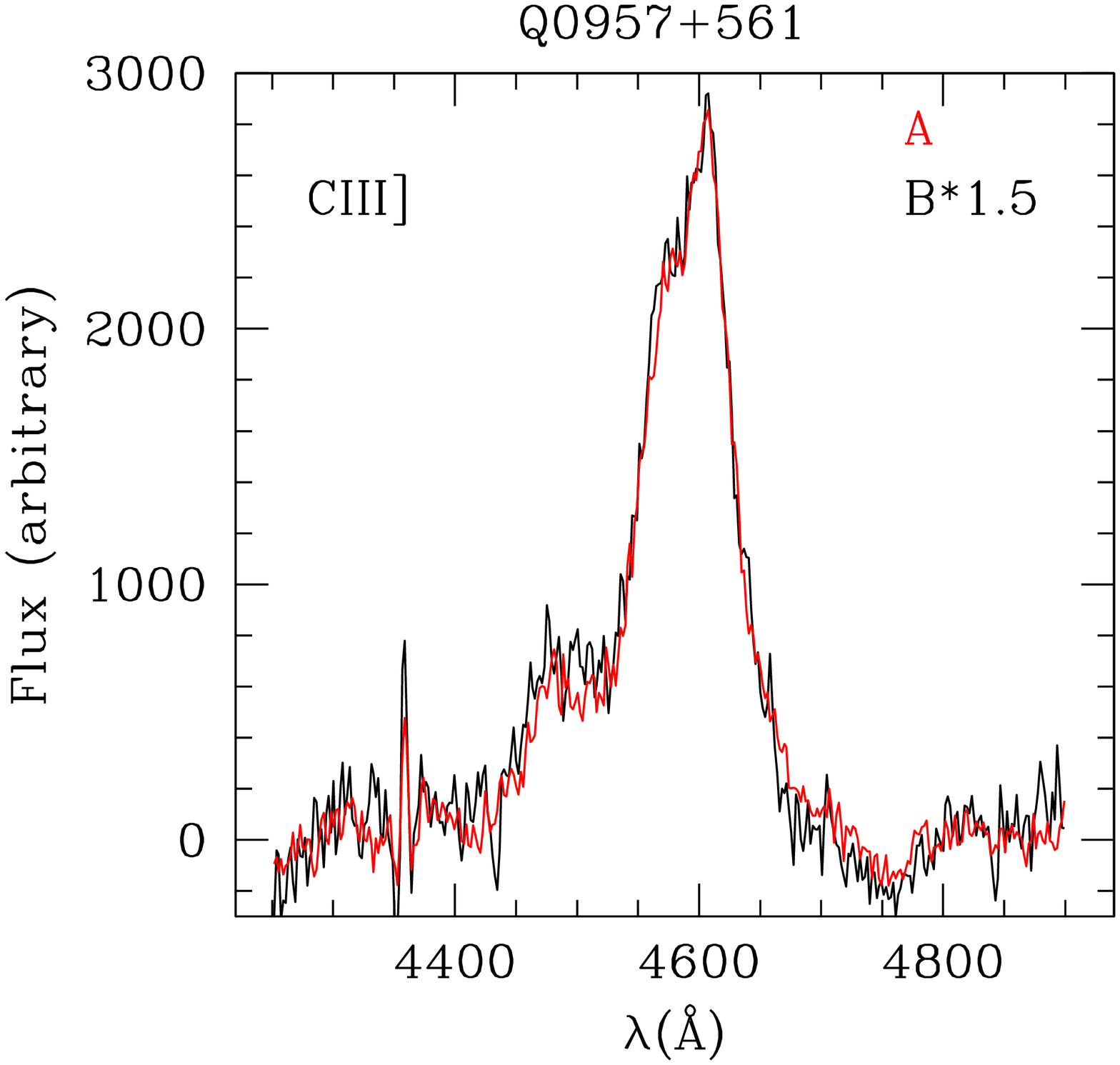,width=5cm,angle=0}
\epsfig{file=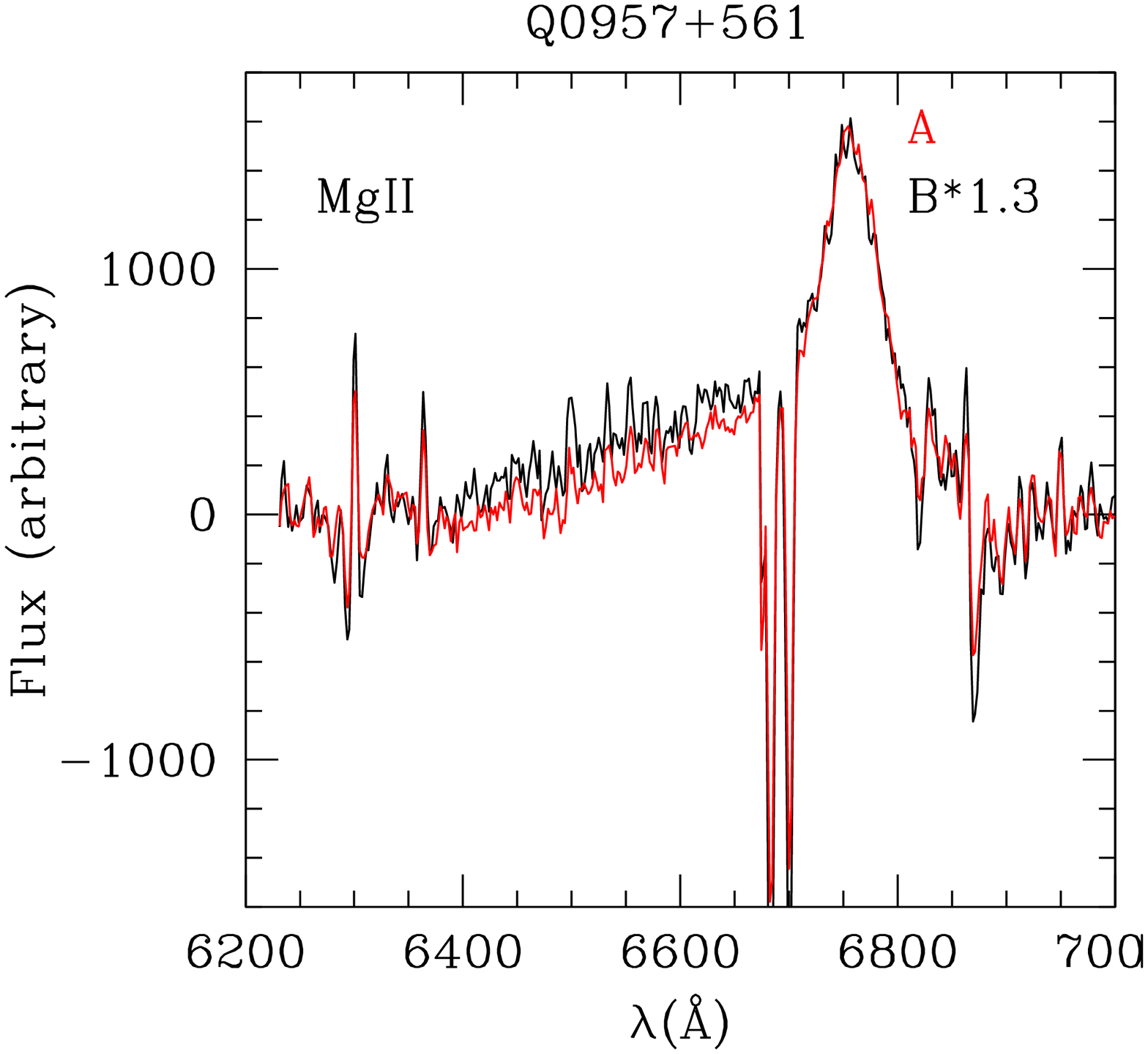,width=5cm,angle=0}
\caption{CIV, CIII], MgII emission line profiles for Q0957+561 vs
 observed $\lambda$. The {\em red line} 
represents the continuum subtracted emission lines for $A$. 
The {\em black line} represents the continuum-subtracted emission lines for $B$ 
multiplied by a factor to match the peak of $A$. 
The factors are shown in each panel.
\label{prof2_q0957}}
\end{center}
\end{figure*}

The first known gravitational lens was discovered by \cite{walsh79};
it has been studied in great detail. It consists of two images $A$ and
$B$ with separation $6\farcs2$.  The source QSO is at $z_s=1.41$ and
the main lens galaxy is at $z_l=0.36$ and is part of a poor cluster of
galaxies.  Comparison between $A$ and $B$ emission lines (Figure
\ref{prof2_q0957}) do not show significant differences between the
emission line profiles of CIV, CIII] and MgII.  This limits the
possible impact of microlensing on the broad component of the
emission lines. To quantify this impact we have compared the $B/A$
flux ratios of the wings and the core of the CIV emission line (that
has the highest S/N ratio) finding differences $<$10\%. In any case
we have computed flux ratios from the cores of the lines.

\begin{figure*}
\begin{center}
\plotone{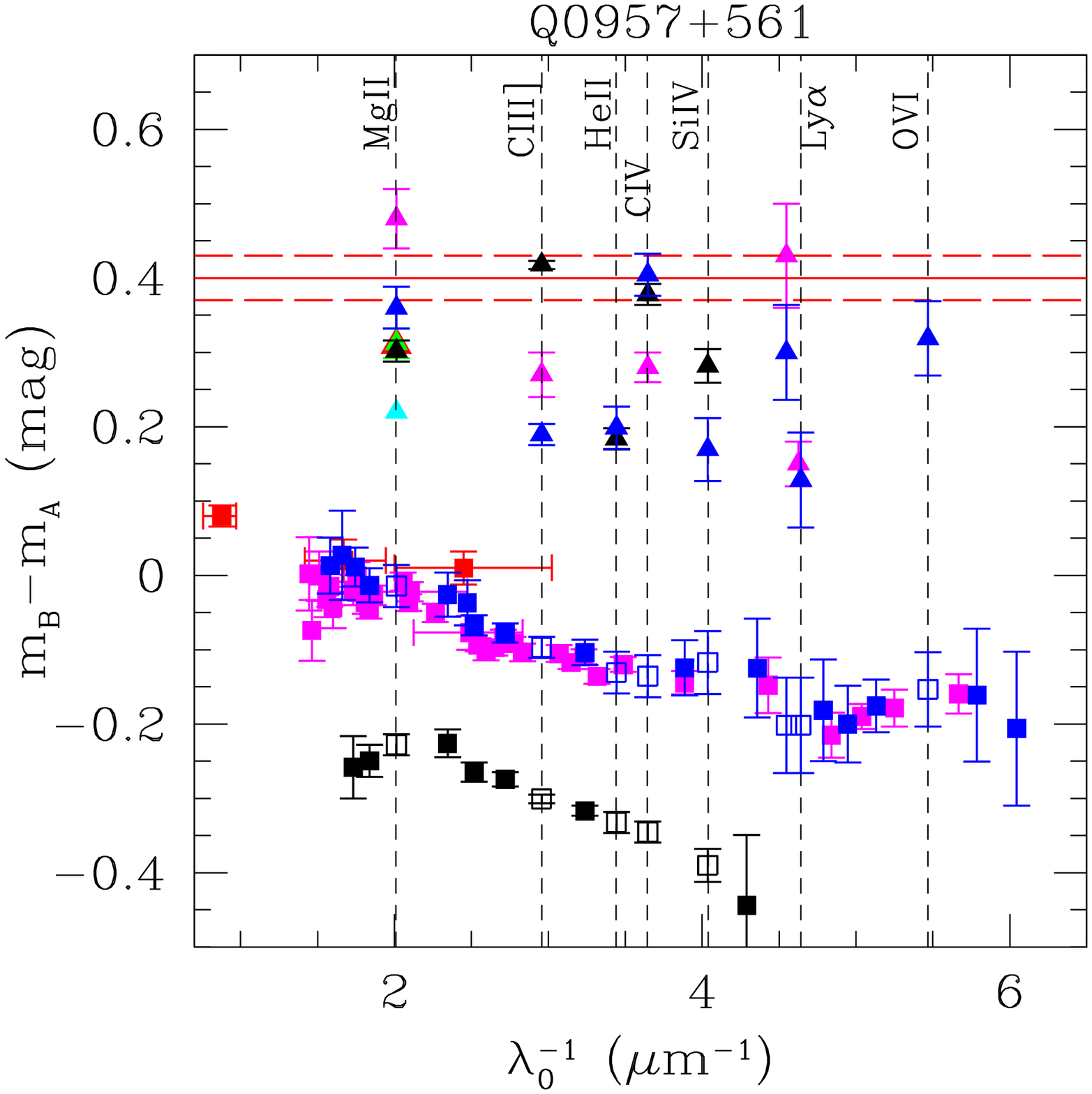}
\caption{\footnotesize{Magnitude differences $m_B-m_A$ vs
    $\lambda_0^{-1}$ ($\lambda$ in the lens galaxy restframe) for
    Q0957+561.  {\em Black and blue} represent the magnitude
    difference obtained from MMT and HST spectra respectively.  {\em
      Solid squares} are the magnitude difference in the integrated
    continuum, {\em open squares} are the integrated continuum under
    the emission lines, and {\em solid triangles} are the integrated
    emission line cores.  The data obtained from other authors are also
    plotted following the previous code: {\em solid squares} are
    broadband data or integrated continuum, and {\em solid triangles}
    are emission line cores. The code for the colors is: {\em red} CASTLES
    data for the continuum \citep{bernstein97} and
    \cite{vanderriest93} data for MgII emission line, {\em cyan}
    estimated from the spectra of Mediavilla et al. (2000), {\em
      magenta} \cite{goicoechea05b}, and {\em green} \cite{schild91}.
    The {\em Red line} is the mean magnitude difference and standard deviation 
of the mean ({\em red
      dashed lines}) at radio wavelengths \citep{conner92,gorenstein88,haschick81}. }
\label{diff_q0957a}}
\end{center}
\end{figure*}

\begin{figure*}
\begin{center}
\plotone{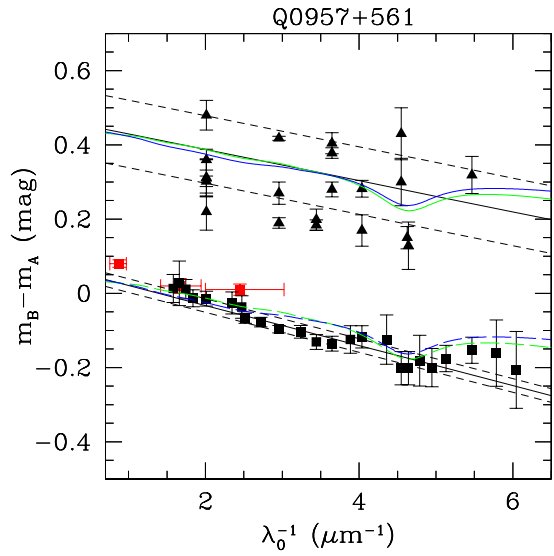}\\
\caption{\footnotesize{Model fitted to the data shown in Figure
    \ref{diff_q0957a}.  The {\em black lines} represent the function
    fitted to the emission line cores ({\em solid triangles}) and the
    continua ({\em solid squares}) obtained with HST respectively.
    {\em Dashed lines} are the standard deviation for each fit.  The
    {\em blue curve} represent the dust extinction function fitted to
    the emission line cores using $R_V=3.1$ ($E(B-V)=0.02\pm0.09$ with
    $\chi ^2_{DOF}=1.6$).  The {\em green line} is the dust extinction
    fitted using variable $R_V$ ($R_V=2.0\pm0.1$, $E(B-V)=0.02\pm0.09$
    with $\chi ^2_{DOF}=1.8$).  The curves shifted $-0.4$~mag ({\em
      dashed blue} and {\em dashed green}) fit the HST continua.}
\label{diff_q0957b}}
\end{center}
\end{figure*}

Figure \ref{diff_q0957a} (see also Table \ref{mag_q0957}) shows the
$B-A$ magnitude differences in the continuum and in the emission lines
for Q0957+561. This figure also includes other data from the
literature and a re-analysis of HST/STIS data by
\cite{goicoechea05b}. Averaging the radio data from \cite{conner92} at $\lambda=6 \rm cm$, 
\cite{gorenstein88} at $\lambda=13 \rm cm$, and \cite{haschick81} at $\lambda=6 \rm cm$, 
the $B-A$ magnitude difference uncontaminated by
the lens galaxy continuum and free from dust extinction is obtained,
$\rm <B-A>_{radio}=0.40\pm0.03$~mag. 
The $B-A$ magnitude differences
corresponding to the emission lines follow a decreasing trend towards
the blue compatible with extinction. A linear fit to the emission line
magnitude differences (see Figure \ref{diff_q0957b}) has a slope of
$-0.04\,\rm mag\, \mu m^{-1}$ and a dispersion of 
0.09~mag. This dispersion is reasonable taking into account the
intrinsic difficulty and the inhomogeneity of the data analysis
procedures followed by the different authors especially regarding the
criteria used to select the continuum. Towards the red this linear fit
is fully consistent with the $B-A$ radio measurements, confirming that
the emission lines are not significantly affected by microlensing.

In Figure \ref{diff_q0957b} we also present an extinction curve fit to
the $B-A$ emission line magnitude differences (both from the
literature and from our own measurements). The best-fit parameters
(obtained fixing the dust redshift to the lens redshift) were: 
$\Delta E(B-V)=0.02\pm0.009$ and $R_V=2.0\pm0.1$ 
($\chi ^2_{DOF} = 1.8$)
The data are also compatible with an extinction curve similar to the 
Milky Way ($\Delta E(B-V)=0.02\pm0.009$, $\chi ^2_{DOF}= 1.6$).
It is remarkable that the fitting
to the narrow emission lines is in agreement with other data
obtained from the continuum $\Delta E(B-V)=0.02\pm0.02$
\citep{falco99}. However our results are not in agreement 
with the values $\Delta E(B-V)=0.068\pm0.005$ and 
$R_V=4.4\pm0.5$ found by \cite{goicoechea05b}.

The continuum data also show a decreasing trend towards the blue with
a slightly steeper slope ($-0.05\pm0.01 \,\rm mag\, \mu m^{-1}$ for
CASTLES and $-0.05\pm0.02 \,\rm mag\, \mu m^{-1}$ for HST/STIS). In
fact, the same extinction curve fitted to the emission line data with
a shift of $B-A\sim -0.4 \, \rm mag$ fits well the continuum flux
ratios from CASTLES and HST/STIS (notice that these continuum flux
ratios obtained from HST data are not affected by lens galaxy
contamination). This global shift between the continuum and the
baseline of no microlensing magnification defined by the emission
lines imply that the continuum is experimenting microlensing of $\sim
-0.4 \, \rm mag$ and allows us to re-examine the microlensing history
in Q0957+561 based in this result. The differential $B-A$ magnitude
lightcurve of Q0957+561 \citep{pelt98,oscoz02} can be described as an
event of 0.25~mag taking place from 1981 to 1986 and a quiet phase of
mean value $\rm <B-A>\sim -0.05$~mag with fluctuations of less than
0.05~mag from 1987 to 1999 (Oscoz et al. 2002). Previous attempts to
model the observed microlensing through simulations have not used the
emission line flux ratios as microlensing zeropoint. These studies
have either accepted any zeropoint for microlensing magnification,
modeling microlensing variability of less than 0.05 mag with respect
to an unrestricted zeropoint value \cite[e.g.][]{refsdal00}, or
implicitly supposed that the zeropoint was placed at the mean value of
the quiet phase ($\rm <B-A>\sim -0.05$~mag) modeling a microlensing
amplitude of less than 0.05~mag \cite[e.g.][]{wambsganss00}.  However,
considering the zeropoint defined by the emission line flux ratios the
correct procedure will be to model fluctuations of 0.05~mag with
respect to a mean microlensing amplitude of $\rm <B-A>\sim -0.4 \, \rm
mag$.

Although detailed microlensing simulations should be made to estimate
physical parameters from the source and/or the microlenses, it seems
that under this new perspective the likelihood of smaller sources (or
bigger microdeflectors) will increase.

On the other hand, there is an offset of $\sim -0.2$ mag between the magnitude
differences in the continuum obtained with the MMT and those obtained with
HST/STIS (data obtained at different epochs to correct for time delay). This
difference between MMT and HST/STIS data might be explained by (i) intrinsic
variability (ii) a microlensing amplitude change, and (iii) continuum
contamination by the lens galaxy.  We plan to examine available
photometric monitoring of Q0957+561 covering 2008 (phased by the lag associated
with component $B$) to ascertain the origin of this offset.

If we compare the slopes of the linear fits corresponding to emission
lines and continua (CASTLES plus HST/STIS data) we found that the
$B-A$ offsets at MgII and at OVI wavelengths have a difference of
$\sim 0.15$~mag, these may be due to a wavelength dependence of
microlensing (chromatic microlensing). This estimate is, however,
greatly affected by the uncertainties in the determination of the
emission line flux ratios.

In summary, for Q0957+561 our results indicate that: (i) there is no
significant variation in the broad component of the emission line
profiles, (ii) there is dust extinction affecting the emission lines
and the continuum produced by dust likely at the same redshift as the
lens galaxy and (iii) there is microlensing with amplitude $\rm
B-A\sim -0.4 \, \rm mag$ affecting the continuum.

\subsection{HS0818+1227}

\begin{figure*}
\begin{center}
\epsfig{file=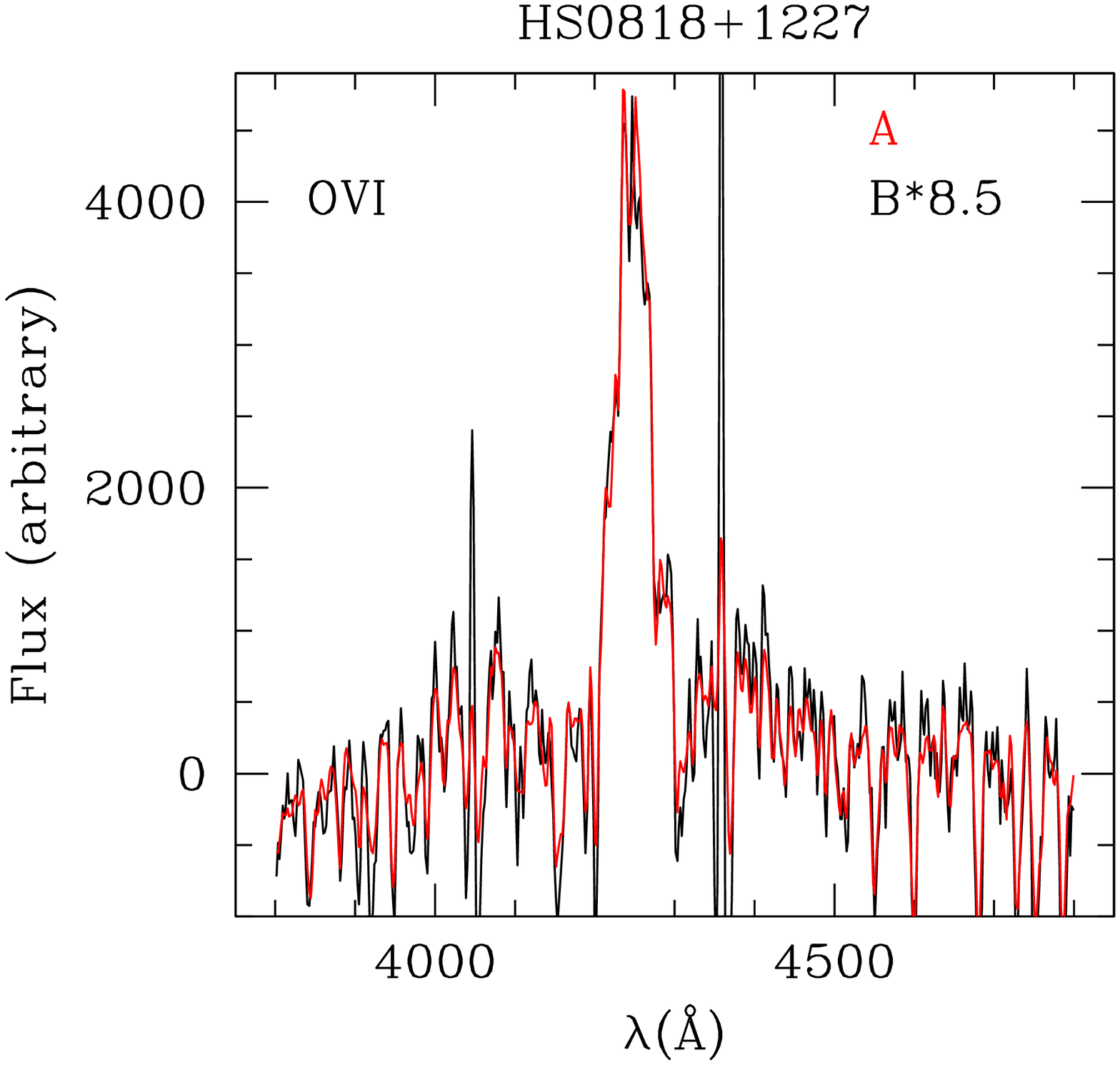,width=5cm,angle=0}
\epsfig{file=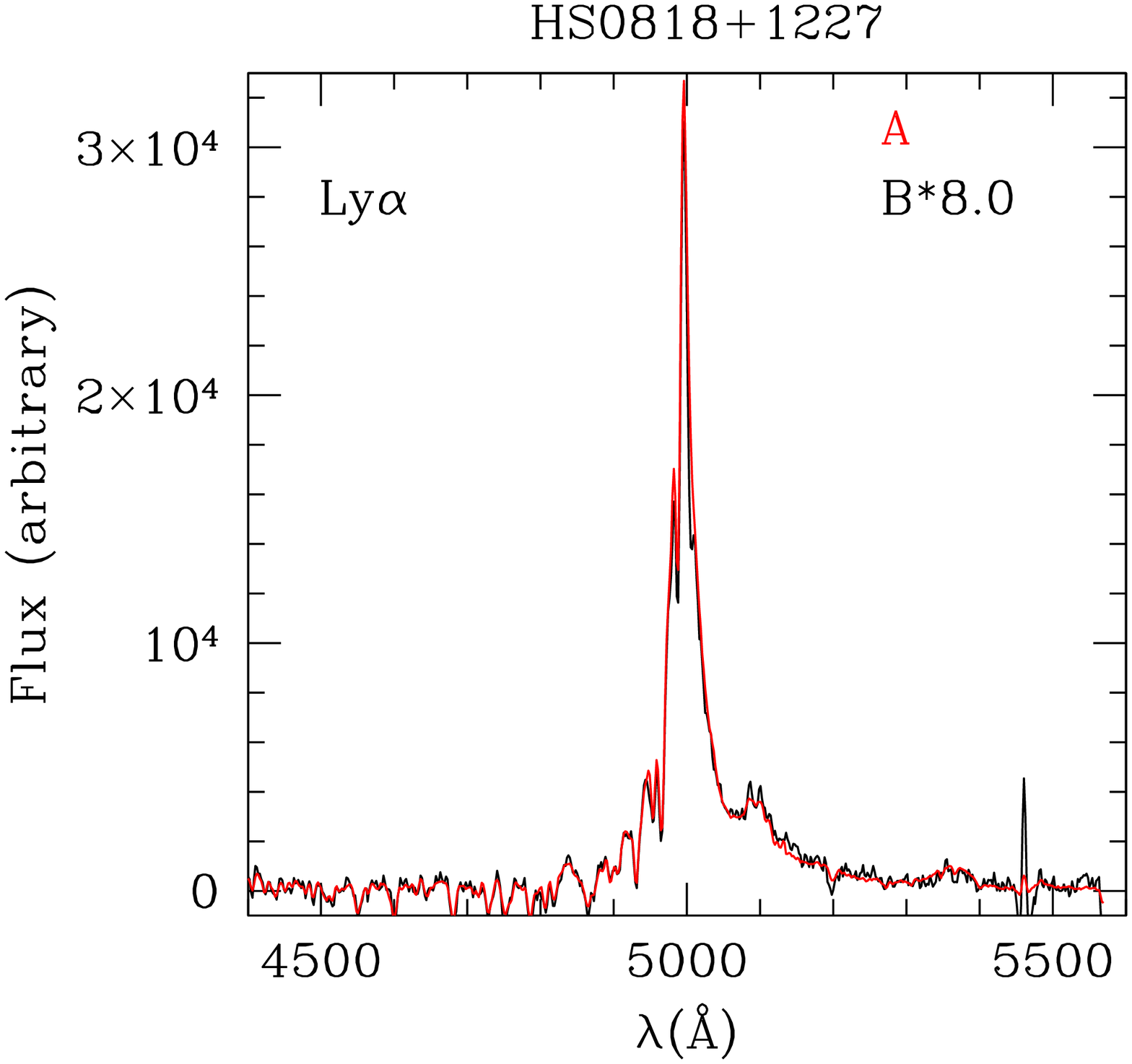,width=5cm,angle=0}
\epsfig{file=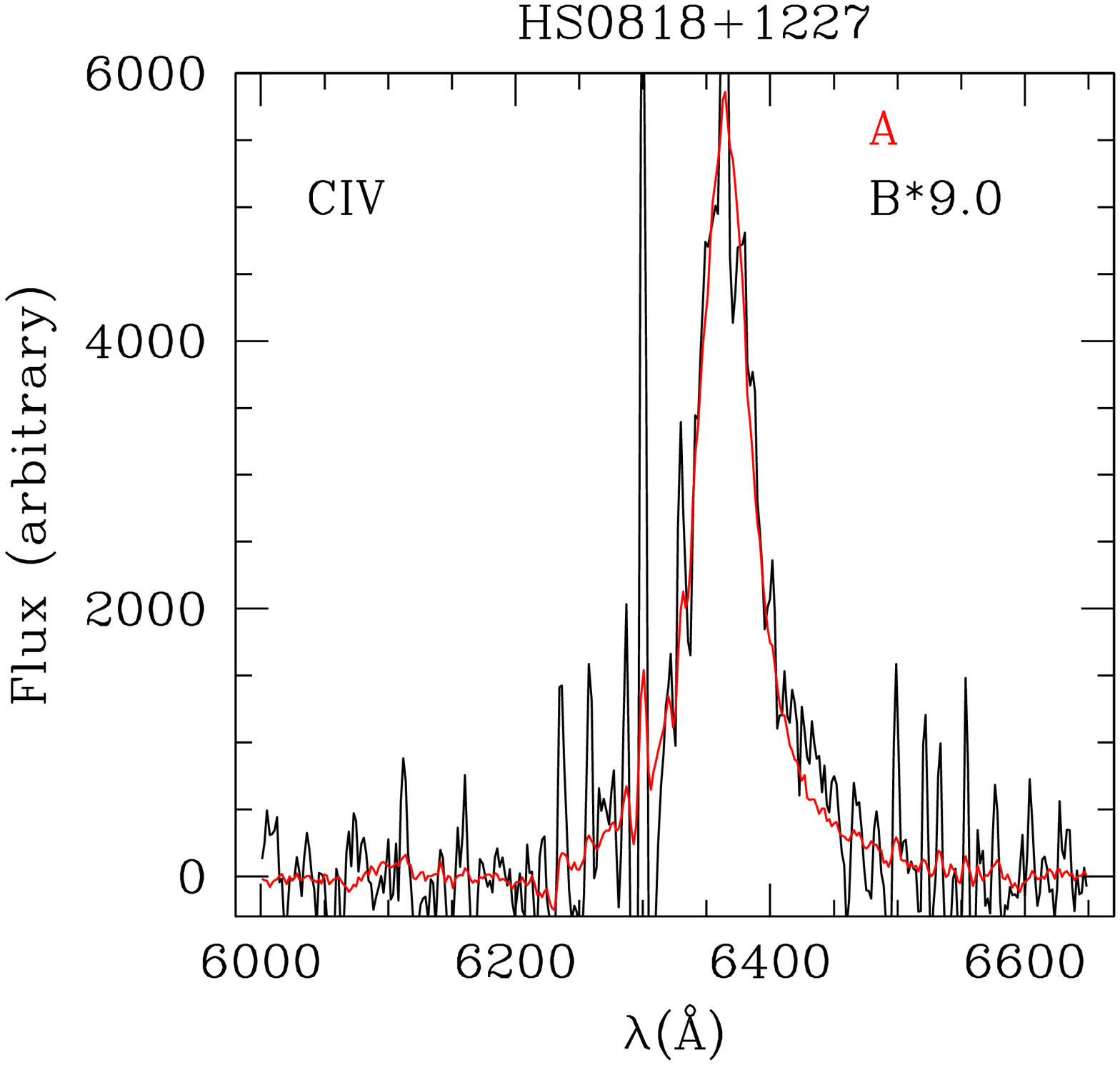,width=5cm,angle=0}\\
\caption{OVI, Ly$\alpha$, and CIV emission line profiles for
  HS0818+1227 vs. observed $\lambda$.  The {\em red line} represents
  the continuum-subtracted emission lines for image $A$.  The {\em
    black lines} represent the continuum subtracted emission lines for
  $B$ multiplied by factors to match the peak of $A$.  The factors are
  shown in each panel.
\label{prof2_hs0818}}
\end{center}
\end{figure*}

HS0818+1227 was discovered by \cite{hagen00}; it consists of two
lensed images $A$, $B$ separated by $2.6 \arcsec$, with $z_s=3.115$
and $z_l=0.39$.  Image $B$ in this case is $\sim 2$ mag fainter than
$A$. We scaled the continuum-subtracted $B$ spectrum to match the
emission line peaks in A (Figure \ref{prof2_hs0818}).  The $A$ and $B$
emission line profiles are very similar to each other and do not show
significant differences in the BLR.

\begin{figure*}
\begin{center}
\plotone{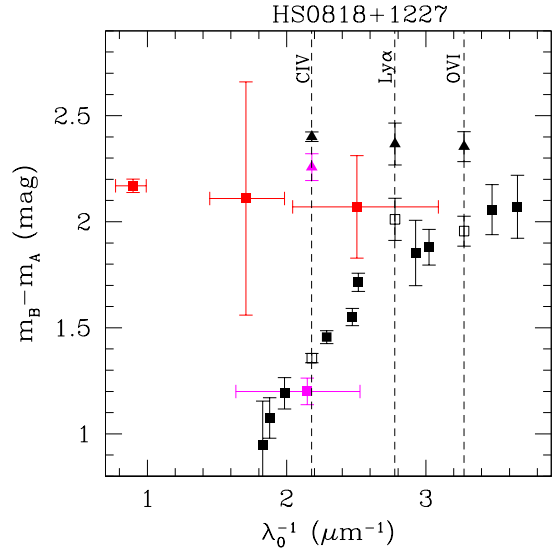}
\caption{Magnitude differences $m_B-m_A$ vs $\lambda_0^{-1}$
  ($\lambda$ in the lens galaxy restframe) for HS0818+1227.  {\em
    Solid squares} represent the integrated continuum obtained by
  CASTLES ({\em red}), \cite{hagen00} ({\em magenta}), and from our
  spectra ({\em black}). The {\em magenta triangle} is the magnitude
  difference in CIV obtained by \cite{hagen00}.  {\em Open black
    squares} represent the integrated fitted continua under the
  emission lines. {\em Black solid triangles} are the magnitude
  difference in the emission line cores.
\label{diff_hs0818a}}
\end{center}
\end{figure*}

Figure \ref{diff_hs0818a} shows the $B-A$ magnitude differences we
calculated from the continuum (solid black squares) and from the cores
of the emission-lines (solid black triangles) integrating our MMT
spectra.  The $B-A$ magnitude differences corresponding to the narrow
emission lines (Table \ref{mag_hs0818}) will define a zero
microlensing baseline of $<m_B-m_A>= 2.34\pm0.03$~mag.  The average of
CASTLES broadband data ($<m_B-m_A>=2.12\pm0.03$~mag) shows an offset
of 0.22~mag with respect to this baseline. Note, however that the
significance of this offset is dominated by the F555W data taken by
CASTLES. According to section 3.5, part of this offset (0.1 mag)
  may arise from intrinsic variability.

Our continuum data agree with \cite{hagen00}, but do not match CASTLES
(Table \ref{lit}), especially in the reddest part. This discrepancy is
due to the lens galaxy continuum. We estimate, using the
  integrated broad-band magnitudes obtained by CASTLES for the lens
  galaxy, that the contamination is $\sim 40$ \% of the lens galaxy
flux.

Considering the emission lines, our results indicate 
negligible dust extinction and posible evidence of microlensing.

\section{Conclusions}

The method we use in this paper allows us to separate microlensing
from dust extinction without a theoretical model for the lens
system. We have demonstrated the method for the most complicated
cases: doubly-imaged quasars.

We tested the hypothesis that the cores of the emission lines do not
vary with time by comparing our own magnitude differences in the
emission lines with values from the literature that were obtained at
different epochs, including values corrected for measured time-delays,
and we conclude that they are nearly constant in time. Thus, except in
cases where extinction is significant, the magnitude differences in
the emission line cores are reliable estimators of the intrinsic
magnitude differences unaffected by microlensing.

Following \cite{yonehara08} we have estimated the impact of time
  delays in our microlensing measurements for our objects. In the worst
  case scenario (Q0957+561, SDSS1004+4142, and SDSS1029+2623) a time
  delay can introduce variabilities $\lesssim 0.2$ mag and
  chromaticities $\lesssim 0.1$ mag. The measurements we obtain for
  those objects are at least twice the estimated values. Although more
  data are needed for confirmation, it appears that time-delay induced
  variability has a modest impact.  

Differences in the wings of the CIV and SiIV broad emission line
profiles are found in $A$ and $B$ images of SDSS1004+4112, as detected
previously by \cite{richards04}, but the enhancement in the blue wing
is smaller than observed in 2004.  In HE1104-1805 we also have 
detected a slight enhancement in the red wings of CIV and SiIV in 
image $A$ with respect to image $B$.

The average microlensing magnification free from extinction was
obtained as the difference in magnitudes between the emission lines
and the continuum. The latter was obtained directly from the spectra
in those cases where there is no contamination by the lens galaxy (all
systems except HS0818+1227 and Q0957+561), otherwise we used 
HST continuum
data free from lens galaxy contamination available in the literature. 
Significant chromatic microlensing
was detected in SDSS1004+4112, SDSS1029+2623, and HE1104-1805.

Below is a summary of the results for each system:

\begin{enumerate}

\item We detected a blue wing enhancement in the high ionization lines
  of SDSS1004+4112 that are qualitatively similar to the effect
  described by previous authors but of smaller amplitude. We have also
  detected strong chromatic variability in the continuum. The presence
  of variability in both lines and continuum supports the hypothesis
  of microlensing to explain the blue wing enhancements in the lines.
  Our data indicate negligible dust extinction.  We infer an accretion
  disk size of $r_s=7\pm 3 \, \rm light days = 18.1\pm 7.8 \times 10^{15} \, \rm cm$ 
 at  $\lambda_{rest}=3363$\AA\ and a wavelength dependence of the size
  with exponent $p=1.1\pm0.4$.

\item In HE1104-1805 we find no extinction but we detected
  chromatic microlensing with large variations between two epochs that
  change the sign of the continuum slope.  We estimate 
$r_s=6\pm 2 \, \rm light days = 15.5\pm 5.2 \times 10^{15} \, \rm cm$ 
 at $\lambda_{rest}=3363$\AA\ and $p=0.7\pm0.1$. This size
  is greater than those inferred from the thin disk theory and either
  the intrinsic flux or the central black hole mass by
  \cite{poindexter08}.
 However, our results agree with 
recent values presented in \cite{munoz11} 
$r_s(\lambda3363)=7\pm4 \, \rm light-days$ and $p=1.1\pm0.6$
The value of $p$ we obtain is significantly smaller than
  the 4/3 value predicted by the standard model. Similar
  discrepancies have been found in other objects
  \citep{poindexter08,floyd09,morgan10,blackburne11,mediavilla11,munoz11}.

\item SDSS1029+2623 is affected by strong chromaticity of about 0.4
  mag that can be explained neither by dust extinction nor by
  intrinsic variability.  Chromatic microlensing is the most probable
  explanation although not all the available data fit well within this
  hypothesis.

\item A \cite{cardelli89} extinction law with 
$\Delta E(B-V)=0.02\pm0.09$ and $R_V=2.0\pm0.1$ 
($\chi ^2_{DOF}=1.8$) 
  can be used to fit, within uncertainties, both the continuum and the
  emission line flux ratios in Q0957+561. There is a global offset
  between the continuum and emission line $B-A$ magnitude differences
  that implies a microlensing amplitude of $<B-A>\sim -0.4$~mag.  There
  are marginal indications of chromatic microlensing.

\item We detect evidence of microlensing but not extinction (within
  uncertainties) in HS0818+1227.

\end{enumerate}

\acknowledgments

We thank the anonymous referee for thoughtful suggestions.  We thank
G.T. Richards for kindly providing us the Keck spectra of
SDSS1004+4112, and N. Inada for kindly confirming us the infrared
measurements for SDSS1029+2623.  V.M. gratefully acknowledges support
from FONDECYT through grants 1090673 and 1120741.  E.M. and J.A.M are
supported by the Spanish Ministerio de Educaci\'{o}n y Ciencias
through the grants AYA2007-67342-C03-01/03 and AYA2010-21741-C03/02.
J.A.M. is also supported by the Generalitat Valenciana with the grant
PROMETEO/2009/64.  This research has made use of NASA's Astrophysics
Data System.



{\it Facilities:} \facility{MMT (Blue-Channel)}, \facility{HST (STIS)}, 
\facility{VLT (FORS2)}.

\clearpage


\clearpage

\clearpage

\clearpage

\clearpage

\clearpage

\clearpage

\clearpage

\clearpage

\clearpage

\clearpage

\clearpage

\clearpage

\clearpage

\clearpage

\clearpage

\clearpage

\clearpage

\clearpage
\begin{deluxetable}{lccccccccr}
\tabletypesize{\scriptsize}
\tablecolumns{10} 
\tablewidth{0pt}
\tablecaption{Log of observations \label{obs}}
\tablehead{
  \colhead{Objects} &
  \colhead{Pair\tablenotemark{a}} &
  \colhead{$\Delta$\tablenotemark{b} (\arcsec)} &
  \colhead{Instrument} &
  \colhead{Grating} &
  \colhead{Date} &
  \colhead{Airmass} &
  \colhead{P.A.\tablenotemark{c}} &
  \colhead{Seeing\tablenotemark{d}} &
  \colhead{Exposure\tablenotemark{e}}
}
\startdata
HS0818+1227   & AB & 2.6  & MMT/Blue-Channel & 300   & 2008/01/12 & 1.168 & -36.10 & 0.59 & 1800  \\
Q0957+561     & AB & 6.2  & MMT/Blue-Channel & 300   & 2008/01/12 & 1.096 & 168.51 & 0.61 & 900   \\
              & A  & \nodata & HST/STIS     & G230L & 1999/04/15 & \nodata & \nodata & \nodata & 1900  \\
              & B  & \nodata & HST/STIS     & G230L & 2000/06/02 & \nodata & \nodata & \nodata & 1900  \\
              & A  & \nodata & HST/STIS     & G430L & 1999/04/15 & \nodata & \nodata & \nodata & 900   \\
              & B  & \nodata & HST/STIS     & G430L & 2000/06/03 & \nodata & \nodata & \nodata & 900   \\
              & A  & \nodata & HST/STIS     & G750L & 1999/04/15 & \nodata & \nodata & \nodata & 660   \\
              & B  & \nodata & HST/STIS     & G750L & 2000/06/03 & \nodata & \nodata & \nodata & 660   \\
SBSS1004+4112 & AB & 3.8  & MMT/Blue-Channel & 300   & 2008/01/12 & 1.028 & 200.40 & 0.61 & $2\times 900$   \\
SBSS1029+2623 & AB & 22.6 & MMT/Blue-Channel & 300   & 2008/01/11 & 1.072 & 11.12  & 0.67 & 1800  \\
HE1104-1805   & AB & 3.2  & MMT/Blue-Channel & 300   & 2008/01/11 & 1.766 & 114.66 & 0.67 & 1000  \\
              & AB & 3.2  & VLT/FORS2       & 300V  & 2008/04/07 & 1.315 & 64.18  & 0.60 & $3\times 250$  \\
\enddata
\tablenotetext{a}{Pair or image observed}
\tablenotetext{b}{Separation between images in arcsec}
\tablenotetext{c}{Position angle in degrees E of N}
\tablenotetext{d}{Seeing in arcsec}
\tablenotetext{e}{Seconds of time}
\end{deluxetable}

\clearpage
\begin{deluxetable}{lcccccr}
\tablecaption{Summary of Known Quasar Image Properties \label{lit}}
\tablewidth{0pt} 
\tablehead{ 
  \colhead{Lens Name} &
  \colhead{z$_L$\tablenotemark{a}} &
  \colhead{z$_S$\tablenotemark{b}} &
  \colhead{Filter\tablenotemark{c}} &
  \colhead{$1/\lambda$  \tablenotemark{d} ($\mu$m$^{-1}$)} &
  \colhead{$\Delta m$ (mag) \tablenotemark{e}} &
  \colhead{Source \tablenotemark{f}}
}
\startdata
HS0818+1227   & 0.39 & 1.3115 & F160W        & 0.65 &  $2.17 \pm 0.03$    & 1 \\
              &      &        & F814W        & 1.23 &  $2.11 \pm 0.00$    & 1 \\
              &      &        & R            & 1.55 &  $1.2$              & 2 \\
              &      &        & (CIV)        & 1.57 &  $2.26$             & 2 \\
              &      &        & F555W        & 1.80 &  $2.07 \pm 0.22$    & 1 \\
Q0957+561     & 0.36 & 1.41   & radio        & 0.00 &  $0.32 \pm 0.02$    & 3 \\
	      &      &        & radio        & 0.00 &  $0.4 \pm 0.2$      & 22 \\
	      &      &        & radio        & 0.00 &  $0.48 \pm 0.03$    & 23 \\
	      &      &        & F160W        & 0.65 &  $0.08 \pm 0.01$    & 1 \\
	      &      &        & F814W        & 1.23 &  $0.02 \pm 0.02$    & 1, 4 \\
              &      &        & (MgII)       & 1.49 &  $0.48 \pm 0.04$    & 5 \\
              &      &        & (MgII)       & 1.49 &  $0.31 \pm 0.02$    & 6,7 \\
              &      &        & (MgII)       & 1.49 &  $0.22$             & 8 \\
              &      &        & R            & 1.55 &  $-0.022 \pm 0.013$ & 9 \\
	      &      &        & F555W        & 1.80 &  $0.01 \pm 0.02$    & 1, 4 \\
              &      &        & V            & 1.83 &  $-0.077 \pm 0.023$ & 9 \\
              &      &        & (CIII])      & 2.17 &  $0.27 \pm 0.03$    & 5 \\
              &      &        & (CIV)        & 2.68 &  $0.28 \pm 0.02$    & 5 \\
              &      &        & (NV)         & 3.34 &  $0.43 \pm 0.07$    & 5 \\
              &      &        & (Ly$\alpha$) & 3.40 &  $0.15 \pm 0.03$    & 5 \\
              &      &        & F284M        & 3.52 &  $0.06 \pm 0.04$    & 10 \\
              &      &        & F277M        & 3.61 &  $0.02 \pm 0.03$    & 10 \\
              &      &        & F248M        & 4.03 &  $0.06 \pm 0.04$    & 10 \\
              &      &        & F140LP       & 4.55 &  $0.00 \pm 0.06$    & 10 \\
SBSS1004+4112 & 0.68 & 1.734  & F160W        & 0.65 &  $-0.47 \pm 0.04$   & 1 \\
              &      &        & z            & 1.10 &  $-0.45 \pm 0.09$   & 11 \\
              &      &        & z            & 1.10 &  $-0.45 \pm 0.06$   & 12 \\
              &      &        & F814W        & 1.23 &  $-0.31 \pm 0.15$   & 1 \\
              &      &        & F814W        & 1.23 &  $-0.34 \pm 0.12$   & 13 \\
              &      &        & i            & 1.30 &  $-0.40 \pm 0.06$   & 11 \\
              &      &        & i            & 1.30 &  $-0.40 \pm 0.06$   & 12 \\
              &      &        & r            & 1.60 &  $-0.460 \pm 0.005$ & 14 \\
              &      &        & r            & 1.60 &  $-0.283 \pm 0.007$ & 14 \\
              &      &        & r            & 1.60 &  $-0.339 \pm 0.005$ & 14 \\
              &      &        & r            & 1.60 &  $-0.381 \pm 0.007$ & 14 \\
              &      &        & r            & 1.60 &  $-0.39 \pm 0.08$   & 11 \\
              &      &        & r            & 1.60 &  $-0.39 \pm 0.06$   & 12 \\
              &      &        & F555W        & 1.80 &  $0.09 \pm 0.17$    & 1 \\
\hline
\tablebreak
              &      &        & g            & 2.08 &  $-0.37 \pm 0.08$   & 11 \\
              &      &        & g            & 2.08 &  $-0.37 \pm 0.06$   & 12 \\
              &      &        & u            & 2.84 &  $-0.40 \pm 0.08$   & 12 \\
SBSS1029+2623 & 0.55 & 2.197  & radio        & 0.0  &  $-0.272\pm0.013$   & 24 \\
              &      &        & z            & 1.10 &  $0.44 \pm 0.09$    & 15 \\
              &      &        & I            & 1.24 &  $-0.06 \pm 0.02$   & 16 \\
              &      &        & i            & 1.30 &  $0.02 \pm 0.04$    & 15 \\
              &      &        & R            & 1.52 &  $-0.04 \pm 0.02$   & 16 \\
              &      &        & R            & 1.52 &  $0.12 \pm 0.02$    & 16 \\
              &      &        & r            & 1.60 &  $0.03 \pm 0.01$    & 15 \\
              &      &        & V            & 1.82 &  $-0.05 \pm 0.02$   & 16 \\
              &      &        & g            & 2.08 &  $-0.01 \pm 0.01$   & 15 \\
              &      &        & g            & 2.08 &  $0.09 \pm 0.02$    & 16 \\
              &      &        & B            & 2.25 &  $-0.17 \pm 0.02$   & 16 \\
              &      &        & u            & 2.84 &  $-0.06 \pm 0.04$   & 15 \\
HE1104-1805   & 0.73 & 2.32   & IRAC 8.0$\mu$m & 0.13 &  $-1.15 \pm 0.01$  & 17 \\
              &      &        & IRAC 8.0$\mu$m & 0.13 &  $-1.11 \pm 0.01$  & 17 \\
              &      &        & IRAC 5.8$\mu$m & 0.17 &  $-1.13 \pm 0.02$  & 17 \\
              &      &        & IRAC 5.8$\mu$m & 0.17 &  $-1.09 \pm 0.02$  & 17 \\
              &      &        & IRAC 4.5$\mu$m & 0.22 &  $-1.37 \pm 0.04$  & 17 \\
              &      &        & IRAC 4.5$\mu$m & 0.22 &  $-1.13 \pm 0.03$  & 17 \\
              &      &        & IRAC 3.6$\mu$m & 0.28 &  $-1.44 \pm 0.02$  & 17 \\
              &      &        & IRAC 3.6$\mu$m & 0.28 &  $-1.15 \pm 0.02$  & 17 \\
              &      &        & K            & 0.45 &  $-1.35 \pm 0.11$    & 18 \\
	      &      &        & F160W        & 0.65 &  $-1.44 \pm 0.03$    & 1 \\
	      &      &        & F160W        & 0.65 &  $-1.47 \pm 0.03$    & 19,20 \\
              &      &        & J            & 0.77 &  $-1.53 \pm 0.08$    & 18 \\
              &      &        & J            & 0.80 &  $-1.45$             & 17 \tablenotemark{h}\\
              &      &        & J            & 0.80 &  $-1.38$             & 17 \tablenotemark{i}\\
	      &      &        & F814W        & 1.23 &  $-1.55 \pm 0.03$    & 1 \\
	      &      &        & F814W        & 1.23 &  $-1.61 \pm 0.02$    & 20 \\
              &      &        & F814W        & 1.23 &  $-1.63 \pm 0.06$    & 19 \\
              &      &        & I            & 1.27 &  $-1.25$             & 17 \tablenotemark{i}\\
              &      &        & R            & 1.55 &  $-1.35$             & 17 \tablenotemark{h}\\
              &      &        & R            & 1.55 &  $-1.23$             & 17 \tablenotemark{i}\\
              &      &        & F555W        & 1.80 &  $-1.78 \pm 0.03$    & 1 \\
              &      &        & F555W        & 1.80 &  $-1.76 \pm 0.03$    & 20 \\
              &      &        & F555W        & 1.80 &  $-1.82 \pm 0.05$    & 19 \\
\hline
\tablebreak
              &      &        & V            & 1.83 &  $-1.748 \pm 0.03$   & 21 \\
              &      &        & B            & 2.28 &  $-1.23$             & 17 \tablenotemark{h}\\
              &      &        & B            & 2.28 &  $-1.13$             & 17 \tablenotemark{i}\\
\enddata
\tablenotetext{a}{~Lens galaxy redshift}
\tablenotetext{b}{~Lensed quasar redshift}
\tablenotetext{c}{ Filter or, when available, the line emission flux 
between parentheses}
\tablenotetext{d}{ Inverse of the central wavelength (rest frame) 
Radio wavelengths are approximated as 0 $\mu$ m$^{-1}$ in our plots}
\tablenotetext{e}{ Magnitude difference of B-A. 
Except for SDSS1004+4112 and HE1104-1805 where we show A-B}
\tablenotetext{h}{ Magnitude difference with time delay correction}
\tablenotetext{i}{ Magnitude difference without time delay correction}
\tablenotetext{f}{ REFERENCES: 
(1) CASTLES; 
(2) \cite{hagen00}; 
(3) \cite{conner92}; 
(4) \cite{bernstein97}; 
(5) \cite{goicoechea05b}; 
(6) \cite{schild91}; 
(7) \cite{vanderriest93}; 
(8) Mediavilla (private comunication based on data taken in 1997); 
(9) \cite{goicoechea05a}; 
(10) \cite{dolan95}; 
(11) \cite{inada03}; 
(12) \cite{oguri04}, 
(13) \cite{inada05}; 
(14) \cite{fohlmeister08}; 
(15) \cite{inada06}; 
(16) \cite{oguri08}; 
(17) \cite{poindexter07};
(18) \cite{courbin98}; 
(19) \cite{lehar00};  
(20) \cite{falco99}; 
(21) \cite{schechter03}; 
(22) \cite{haschick81}; 
(23) \cite{gorenstein88}; 
(24) \cite{kratzer11}. 
}
\end{deluxetable}

\clearpage
\begin{deluxetable}{llcrr}
\tablecolumns{3}
\tablewidth{0pt} 
\tablecaption{SDSS1004+4112 magnitude differences \label{mag_sdss1004}}
\tablehead{ 
  \colhead{Region} &
  \colhead{$\lambda_c$ (\AA)} &
  \colhead{Window\tablenotemark{c} (\AA)} &
  \colhead{$m_A-m_B$\tablenotemark{a} (mag)} &
  \colhead{$m_A-m_B$\tablenotemark{b} (mag)} 
}
\startdata
Continuum & 3320     & 3100-3700  & $0.07\pm0.05$  & $0.03\pm0.08$ \\
	  & 3820     & 3600-4040  & $0.05\pm0.04$  & $0.01\pm0.05$ \\
	  & 4230     & 3970-4450  & $-0.05\pm0.03$ & $-0.12\pm0.03$ \\
          & 4500     & 4350-4750  & $-0.04\pm0.04$ & $-0.08\pm0.04$ \\
	  & 5215     & 4600-5550  & $-0.01\pm0.03$ & $-0.09\pm0.04$ \\
          & 6650     & 6400-6870  & $-0.27\pm0.03$ & $-0.28\pm0.04$ \\
          & 7680     & 7150-8100  & $-0.25\pm0.05$ & $-0.25\pm0.05$ \\
\hline		   
Line  & Ly$\alpha$1216     & 3310-3345  & $-0.45\pm0.05$ & $-0.54\pm0.08$ \\
      & SiIV$\lambda$1400  & 3790-3870  & $-0.56\pm0.04$ & $-0.57\pm0.05$ \\
      & CIV$\lambda$1549   & 4220-4260  & $-0.57\pm0.03$ & $-0.61\pm0.03$ \\
      & HeII$\lambda$1640  & 4470-4510  & $-0.66\pm0.04$ & $-0.53\pm0.04$ \\
      & CIII]$\lambda$1909 & 5210-5250  & $-0.38\pm0.03$ & $-0.41\pm0.04$ \\
      & CII$\lambda$2326   & 6630-6680  & $-0.52\pm0.03$ & $-0.48\pm0.04$ \\
      & MgII$\lambda$2800  & 7640-7760  & $-0.30\pm0.05$ & $-0.33\pm0.05$\\
\enddata
\tablenotetext{a}{Exposure 1} 
\tablenotetext{b}{Exposure 2} 
\tablenotetext{c}{Integration window.}
\end{deluxetable}

\clearpage
\begin{deluxetable}{rc}
\tablecolumns{3}
\tablewidth{0pt} 
\tablecaption{SDSS1004+4112 chromatic microlensing \label{map_sdss1004}}
\tablehead{ 
  \colhead{$\lambda_c$ (\AA)} &
  \colhead{$\Delta m_C - \Delta m_L$\tablenotemark{a} (mag)} 
}
\startdata
 3700      & $0.60\pm0.02$ \\
 6338      & $0.40\pm0.02$ \\
 12500     & $0.08\pm0.04$ \\
\enddata
\tablenotetext{a}{Difference between the magnitude difference in the continuum and in 
the emission lines $(m_B-m_A)_C-(m_B-m_A)_L$}
\end{deluxetable}

\clearpage
\begin{deluxetable}{llcrr}
\tablecolumns{3}
\tablewidth{0pt} 
\tablecaption{HE1104-1805 magnitude differences \label{mag_he1104}}
\tablehead{ 
  \colhead{Region} &
  \colhead{$\lambda_c$ (\AA)} &
  \colhead{Window\tablenotemark{c} (\AA)} &
  \colhead{$m_A-m_B$\tablenotemark{a} (mag)} &
  \colhead{$m_A-m_B$\tablenotemark{b} (mag)} 
} 
\startdata
Continuum & 4037          & 3800-4350 & $-1.10\pm0.04$ & \nodata \\
          & 4638          & 4400-4900 & $-1.13\pm0.01$ & $-1.11\pm0.02$ \\
          & 5143          & 4850-5400 & $-1.22\pm0.01$ & $-1.16\pm0.02$ \\
          & 6338          & 5600-6800 & $-1.15\pm0.02$ & $-1.24\pm0.01$ \\
          & 9293          & 8700-9600 & \nodata        & $-1.32\pm0.03$ \\ \hline		   
Line & Ly$\alpha$1216     & 4013-4050 & $-1.02\pm0.04$ & \nodata \\
     & SiIV$\lambda$1400  & 4600-4670 & $-1.12\pm0.01$ & $-1.12\pm0.02$ \\
     & CIV$\lambda$1549   & 5080-5160 & $-1.22\pm0.01$ & $-1.18\pm0.02$ \\
     & CIII]$\lambda$1909 & 6250-6360 & $-1.20\pm0.02$ & $-1.04\pm0.01$ \\
     & MgII$\lambda$2800  & 9240-9290 & \nodata        & $-1.21\pm0.03$ \\ 
\enddata
\tablenotetext{a}{MMT data}
\tablenotetext{b}{averaged VLT data}
\tablenotetext{c}{Integration window.}
\end{deluxetable}

\clearpage
\begin{deluxetable}{rc}
\tablecolumns{3}
\tablewidth{0pt} 
\tablecaption{HE1104-1805 chromatic microlensing \label{map_he1104}}
\tablehead{ 
  \colhead{$\lambda_c$ (\AA)} &
  \colhead{$\Delta m_C - \Delta m_L$\tablenotemark{a} (mag)} 
}
\startdata
 4380      & $-0.09\pm0.02$ \\
 6470      & $-0.21\pm0.02$  \\
 12500     & $-0.34\pm0.04$  \\
\hline
 5550      & $0.65\pm0.03$ \\
 8140      & $0.46\pm0.03$  \\
 15500     & $0.32\pm0.03$  \\
\hline
 3700      & $-0.15\pm0.02$ \\
 6338      & $0.16\pm0.02$  \\
 12500     & $0.31\pm0.02$  \\
\enddata
\tablenotetext{a}{Difference between the magnitude difference in the continuum and 
in the emission line core $(m_B-m_A)_C-(m_B-m_A)_L$. Measurements corresponding to our 
MMT/VLT data, average data from before 2003 (see text), and 
\cite{poindexter07} data corrected by time delay respectively.}
\end{deluxetable}

\clearpage
\begin{deluxetable}{lcccc}
\tablecolumns{3}
\tablewidth{0pt} 
\tablecaption{HE1104-1805 accretion disk parameters ($1\sigma$ error) \label{pdfs_he1104}}
\tablehead{ 
  \colhead{ } &
  \multicolumn{2}{c}{Linear} &
  \multicolumn{2}{c}{Logarithmic} \\
  \colhead{Data} &
  \colhead{$p$} &
  \colhead{$r_s \times 10^{15}$ (cm)} &
  \colhead{$p$} &
  \colhead{$r_s \times 10^{15}$ (cm)}
}
\startdata
This work  	    & $1.7\pm0.8$ & $ 28.5\pm10.4$ & $1.7\pm0.8$ & $23.3\pm5.2$ \\ 
Average Lit.  	    & $1.1\pm0.7$ & $ 23.3\pm10.4$ & $0.9\pm0.2$ & $10.4\pm5.2$ \\ 
\cite{poindexter07} & $0.8\pm0.3$ & $ 23.3\pm 7.8$ & $1.1\pm0.6$ & $18.1\pm5.2$ \\ 
Intersection  	    & $0.6\pm0.1$ & $ 15.5\pm 5.2$ & $0.7\pm0.1$ & $15.5\pm2.6$ \\ 
\enddata
\end{deluxetable}

\clearpage
\begin{deluxetable}{llcr}
\tablecolumns{3}
\tablewidth{0pt} 
\tablecaption{SDSS1029+2623 magnitude differences \label{mag_sdss1029}}
\tablehead{ 
  \colhead{Region} &
  \colhead{$\lambda_c$ (\AA)} &
  \colhead{Window\tablenotemark{a} (\AA)} &
  \colhead{$m_B-m_A$ (mag)} 
}
\startdata
Continuum & 3888      & 3650-4450 & $-0.07\pm0.04$ \\
          & 4466      & 4250-4670 & $-0.01\pm0.02$ \\
          & 4952      & 4700-5150 &  $0.05\pm0.02$ \\
          & 6103      & 5750-6280 &  $0.10\pm0.01$ \\
\hline		   
Line & Ly$\alpha$1216     & 3890-3905 & $-0.39\pm0.04$ \\
     & SiIV$\lambda$1400  & 4450-4470 & $-0.40\pm0.02$ \\
     & CIV$\lambda$1549   & 4965-4975 & $-0.33\pm0.02$ \\
     & CIII]$\lambda$1909 & 6090-6115 & $-0.03\pm0.02$ \\
\enddata
\tablenotetext{a}{Integration window.}
\end{deluxetable}

\clearpage
\begin{deluxetable}{llcrr}
\tablecolumns{4}
\tablewidth{0pt} 
\tablecaption{Q0957+561 magnitude differences \label{mag_q0957}}
\tablehead{ 
  \colhead{Region} &
  \colhead{$\lambda_c$ (\AA)} &
  \colhead{Window\tablenotemark{c} (\AA)} &
  \colhead{$m_B-m_A$\tablenotemark{a} (mag)} &
  \colhead{$m_B-m_A$\tablenotemark{b} (mag)} 
} 
\startdata
Continuum & 2487        & 2400-2580 & \nodata         & $-0.15\pm0.05$ \\
	  & 2930        & 2750-3100 & \nodata         & $-0.20\pm0.06$ \\
	  & 2990        & 2750-3100 & \nodata         & $-0.20\pm0.06$ \\
	  & 3370        & 3210-3600 & $-0.39\pm0.02$  & $-0.12\pm0.04$ \\
	  & 3730        & 3500-3890 & $-0.35\pm0.01$  & $-0.14\pm0.03$ \\
	  & 3950        & 3850-4130 & $-0.33\pm0.01$  & $-0.13\pm0.03$ \\
	  & 4600        & 4270-4850 & $-0.30\pm0.006$ & $-0.10\pm0.01$ \\
	  & 6760        & 6230-7000 & $-0.23\pm0.01$  & $-0.01\pm0.03$ \\
\hline	   
Line  & OVI$\lambda$1032   & 2490-2515 & \nodata        & $0.32\pm0.05$ \\
      & Ly$\alpha$1216     & 2920-2945 & \nodata        & $0.13\pm0.06$ \\
      & NV$\alpha$1240     & 2980-3000 & \nodata        & $0.30\pm0.06$ \\
      & SiIV$\lambda$1400  & 3355-3400 & $0.28\pm0.02$  & $0.17\pm0.04$ \\
      & CIV$\lambda$1549   & 3710-3755 & $0.38\pm0.01$  & $0.40\pm0.03$ \\
      & He$\lambda$1640    & 3935-3965 & $0.18\pm0.01$  & $0.20\pm0.03$ \\
      & CIII]$\lambda$1909 & 4560-4630 & $0.42\pm0.006$ & $0.19\pm0.01$ \\
      & MgII$\lambda$2800  & 6730-6785 & $0.30\pm0.01$  & $0.26\pm0.03$ \\
\enddata
\tablenotetext{a}{MMT data}
\tablenotetext{b}{HST data}
\tablenotetext{c}{Integration window.}
\end{deluxetable}

\clearpage
\begin{deluxetable}{llcr}
\tablecolumns{3}
\tablewidth{0pt} 
\tablecaption{HS0818+1227 magnitude differences \label{mag_hs0818}}
\tablehead{ 
  \colhead{Region} &
  \colhead{$\lambda_c$ (\AA)} &
  \colhead{Window\tablenotemark{a} (\AA)} &
  \colhead{$m_B-m_A$ (mag)} 
}
\startdata
Continuum & 4245      & 3800-4800 & $1.96\pm0.07$ \\
          & 5004      & 4700-5070 & $2.0\pm0.1$ \\
          & 6374      & 6000-6650 & $1.36\pm0.02$ \\
\hline		   
Line  & OVI$\lambda$1037   & 4220-4270 & $2.35\pm0.07$ \\
      & Ly$\alpha$1216     & 4985-5015 & $2.4\pm0.1$ \\
      & CIV$\lambda$1549   & 6340-6390 & $2.40\pm0.02$ \\
\enddata
\tablenotetext{a}{Integration window.}
\end{deluxetable}

\end{document}